\documentclass[]{aa}                 
\usepackage{amsmath,amsfonts,amsbsy,mathrsfs,amssymb}
\usepackage{graphicx}
\usepackage[T1]{fontenc}
\usepackage{lmodern}
\usepackage[utf8]{inputenc}
\usepackage[english]{babel}
\usepackage{lscape}
\usepackage{supertabular,booktabs}
\usepackage{subcaption}
\usepackage{float}
\usepackage[flushleft]{threeparttable}
\usepackage{array}
\usepackage{longtable}
\usepackage{rotating}
\usepackage{chngcntr}
\usepackage{tikz}
\usetikzlibrary{shapes,arrows}
\usepackage[pdftex=true, colorlinks=true, linkcolor=blue, citecolor=blue, pdfborder={0 0 0}, bookmarks=false, urlcolor=blue, breaklinks=true, pdftitle={The physical and chemical structure of Sagittarius B2,  VIIa. Dust and ionized gas contributions to the full molecular line survey of 47 hot cores}, pdfsubject={}, pdfauthor={}, pdfkeywords={}]{hyperref}

\usepackage{array}
\newsavebox\dummy
\newcolumntype{H}{>{\begin{lrbox}{\dummy}}c<{\end{lrbox}}@{}}

\usepackage{natbib,twoopt}
\bibpunct{(}{)}{;}{a}{}{,}          

\makeatletter
\newcommandtwoopt{\citeads}[3][][]{\href{http://adsabs.harvard.edu/abs/#3}%
{\def\hyper@linkstart##1##2{}%
\let\hyper@linkend\@empty\citealp[#1][#2]{#3}}}
\newcommandtwoopt{\citepads}[3][][]{\href{http://adsabs.harvard.edu/abs/#3}%
{\def\hyper@linkstart##1##2{}%
\let\hyper@linkend\@empty\citep[#1][#2]{#3}}}
\newcommandtwoopt{\citetads}[3][][]{\href{http://adsabs.harvard.edu/abs/#3}%
{\def\hyper@linkstart##1##2{}%
\let\hyper@linkend\@empty\citet[#1][#2]{#3}}}
\newcommandtwoopt{\citeyearads}[3][][]%
{\href{http://adsabs.harvard.edu/abs/#3}
{\def\hyper@linkstart##1##2{}%
\let\hyper@linkend\@empty\citeyear[#1][#2]{#3}}}
\makeatother

\newcolumntype{C}[1]{>{\centering\arraybackslash}p{#1}}
\newcolumntype{R}[1]{>{\flushright\arraybackslash}p{#1}}

\newcommand{\hii}     {\hbox{H\,{\sc ii}}}

\begin{document}
    \title{The physical and chemical structure of Sagittarius B2\\
           VIIa. Dust and ionized gas contributions to the full molecular line survey of 47 hot cores}


    \author{T.~M\"{o}ller\inst{1}
            \and
            P.~Schilke\inst{1}
            \and
            \'{A}.~S\'{a}nchez-Monge\inst{1,2,3}
            \and
            A.~Schmiedeke\inst{1,4}
            \and
            F.~Meng\inst{1,5}
            }

    \institute{I. Physikalisches Institut, Universit\"{a}t zu K\"{o}ln,
               Z\"{u}lpicher Str. 77, D-50937 K\"{o}ln, Germany\\
               \email{moeller@ph1.uni-koeln.de}
               \and
               Institut de Ci\`encies de l'Espai (ICE, CSIC), Can Magrans s/n, E-08193, Bellaterra, Barcelona, Spain
               \and
               Institut d'Estudis Espacials de Catalunya (IEEC), Barcelona, Spain
               \and
               Green Bank Observatory, 155 Observatory Rd,
               Green Bank, WV 24944 (USA)
               \and
               University of Chinese Academy of Sciences, Beijing 100049, People’s Republic of China
              }

    \date{Received May 01, 2023 / Accepted May 31, 2023}

    \abstract
    {Sagittarius~B2 (Sgr~B2) is a giant molecular cloud complex in the central molecular zone of our Galaxy hosting several sites of high-mass star-formation. The two main centers of activity are Sgr~B2(M) and Sgr~B2(N) containing 27 continuum sources in Sgr~B2(M) and 20 sources in Sgr~B2(N), respectively. Our analysis aims to be a comprehensive modelling of each core spectrum, where we take the complex interaction between molecular lines, dust attenuation, and free-free emission arising from \hii~regions into account. In this work, which is the first of two papers on the complete analysis, we determine the dust and, if \hii~regions are contained, the parameters of the free-free thermal emission of the ionized gas for each core and derive a self-consistent description of the continuum levels of each core.}
    {Using the high sensitivity of ALMA we seek to characterize the physical and chemical structure of these continuum sources and gain better insight into the star formation process within the cores.}
    {We used ALMA to perform an unbiased spectral line survey  of all 47 sources in the ALMA band 6 with a frequency coverage from 211 GHz to 275~GHz. In order to model the free-free continuum contribution of a specific core we fit the contained recombination lines to obtain the electron temperatures and the emission measures, where we use an extended XCLASS program to describe recombination lines and free-free continuum simultaneously. In contrast to previous analyses, we derived the corresponding parameters here not only for each core but also for their local surrounding envelope and determined their physical properties.}
    {The distribution of recombination lines we found in the core spectra fits well with the distribution of \hii~regions described in previous analyses. In Sgr~B2(M), the three inner sources are the most massive, whereas in Sgr~B2(N) the innermost core A01 dominates all other sources in mass and size. For the cores we determine average dust temperatures around 236~K (Sgr~B2(M)) and 225~K (Sgr~B2(N)), while the electronic temperatures are located in a range between 3800~K and 23800~K.}
    {The self-consistent description of the continuum levels and the quantitative description of the dust and free-free contributions form the basis for the further analysis of the chemical composition of the individual sources, which is continued in the next paper. This detailed modeling will give us a more complete picture of the star formation process in this exciting environment.}


    \titlerunning{Dust and ionized gas contributions to the full molecular line survey of 47 hot cores}
    \authorrunning{T.~M\"{o}ller \textit{et al.}}

    \maketitle

\section{Introduction}\label{sec:Introduction}

Sagittarius~B2 (Sgr~B2) is a giant molecular cloud complex in the central molecular zone (CMZ) of our Galaxy and hosts several sites of high-mass star-formation. Situated at a distance of $8.34 \pm 0.16$~kpc \citepads{2019A&A...625L..10G}, Sgr~B2 is one of the most massive molecular clouds in the Galaxy with a mass of 10$^7 \, M_\odot$ and H$_2$ densities of $10^3$ -- $10^5$~cm$^{-3}$ (\citetads{2016A&A...588A.143S}, \citetads{1995A&A...294..667H}, \citetads{1989ApJ...337..704L}).

The Sgr~B2 complex has a diameter of 36~pc \citepads{2016A&A...588A.143S} and contains two main sites of active high-mass star formation, Sgr~B2 Main (M) and North (N), which are separated by $\sim$48$\arcsec$ ($\sim$1.9~pc in projection). Both sites have comparable luminosities of $2 - 10 \times 10^6 \, L_\odot$, masses of $5 \times 10^4 \, M_\odot$ and sizes of $\sim$0.5~pc \citepads[see][]{2016A&A...588A.143S} and are surrounded by an envelope, which occupies an area of around 2~pc in radius. This envelope contains at least $\sim$70 high-mass stars with spectral types in the range from O5 to B0 (see e.g.\,  \citetads{1995ApJ...449..663G}, \citetads{2014ApJ...781L..36D}). All together is embedded in another envelope with a radius of 20~pc, which contains more than 99~\% of the total mass of Sgr~B2, although it has a much lower density ($n_{H_2} \sim 10^3$~cm$^{-3}$) and hydrogen column density (N$_H \sim 10^{23}$~cm$^{-2}$) compared to the inner envelope, whose density ($n_{H_2} \sim 10^5$~cm$^{-3}$) and hydrogen column density (N$_H \sim 10^{24}$~cm$^{-2}$) are significantly higher.

The greater number of \hii~regions and the higher degree of fragmentation observed in Sgr~B2(M) suggests a more evolved stage and a greater amount of feedback compared to Sgr~B2(N) (see e.g.\,  \citetads{1992ApJ...389..338G}, \citetads{2011A&A...530L...9Q}, \citetads{2017A&A...604A...6S}, \citetads{2019A&A...628A...6S}, \citetads{2019A&A...630A..73M}, \citetads{2022A&A...666A..31M}). Furthermore, M is very rich in sulfur-bearing molecules, while N is dominated by organic matter (\citetads {1991ApJS...77..255S}, \citetads{1998ApJS..117..427N}, \citetads{2004ApJ...600..234F}, \citetads{2013A&A...559A..47B}, \citetads{2014ApJ...789....8N}, \citetads{2021A&A...651A...9M}). Since both sites have large luminosities indicating ongoing high-mass star formation, the age difference between M and N is not very large, as shown by \citetads{1995ApJ...451..284D},  (\citeyearads{1996ApJ...464..788D}), who used the photon flux of the exciting stars and the ambient gas density to estimate an age of $\sim10^4$~yr for both regions.

High mass proto-clusters such as Sgr~B2 have complex, multi-layered structures that require an extensive analysis. Sgr~B2 provides a unique opportunity to study in detail the nearest counterpart of the extreme environments that dominate star formation in the Universe. The high density of molecular lines and the continuum emission detected toward the two main sites indicate the presence of a large amount of material to form new stars. Spectral line surveys give the possibility to obtain a census of all atoms and molecules and give insights into their thermal excitation conditions and dynamics by studying line intensities and profiles, which allows one to separate different physical components and to identify chemical patterns.

\begin{figure*}[ht!]
   \centering
   \includegraphics[width=0.92\textwidth]{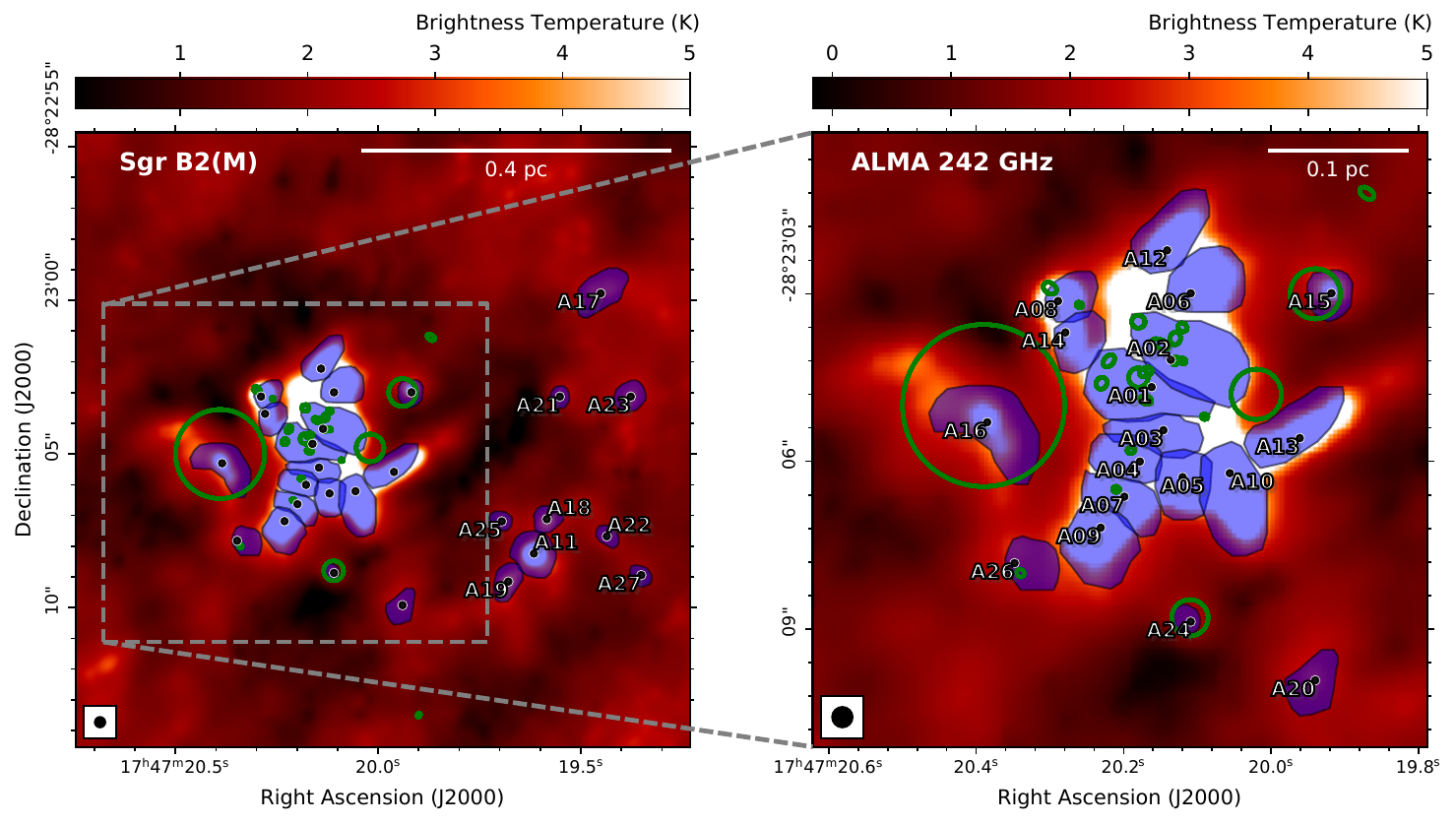}\\
   \caption{Continuum emission towards Sgr~B2(M) at 242~GHz. A close-up of the central part is presented in the right panel. The identified sources are marked with blue shaded polygons and indicated with the corresponding source ID. The black points indicate the position of each core, described by \citetads{2017A&A...604A...6S}. The intensity color scale is shown in units of brightness temperature, and the synthesized beam of $0 \farcs 4$ is described in the lower left corner. The green ellipses describe the \hii~regions identified by \citetads{2015ApJ...815..123D}.}
   \label{fig:CorePosSgrB2M}
\end{figure*}

\begin{figure*}[ht!]
   \centering
   \includegraphics[width=0.92\textwidth]{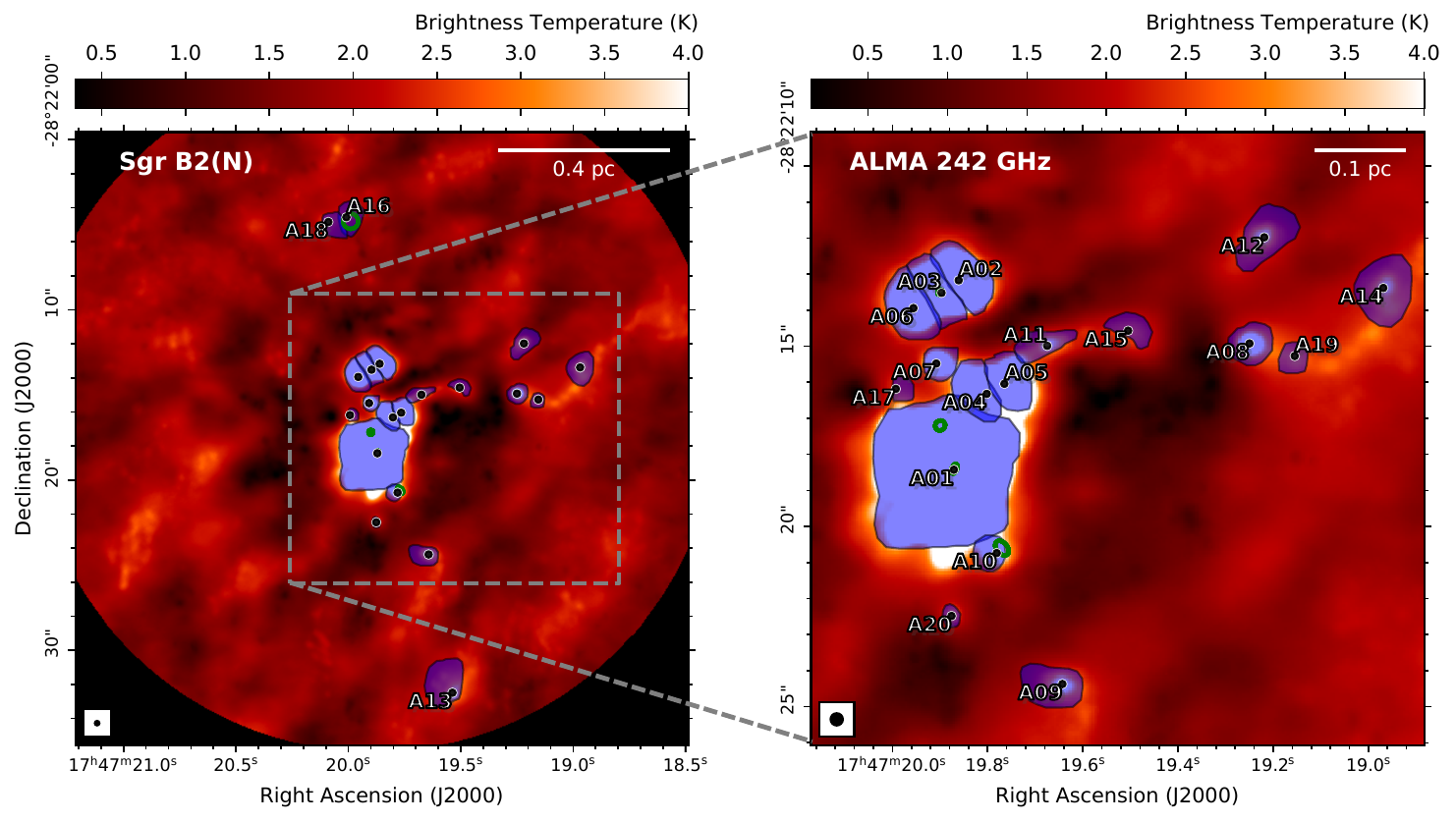}\\
   \caption{Continuum emission towards Sgr~B2(N) at 242~GHz. The right panel describes a close-up of the central part. The blue shaded polygons together with the corresponding source ID indicate the identified hot cores. The black points indicate the position of each core described by \citetads{2017A&A...604A...6S}. The intensity color scale is shown in units of brightness temperature, while the synthesized beam of $0 \farcs 4$ is indicated in the lower left corner. The green ellipses describe the \hii~regions identified by \citetads{2015ApJ...815..123D}.}
   \label{fig:CorePosSgrB2N}
\end{figure*}

Although the molecular content of Sgr~B2 was analyzed in many line surveys before (see e.g.\,  \citetads{1986ApJS...60..819C}, \citetads{1989ApJS...70..539T}, \citetads{1991ApJS...77..255S}, \citetads{1998ApJS..117..427N}, \citetads{2004ApJ...600..234F}, \citetads{2013A&A...559A..47B}, \citetads{2014ApJ...789....8N}, \citetads{2021A&A...651A...9M}), the high sensitivity of the Atacama Large Millimeter/submillimeter Array (ALMA) offers the possibility to gain better insight into the star formation process. Our analysis, which we will describe in the following, is a continuation of the paper by \citetads{2017A&A...604A...6S}, where 47 hot-cores in the continuum emission maps of Sgr~B2(M) and N were identified, see Figs.~\ref{fig:CorePosSgrB2M}~-~\ref{fig:CorePosSgrB2N}.

This paper is the first of two papers describing the full analysis of broadband spectral line surveys towards these 47 hot-cores to characterize the hot core population in Sgr~B2(M) and N. This analysis aims to be a comprehensive modelling of each core spectrum, where we take the complex interaction between molecular lines, dust attenuation, and free-free emission arising from \hii~regions into account.

As shown by \citetads{2017A&A...604A...6S}, many of the identified cores contain large amounts of dust. Additionally, some cores were associated with \hii~regions. However, \citetads{2017A&A...604A...6S} do not distinguish between the contributions from the cores and the envelope, which did not allow to isolate the core mass, especially for the weaker cores. In addition, the extinction due to dust and ionized gas must be determined properly to get reliable results for massive sources such as Sgr~B2(M) and N. In this paper we quantify the dust and, if present, the free-free contributions to the continuum by deriving the appropriate parameters for each core and the local surrounding envelope and determine the corresponding physical properties. Here, we obtain the dust temperatures from the results of the line surveys by assuming that the dust temperature equals the gas temperature following \citetads{1984A&A...130....5K} and \citetads{2001ApJ...557..736G}, who showed that gas and dust are thermally coupled at high densities ($n_{H_2} > 10^5$~cm$^{-3}$), which can be found at the inner parts of the Sgr~B2 complex.

In the second paper (Möller et al. in prep.), we describe the analysis of the molecular content of each hot core, where we identify the chemical composition of the detected sources and derive column densities and temperatures.

This paper is structured as follows: We start with Section~\ref{sec:DataRed}, where we describe the observations and outline the data reduction procedure, followed by Section~\ref{sec:DataAnalysis}, where we present the modeling methodology used to analyze the data set. Afterwards, our results are described and discussed in Section~\ref{sec:Results}. Finally, we present our conclusions in Section~\ref{sec:Conclusions}.

\section{Observations and data reduction}\label{sec:DataRed}

Sgr~B2 was observed with ALMA \citepads[Atacama Large Millimeter/submillimeter Array;][]{2015ApJ...808L...1A} during Cycle~2 in June~2014 and June~2015, using 34 -- 36 antennas in an extended configuration with baselines in the range from 30~m to 650~m, which results in an angular resolution of $0 \farcs 3 - 0 \farcs 7$ (corresponding to $\sim$3300~au). The observations were carried out in the spectral scan mode covering the whole ALMA band 6 (211 to 275~GHz) with 10 different spectral tunings, providing a resolution of 0.5 -- 0.7~km~s$^{-1}$ across the full frequency band. The two sources Sgr~B2(M) and Sgr~B2(N) were observed in track-sharing mode, with phase centers at $\alpha_{\rm J2000} = 17^{\rm h} 47^{\rm m} 20 \fs 157$, $\delta_{\rm J2000} = -28^\circ 23' 04 \farcs 53$ for Sgr~B2(M), and at $\alpha_{\rm J2000} = 17^{\rm h} 47^{\rm m} 19 \fs 887$, $\delta_{\rm J2000} = -28^\circ 22' 15 \farcs 76$ for Sgr~B2(N). Calibration and imaging were carried out with CASA\footnote{The Common Astronomy Software Applications \citepads[CASA,][]{2007ASPC..376..127M} is available at \url{https://casa.nrao.edu}.} version 4.4.0. Finally, all images were restored with a common Gaussian beam of $0 \farcs 4$. Details of the observations, calibration and imaging procedures are described in \citetads{2017A&A...604A...6S} and \citetads{2019A&A...628A...6S}.


\section{Data analysis}\label{sec:DataAnalysis}

The spectra of each hot core and the corresponding surrounding envelope were modeled using the eXtended CASA Line Analysis Software Suite \citepads[XCLASS\footnote{\url{https://xclass.astro.uni-koeln.de/}},][]{2017A&A...598A...7M} with additional extensions (M\"{o}ller in prep.). By solving the 1D radiative transfer equation assuming local thermal equilibrium (LTE) conditions and an isothermal source, XCLASS enables the modeling and fitting of molecular lines
\begin{align}\label{myXCLASS:modelFirstDist}
  T_{\rm mb}(\nu) = \sum_{m,c \in i} &\Bigg[\eta \left(\theta_{\rm source}^{m,c}\right) \left[S^{m,c}(\nu) \left(1 - e^{-\tau_{\rm total}^{m,c}(\nu)}\right)\right. \\
  &\left. \,\, + I_{\rm bg} (\nu) \left(e^{-\tau_{\rm total}^{m,c}(\nu)} - 1\right) \right] \Bigg]\nonumber\\
  & \,\, + \left(I_{\rm bg}(\nu) - J_\mathrm{CMB} \right),\nonumber
\end{align}
where the sums go over the indices $m$ for molecule, and $c$ for component, respectively. In Eq.~\eqref{myXCLASS:modelFirstDist}, $T_{\rm mb}(\nu)$ represents the intensity in Kelvin, $\eta(\theta^{m,c})$ the beam filling (dilution) factor, $S^{m,c}(\nu)$ the source function, see Eq.~\eqref{myxclass:LocalOverlap}, and $\tau_{\rm total}^{m,c}(\nu)$ the total optical depth of each molecule $m$ and component $c$. Additionally, $I_{\rm bg}$ indicates the background intensity and $J_\mathrm{CMB}$ the intensity of the cosmic microwave background. As Sgr~B2(M) and N have high H$_2$ densities ($n_{H_2} > 10^6$~cm$^{-3}$), LTE conditions can be assumed \citepads{2015PASP..127..266M} and the kinetic temperature of the gas can be estimated from the rotation temperature: $T_{\rm rot} \approx T_{\rm kin}$. For molecules, we assume Gaussian line profiles, whereas Voigt line profiles are used for radio recombination lines (RRLs), see Sect.~\ref{subsec:RRLs}. Additionally, finite source size, dust attenuation, and optical depth effects are taken into account as well.

All molecular parameters (e.g.\, transition frequencies, Einstein A coefficients) are taken from an embedded SQLite database containing entries from the Cologne Database for Molecular Spectroscopy (CDMS, \citetads{2001A&A...370L..49M}, \citetads{2005JMoSt.742..215M}) and Jet Propulsion Laboratory database (JPL, \citetads{1998JQSRT..60..883P}) using the Virtual Atomic and Molecular Data Center (VAMDC, \citetads{2016JMoSp.327...95E}). Additionally, the database used by XCLASS describes partition functions for more than 2500 molecules between 1.07~and 1000~K.

The contribution of each molecule is described by multiple emission and absorption components, where each component is specified by the source size $\theta_{\rm source}$, the rotation temperature $T_{\rm rot}$, the column density $N_{\rm tot}$, the line width~$\Delta v$, and the velocity offset ${\rm v}_{\rm offset}$ from the source velocity (${\rm v}_{\rm LSR}$). Moreover, XCLASS offers the possibility to locate each component at a certain distance $l$ along the line of sight. All model parameters can be fitted to observational data by using different optimization algorithms provided by the optimization package MAGIX \citepads{2013A&A...549A..21M}. In order to reduce the number of fit parameters, the modeling can be done simultaneously with corresponding isotopologues and vibrationally excited states. The ratio with respect to the main species can be either fixed or used as an additional fit parameter. Details are described in \citetads{2017A&A...598A...7M}.

\subsection{Recombination lines}\label{subsec:RRLs}

In addition to molecules, \textsc{XCLASS} can analyze radio recombination lines as well. According to \citetads{2006ApJ...653.1226Q}, who analyzed a large number of galactic \hii~regions, deviations from LTE are small, making LTE a reasonable assumption. Similar to molecules, the contribution of each RRL is described by a certain number of components, where each component is, in addition to source size $\theta_{\rm source}$ (in~arcsec) and distance $l$ (stacking parameter), defined by the electronic temperature T$_{\rm e}$ (in~K), the emission measure EM (in pc~cm$^{-6}$), the line width(s) $\Delta v$ (in km~s$^{-1}$), and the velocity offset ${\rm v}_{\rm offset}$ (in km~s$^{-1}$).

The optical depth of RRLs (in LTE) is given by \citepads{2002ASSL..282.....G}
\begin{align}\label{myxclass:RRLopticalDepthLine}
  \tau_{{\rm RRL}, \nu} &= \int \kappa_{n_1, n_2, \nu}^{\rm ext} \, ds \nonumber \\
  &= \frac{\pi \, h^3 \, e^2}{(2 \pi \, m_e \, k_B)^{3/2} \, m_e \, c} \cdot {\rm EM} \cdot \frac{n_1^2 \, f_{n_1, n_2}}{T_e^{3/2}} \nonumber \\
  & \quad \times \exp \left[\frac{Z^2 \, E_{n_1}}{k_B T_e} \right] \, \left(1 - e^{-h \nu_{n_1, n_2} / k_B T_e} \right) \, \phi_\nu.
\end{align}
Here, $f_{n_1, n_2}$ indicates the oscillator strength $\nu_{n_1, n_2}$, the transition frequency, $n_1$ the main quantum number, and $E_{n_1}$ the energy of the lower state, which are taken from the embedded database. For each RRL the database contains oscillator strengths up to $\Delta n = 6$ ($\zeta$-transitions), i.e.\ all transitions of the RRL within the given frequency range up to $\zeta$ transitions are taken into account. In Eq.~\eqref{myxclass:RRLopticalDepthLine}, the term $\phi_\nu$ represents the line profile function. \textsc{XCLASS} offers the possibility to use a Gaussian $G(x)$ or a Voigt $V(x, \sigma, \gamma)$ line profile function, which is a convolution of the Gaussian and the Lorentzian function, i.e.\ \citepads{2002ASSL..282.....G}
\begin{equation}
  V(x, \sigma, \gamma) = \int_{-\infty}^{\infty} G(x; \sigma) \, L(x - x', \gamma) \, dx'.
\end{equation}
Due to the fact that the computation of the Voigt profile is computationally quite expensive, \textsc{XCLASS} uses the pseudo-Voigt profile $\phi^{m,c,t}_{\rm pseudo-Voigt}(\nu)$ which is an approximation of the Voigt profile $V(x)$ using a linear combination of a Gaussian $G(x)$ and a Lorentzian line profile function $L(x)$ instead of their convolution. The mathematical definition of the normalized pseudo-Voigt profile, i.e.\ $\int_0^{\infty} \phi^{m,c,t}_{\rm pseudo-Voigt}(\nu) \, d\nu$ = 1, is given by
\begin{equation}
    \phi^{m,c,t}_{\rm pseudo-Voigt}(\nu) = \eta \cdot L (\nu, f) + (1 - \eta) \cdot G (\nu, f),
\end{equation}
with $0 < \eta < 1$. There are several possible choices for the $\eta$ parameter. \textsc{XCLASS} use the expression derived by \citet{Thompson:a27720},
\begin{equation}
    \eta = 1.36603 \, \left(\frac{f_L}{f}\right) - 0.47719 \, \left(\frac{f_L}{f}\right)^2 + 0.11116 \, \left(\frac{f_L}{f}\right)^3,
\end{equation}
which is accurate to 1~\%. Here,
\begin{align}
    f &= \left[f_G^5 + 2.69269 \, f_G^4 \, f_L + 2.42843 \, f_G^3 \, f_L^2 \right. \\
    &\qquad\quad  + 4.47163 \left. f_G^2 \, f_L^3 + 0.07842 \, f_G \, f_L^4 + f_L^5\right]^{1/5} \nonumber ,
\end{align}
indicates the total full width at half maximum (FWHM), where $f_L$ and $f_G$ represents the Lorentzian and Gaussian full width at half-maximum, respectively. The application of a Voigt line profile requires an additional parameter for each RRL and component: In addition to the Gaussian line width $\Delta v_G^{m,c}$, the Lorentzian line width $\Delta v_L^{m,c}$ (in km~s$^{-1}$) has to be specified as well. These line widths are related to the full width at half-maxima $f_L$ and $f_G$ by
\begin{align}\label{myXCLASS:FWHM}
  f_G &= \frac{\Delta v_G^{m,c}}{c_{\rm light}} \cdot \nu_t^{m,c} \cdot \left(1 - \frac{\left( {\rm v}_{\rm offset}^{m,c} + {\rm v}_{\rm LSR}\right)}{c_{\rm light}} \right), \\
  f_L &= \frac{\Delta v_L^{m,c}}{c_{\rm light}} \cdot \nu_t^{m,c} \cdot \left(1 - \frac{\left( {\rm v}_{\rm offset}^{m,c} + {\rm v}_{\rm LSR}\right)}{c_{\rm light}} \right), \nonumber
\end{align}
where $\nu_t^{m,c}$ indicates the transition frequency of RRL $m$, component $c$, and transition $t$, $ {\rm v}_{\rm offset}^{m,c}$ the velocity offset, and ${\rm v}_{\rm LSR}$ the source velocity, respectively.

%

\subsection{Free-free continuum}\label{subsec:freefree}

The hot plasma in \hii~regions gives rise to the emission of thermal bremsstrahlung, which causes a continuum opacity. The optical depth $\tau_{\rm ff}$ of this free-free contribution in terms of the classical electron radius $r_e = \frac{\alpha \, \hbar \, c}{m_e \, c^2} = \alpha^2 \, a_0$ is given by \citepads{2000A&A...356.1149B}


\clearpage
\begin{figure*}[!htb]
   \centering
   \includegraphics[width=0.85\textwidth]{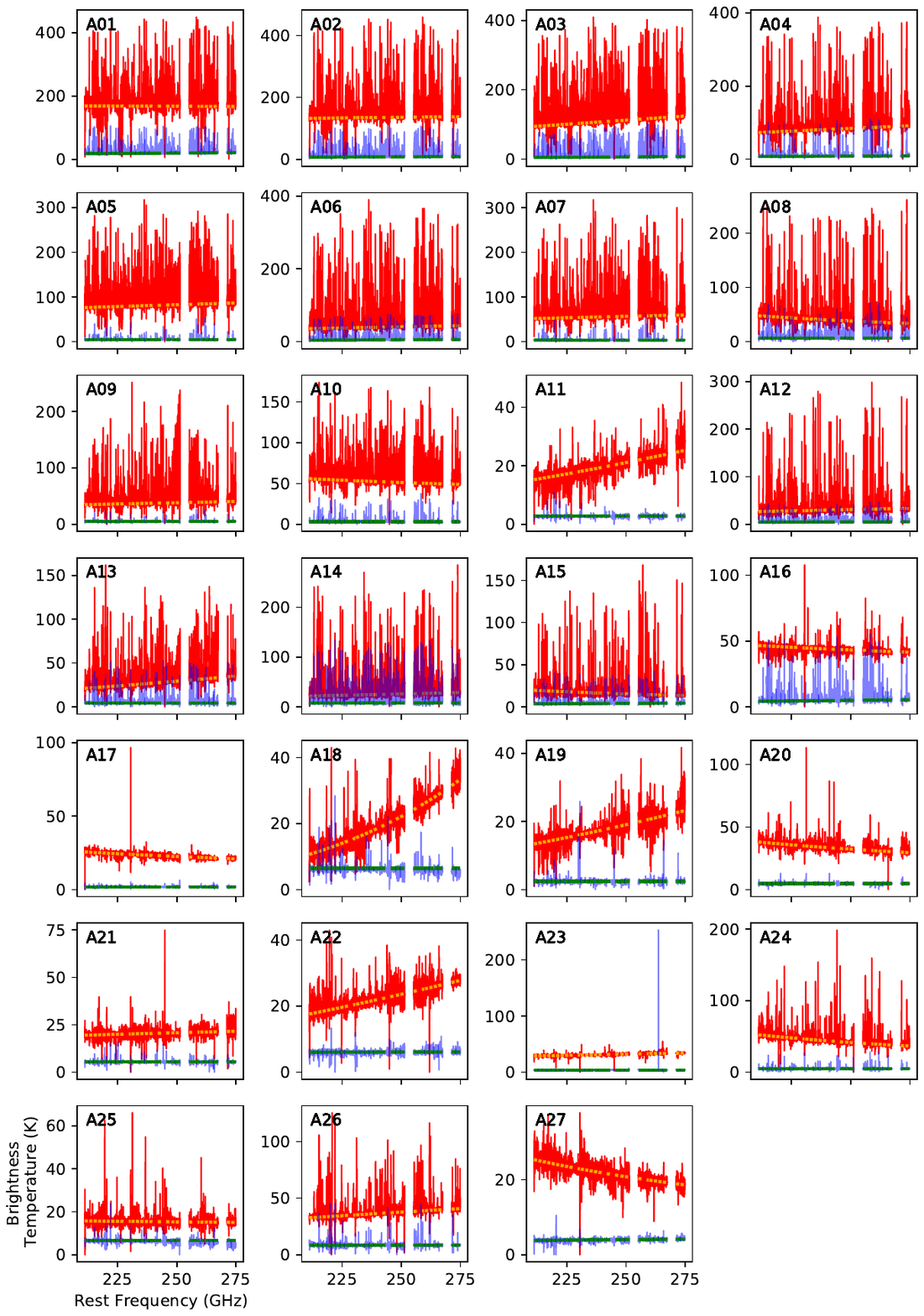}\\
   \caption{Core (red line) and envelope (blue line) spectra for each core toward Sgr~B2(M). The procedure used to calculate the envelope spectra is described in Sect.~\ref{subsusbsec:envspectra}. The continuum level for each core spectrum is indicated by the orange dotted line, while the continuum level of each envelope spectrum is described by the green dotted line.}
   \label{fig:SpectraSgrB2M}
\end{figure*}
\newpage

\clearpage
\begin{figure*}[!htb]
   \centering
   \includegraphics[width=0.85\textwidth]{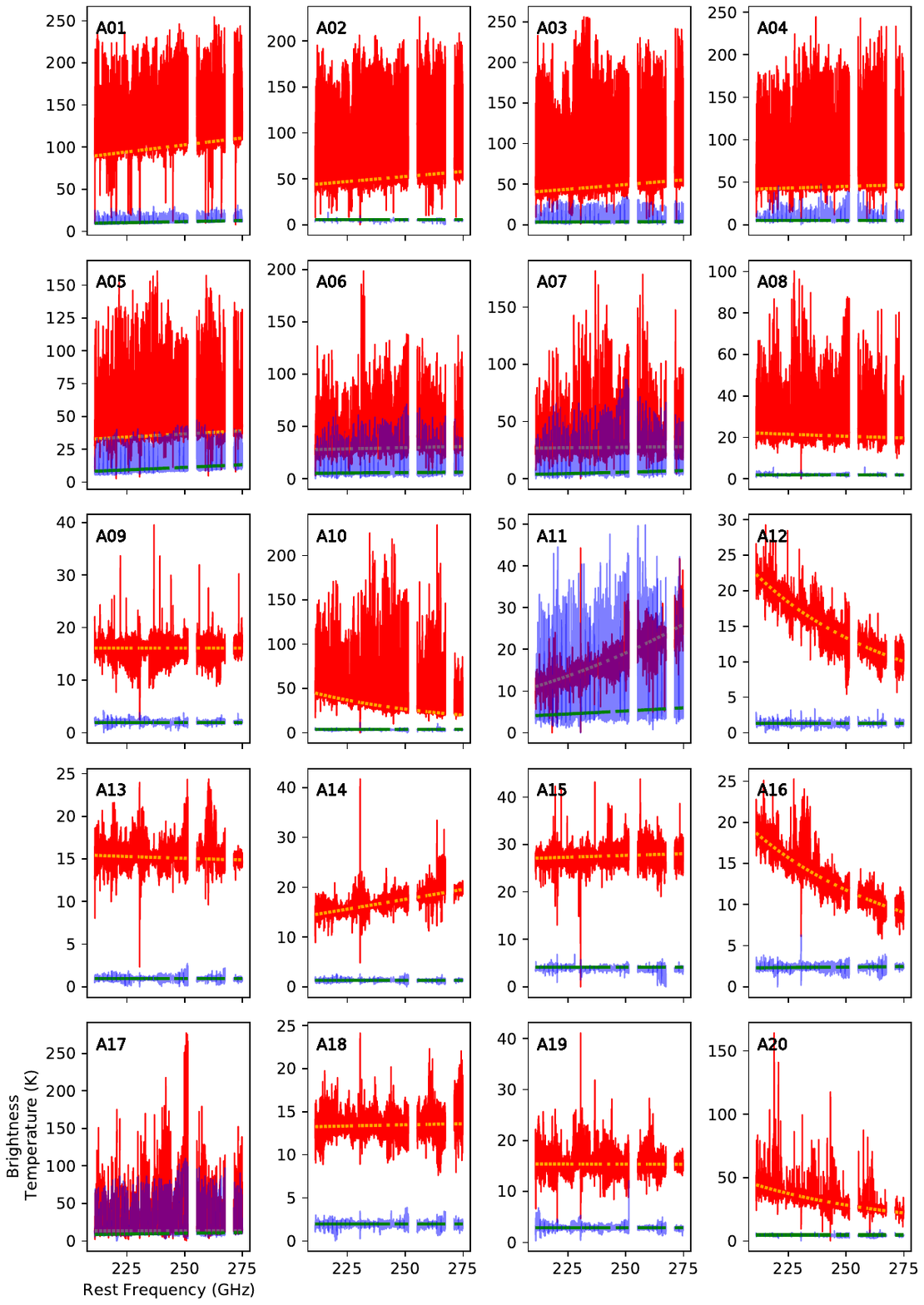}\\
   \caption{Core (red line) and envelope (blue line) spectra for each core toward Sgr~B2(N). The procedure used to calculate the envelope spectra is described in  Sect.~\ref{subsusbsec:envspectra}. The continuum level for each core spectrum is indicated by the orange dotted line, while the continuum level of each envelope spectrum is described by the green dotted line.}
   \label{fig:SpectraSgrB2N}
\end{figure*}
\newpage
\clearpage

\begin{figure*}[!htb]
    \footnotesize

    \begin{tikzpicture}[auto,
        line/.style = {draw, thick, -latex', shorten >= 2pt},
        lblock/.style = {rectangle, draw = black, thick, fill = white,
                         text width = 30em, align = left,
                         minimum height = 2em},
        cblock/.style = {rectangle, draw = black, thick, fill = white,
                         align = flush center,
                         minimum height = 2em},
        sblock/.style = {rectangle, draw = black, thick, fill = white,
                           text width = 19em, align = flush center,
                           minimum height = 2em},
        decision/.style = {diamond, draw = black, thick, fill = white,
                           text width = 4.5em, align = flush center,
                           inner sep=1pt}]

        \matrix [row sep = 0.6cm, column sep = 1cm, nodes = draw]
        {
            \node [lblock] (step1) {Perform a full molecular line survey analysis of each core spectrum using a phenomenological description of the corresponding continuum level and neglecting dust and free-free contributions. (If contained, fit RRLs as well).}; & \\
            \node [cblock] (step2) {Determine continuum parameters of each envelope spectrum}; & \\
            \node [cblock] (step3) {Compute envelope spectrum around each core, see Sect.~\ref{subsusbsec:envspectra}}; & \\
            \node [cblock] (step4) {Subtract continuum using STATCONT}; & \\
            \node [decision] (decide) {Contain RRLs?}; & \\
            \node [cblock] (norrl) {Determine excitation temperatures of selected molecules}; &
            \node [sblock] (rrl) {Fit RRLs in env. spectra to get free-free continuum parameters and compute excitation temperatures of molecules performing a full line survey analysis, see Sect.~\ref{subsusbsec:freefreeenv}}; \\
            \node [cblock] (step7) {Compute average gas temperature of each envelope spectrum}; & \\
            \node [cblock] (step8) {Fit continuum level to get N$_H$ and $\beta$, see Sect.~\ref{subsusbsec:dustenv}}; & \\
            \node [sblock] (step9) {Fit all molecule and RRL parameters together with continuum parameters}; & \\
            \node [cblock] (step10) {Re-compute average gas temperature of each envelope spectrum}; & \\
            \node [cblock] (step11) {Fit continuum level to get N$_H$ and $\beta$}; & \\
            \node [cblock] (step12) {Determine continuum parameters of each core spectrum, see Sect.~\ref{subsusbsec:contcore}}; & \\
            \node [sblock] (step13) {Compute average gas temperatures using excitation temperatures from the full line survey analysis of each core}; & \\
            \node [sblock] (step14) {Fit continuum levels to get N$_H$ and $\beta$ taking continuum parameters of envelope into account}; & \\
            \node [sblock] (step15) {Fit all molecule and RRL parameters together with core continuum parameters}; & \\
        };

        \begin{scope}[every path/.style=line]
            \path  (step1)  -- (step2);
            \path  (step2)  -- (step3);
            \path  (step3)  -- (step4);
            \path  (step4)  -- (decide);
            \path  (decide) -| node [near start] {Yes} (rrl);
            \path  (decide) -- node [midway] {No} (norrl);
            \path  (norrl)  -- (step7);
            \path  (rrl)    |- (step7);
            \path  (step7)  -- (step8);
            \path  (step8)  -- (step9);
            \path  (step9)  -- (step10);
            \path  (step10) -- (step11);
            \path  (step11) -- (step12);
            \path  (step12) -- (step13);
            \path  (step13) -- (step14);
            \path  (step14) -- (step15);
        \end{scope}
    \end{tikzpicture}
    \caption{Flow chart of fitting process.}
    \label{fig:FlowChart}
\end{figure*}
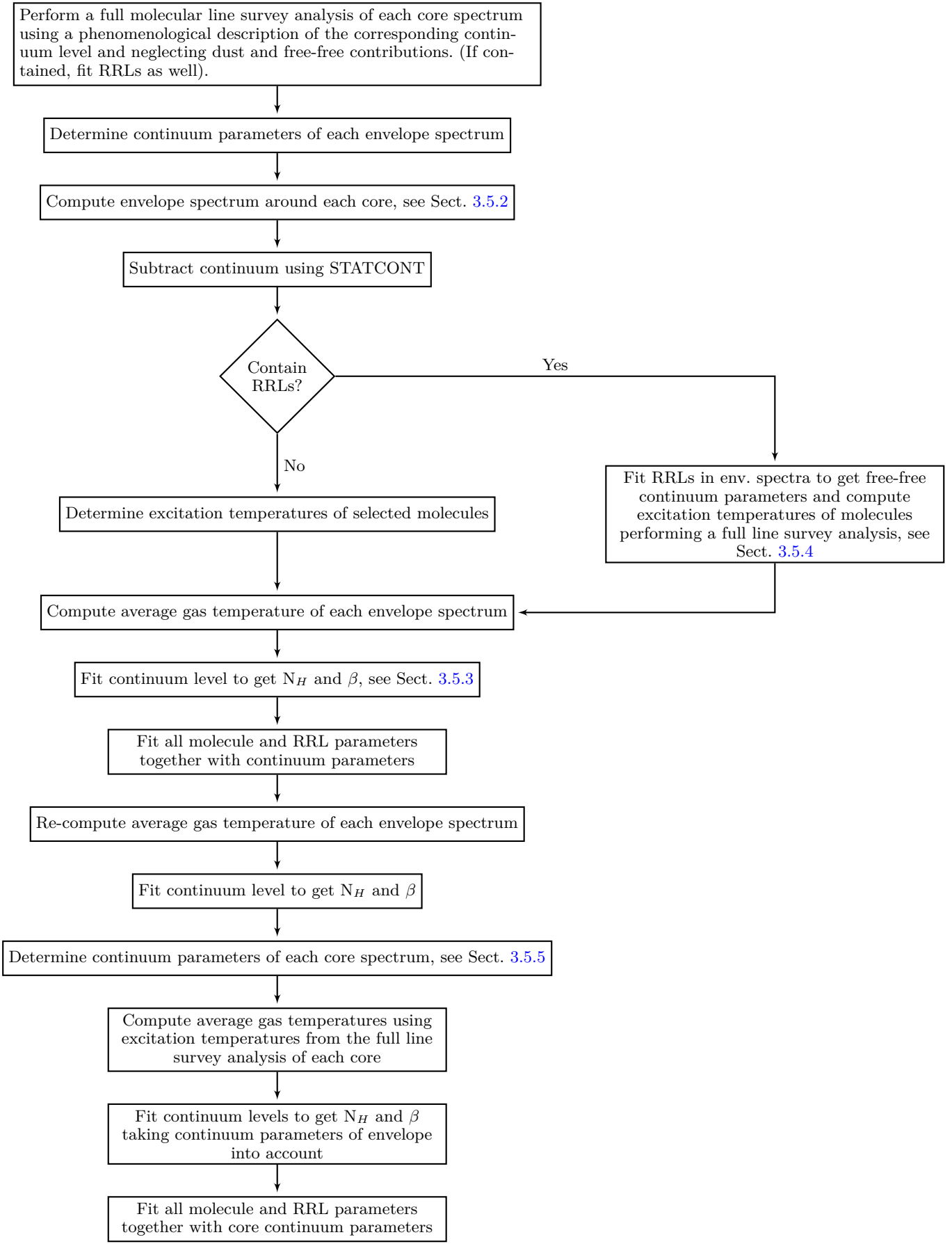

\begin{align}\label{myxclass:FreeFreeTau}
  \tau_{\rm ff} &= \frac{4}{3} \, \left(\frac{2 \pi}{3} \right)^{\frac{1}{2}} r_e^3 \, Z\frac{Z_i^2 \,  m^{3/2} \, c^5}{\sqrt{k_B \, T_e} \, h \, \nu^3} \, \left(1 - e^{-\frac{h \nu}{k_B \, T_e}} \right) \, \langle g_{\rm ff} \rangle \cdot {\rm EM} \nonumber\\
  &= 1.13725 \cdot \left(1 - e^{-\frac{h \nu}{k_B \, T_e}} \right) \cdot \langle g_{\rm ff} \rangle \nonumber\\
  & \qquad \qquad \,\, \times \left[ \frac{T_e}{\rm K}\right]^{-\frac{1}{2}} \, \left[ \frac{\nu}{\rm GHz}\right]^{-3} \, \left[ \frac{\rm EM}{{\rm pc} \, {\rm cm}^{-6}}\right],
\end{align}
where T$_{\rm e}$ indicates the electron temperature, EM the emission measure, and $\langle g_{\rm ff} \rangle$ the thermal averaged free-free Gaunt coefficient, respectively. XCLASS makes use of the tabulated thermal averaged free-free Gaunt coefficients $\langle g_{\rm ff} \rangle$ derived by \citetads{2015MNRAS.449.2112V}, which include relativistic effects as well. In order to model the free-free continuum contribution of a specific core we fit the corresponding RRLs to obtain the electron temperatures T$_{\rm e}$ and the emission measures EM, see Sect.~\ref{subsec:fitting}.

\subsection{Dust extinction}\label{subsec:dust}

Extinction from dust is very important in Sgr~B2 \citepads[see e.g.\, ][]{2021A&A...651A...9M}. Assuming, that dust and gas are well mixed, the dust opacity $\tau_d(\nu)$ used by XCLASS is described by
\begin{align}\label{myxclass:dustOpacity}
  \tau_d(\nu) &= \tau_{d, {\rm ref}} \cdot \left[\frac{\nu}{\nu_{\rm ref}} \right]^{\beta} \\ \nonumber
              &= \left[N_{\rm H} \cdot \kappa_{\nu_{\rm ref}} \cdot m_{{\rm H}_2} \cdot \frac{1}{\chi_{\rm gas-dust}}\right] \cdot \left[\frac{\nu}{\nu_{\rm ref}} \right]^{\beta},
\end{align}
where $N_{\rm H}$ indicates the hydrogen column density (in cm$^{-2}$), $\kappa_{\nu_{\rm ref}}$ the dust mass opacity for a certain type of dust \citepads[in cm$^{2}$~g$^{-1}$, ][]{1994A&A...291..943O}, and $\beta$ the spectral index\footnote{Note, that we use temperature units for the fitting. The spectral indices for flux units $\alpha$ is given by $\alpha = \beta + 2$.}. In addition, $\nu_{\rm ref}$ = 230~GHz represents the reference frequency for $\kappa_{\nu_{\rm ref}}$, $m_{{\rm H}_2}$ the mass of a hydrogen molecule, and $1 / \chi_{\rm gas-dust}$ the ratio of dust to gas, which is set here to $1/100$ \citepads{1983QJRAS..24..267H}.

\subsection{Local overlap}\label{subsec:localoverlap}

In line-crowded sources like Sgr~B2(M) and N line intensities from two neighbouring lines, which have central frequencies with (partly) overlapping width regions, do not simply add up if at least one line is optically thick. Here, photons emitted from one line are absorbed by the other line. XCLASS takes the local line overlap \citepads[described by][]{1991A&A...241..537C} from different components into account, by computing an average source function $S_l (\nu)$ at frequency $\nu$ and distance $l$
\begin{equation}\label{myxclass:LocalOverlap}
  S_l (\nu) = \frac{\varepsilon_l (\nu)}{\alpha_l (\nu)} = \frac{\sum_t \tau_t^c (\nu) \, S_\nu (T_{{\rm rot}}^c)}{\sum_t \tau_t^c (\nu)},
\end{equation}
where $\varepsilon_l$ describes the emission and $\alpha_l$ the absorption function, $T_{\rm rot}^c$ the excitation temperature, and $\tau_t^c$ the optical depth of transition $t$ and component $c$, respectively. Additionally, the optical depths of the individual lines included in Eq.~\eqref{myXCLASS:modelFirstDist} are replaced by their arithmetic mean at distance $l$, that is
\begin{equation}\label{myxclass:LocalOverlapTau}
  \tau_{\rm total}^{l}(\nu) = \sum_c \left[\left[ \sum_t \tau_t^c (\nu) \right] + \tau_d^{c}(\nu)\right],
\end{equation}
where the sums run over both components $c$ and transitions $t$. Here, $\tau_d^{c}(\nu)$ indicates the dust opacity, which is added as well. The iterative treatment of components at different distances, takes also non-local effects into account. Details of this procedure are described in \citetads{2021A&A...651A...9M}.

\begin{figure}[!htb]
   \centering
   \includegraphics[width=0.5\textwidth]{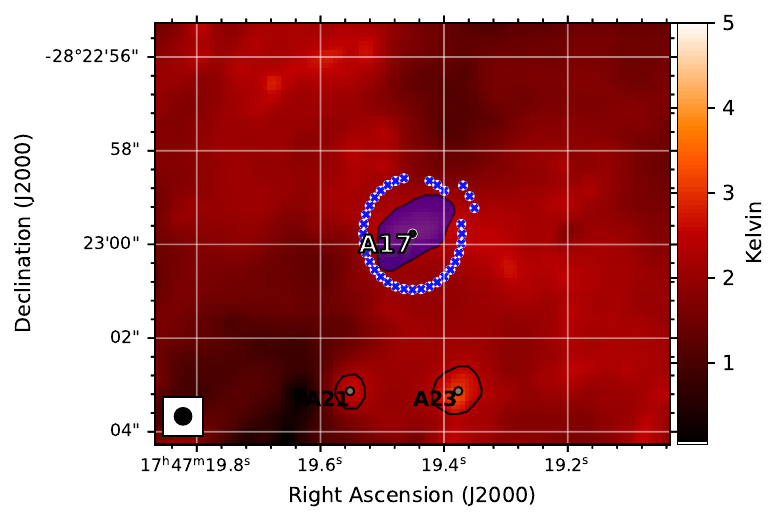}\\
   \caption{Continuum emission toward Sgr~B2(M) at 242~GHz around core A17 (blue shaded polygon). The positions used for the computation of the averaged envelope spectrum are indicated by blue crosses. The neighboring cores are described by black contours. The synthesized beam of $0 \farcs 4$ is described in the lower left corner.}
   \label{fig:MapSgrB2NA14}
\end{figure}

\subsection{Fitting procedure}\label{subsec:fitting}

\subsubsection{General model setup}\label{subsusbsec:setup}

In our analysis, we assume a two-layer model for all cores in Sgr~B2(M) and N, in which the first layer (hereafter called core-layer) describes contributions from the corresponding core. The second layer (envelope-layer) contains features from the local surrounding envelope of Sgr~B2 and is located in front of the core layer. Here, we assume that all components belonging to a layer, have the same distance to the observer. Following \citetads{2017A&A...604A...6S}, a core is identified within the continuum emission maps of Sgr~B2(M) and N, if at least one closed contour (polygon) above the 3$\sigma$ level is found (where $\sigma$ indicates the rms noise level of the map of 8~mJy~beam$^{-1}$ for Sgr~B2(M) and N, respectively), see Figs.~\ref{fig:CorePosSgrB2M}~-~\ref{fig:CorePosSgrB2N}. The spectra for each core, described in Figs.~\ref{fig:SpectraSgrB2M}~-~\ref{fig:SpectraSgrB2N}, are obtained by averaging over all pixels contained in a polygon to improve the signal-to-noise and detection of weak lines.\\

As shown in Figs.~\ref{fig:CorePosSgrB2M}~-~\ref{fig:CorePosSgrB2N}, extended structures are clearly visible in addition to the identified compact sources, i.e.\ the dust is not concentrated on the cores only, but a non-negligible contribution is also contained in the envelope. Here, we cannot distinguish between the contributions of the inner and outer envelope of Sgr~B2. However, the outer envelope will not contribute significantly due to its much lower density. To obtain a more or less consistent description of the continuum for each core spectrum, the dust parameters for each core and the local surrounding envelope have to be determined. In agreement with \citetads{2017A&A...604A...6S}, we assume a dust mass opacity of $\kappa_{1300 \, \mu{\rm m}} = 1.11$~cm$^{2}$~g$^{-1}$ \citepads[agglomerated grains with thin ice mantles in cores of densities 10$^8$ cm$^{-3}$;][]{1994A&A...291..943O} for both layers. For some hot cores, for which we have derived dust temperatures above 300~K, this may not be a good choice. Using a different dust mass opacity of, e.g.\ $\kappa_{1300 \, \mu{\rm m}} = 5.86$~cm$^{2}$~g$^{-1}$ (agglomerated grains without ice mantles in cores of densities 10$^8$ cm$^{-3}$) would result in hydrogen column densities lower by a factor of five.\\

In our analysis, see Fig.~\ref{fig:FlowChart}, we start with identifying molecules and, if contained, recombination lines in each core spectrum and determine a quantitative description of their respective contributions. Here we first use a phenomenological description of the corresponding continuum level and neglecting dust and free-free contributions.

\subsubsection{Envelope spectra}\label{subsusbsec:envspectra}

In the following, we determining the dust parameters of the envelope by selecting pixels around each core that are not too close to another core or \hii~region\footnote{Here we consider the \hii~regions described by \citetads{2015ApJ...815..123D}.}, see Fig.~\ref{fig:MapSgrB2NA14}. Here, we first selected points for each core that have the same distance to the core center, and then successively shifted outward those points that were still within the contour of the source or too close to an \hii~region. The spectra at these positions are used to compute an averaged envelope spectrum for each core, see Figs.~\ref{fig:SpectraSgrB2M}~-~\ref{fig:SpectraSgrB2N}, where we used STATCONT \citepads{2018A&A...609A.101S} to estimate the corresponding continuum level.

\begin{figure*}[!htb]
   \centering
   \includegraphics[width=0.80\textwidth]{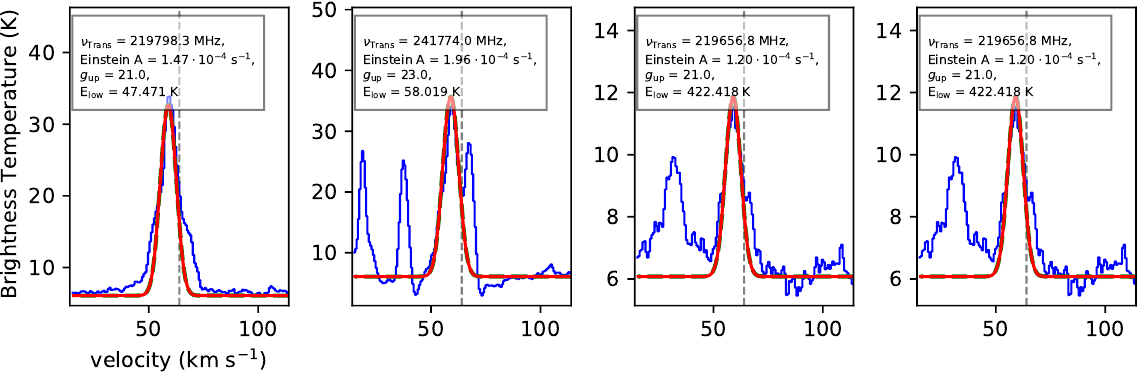}\\
   \caption{Selected transitions of HNCO (red lines) contained in the envelope spectrum (blue lines) around core A08 in Sgr~B2(M). The grey dashed line describes the source velocity v$_{\rm lsr}$~=~64~km~s$^{-1}$ of Sgr~B2.}
   \label{fig:EnvHNCOA08}
\end{figure*}

\subsubsection{Dust parameters of the envelope}\label{subsusbsec:dustenv}

In order to compute the dust parameters for each envelope, we assume that the gas temperature equals the dust temperature and estimate the gas temperature for each envelope by fitting CH$_3$CN, H$_2$CCO, H$_2$CO, H$_2$CS, HNCO, and SO in the corresponding spectra, see Fig.~\ref{fig:EnvHNCOA08}. These molecules were chosen because they show mostly isolated and non-blended transitions, and therefore, they can be used to derive temperatures without requiring a full line survey analysis of the entire envelope spectrum. The final dust temperature for each envelope is computed by averaging over the obtained excitation temperatures. The corresponding hydrogen column density~$N_{\rm H}$ and spectral index~$\beta$ are determined by fitting the continuum level of each envelope spectrum using XCLASS.

\begin{figure*}[!tb]
    \captionsetup[subfigure]{slc=off,margin={1cm,0cm}}
    \centering
    \begin{subfigure}[t]{0.88\textwidth}
       \subcaption{Sgr~B2(M), A07 (envelope):}
       \includegraphics[width=0.88\columnwidth]{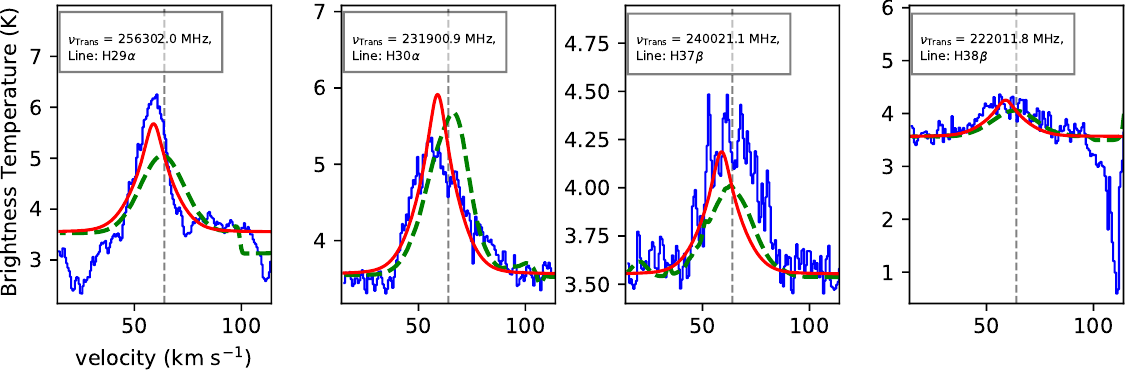}\\
        \label{fig:RRLEnv07}
    \end{subfigure}
\quad
    \begin{subfigure}[t]{0.88\textwidth}
       \subcaption{Sgr~B2(M), A09 (envelope):}
       \includegraphics[width=0.88\columnwidth]{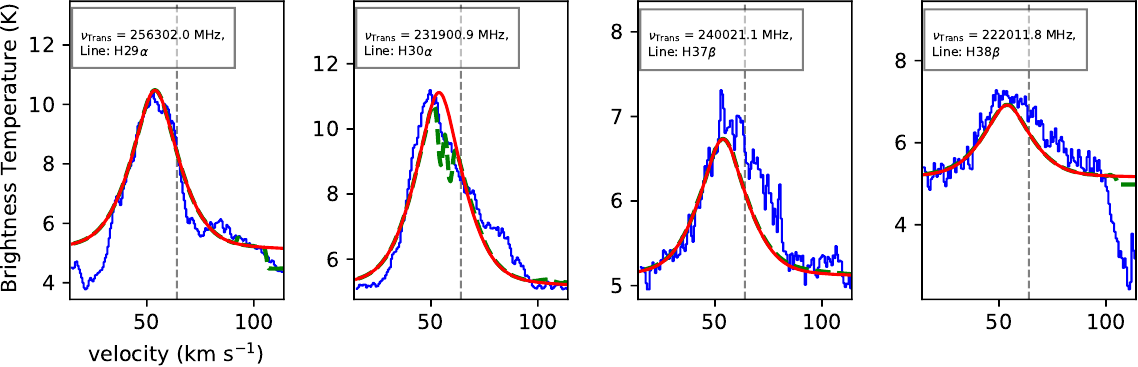}\\
        \label{fig:RRLEnv09}
    \end{subfigure}
\quad
    \begin{subfigure}[t]{0.88\textwidth}
       \subcaption{Sgr~B2(M), A24 (envelope):}
       \includegraphics[width=0.88\columnwidth]{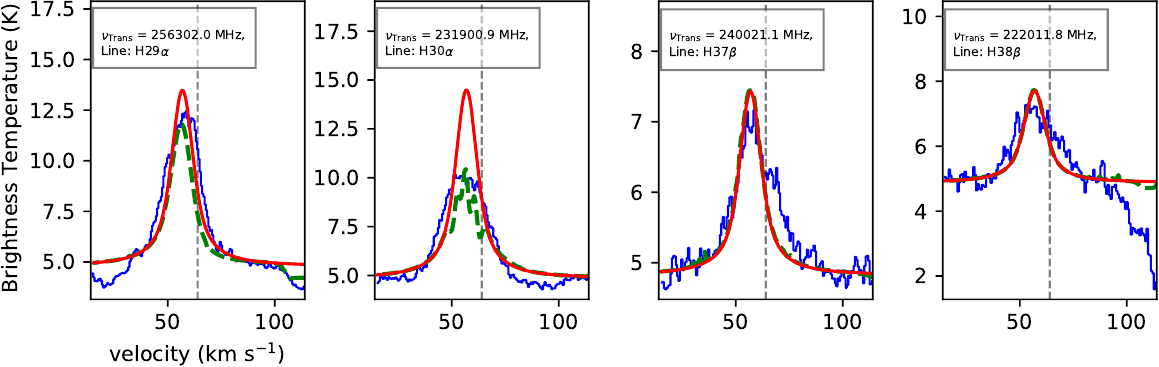}\\
        \label{fig:RRLEnv24}
    \end{subfigure}
\quad
    \begin{subfigure}[t]{0.88\textwidth}
       \subcaption{Sgr~B2(M), A26 (envelope):}
       \includegraphics[width=0.88\columnwidth]{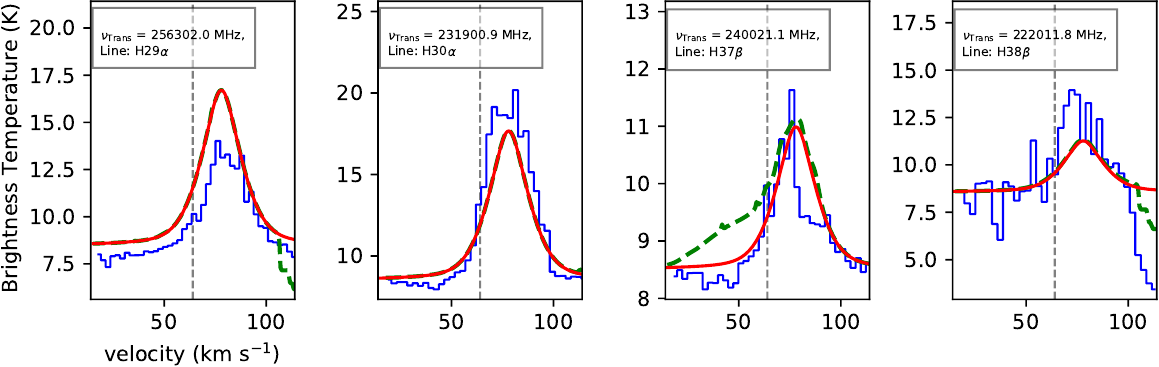}\\
        \label{fig:RRLEnv26}
    \end{subfigure}
   \caption{Selected RRLs in the envelope spectra around core A07~(a), A09~(b), A24~(c), and A26~(d) in Sgr~B2(M). The observational data is described by the blue line, the pure RRL contribution by the red line, and the model spectrum taking all identified molecules into account by the dashed green line. The vertical grey dashed line indicates the source velocity of Sgr~B2(M).}
   \ContinuedFloat
   \label{fig:RRLEnvSgrB2M}
\end{figure*}

\begin{figure*}[!tb]
    \captionsetup[subfigure]{slc=off,margin={1cm,0cm}}
    \centering
    \begin{subfigure}[t]{0.88\textwidth}
       \subcaption{Sgr~B2(M), A01 (core):}
       \includegraphics[width=0.88\columnwidth]{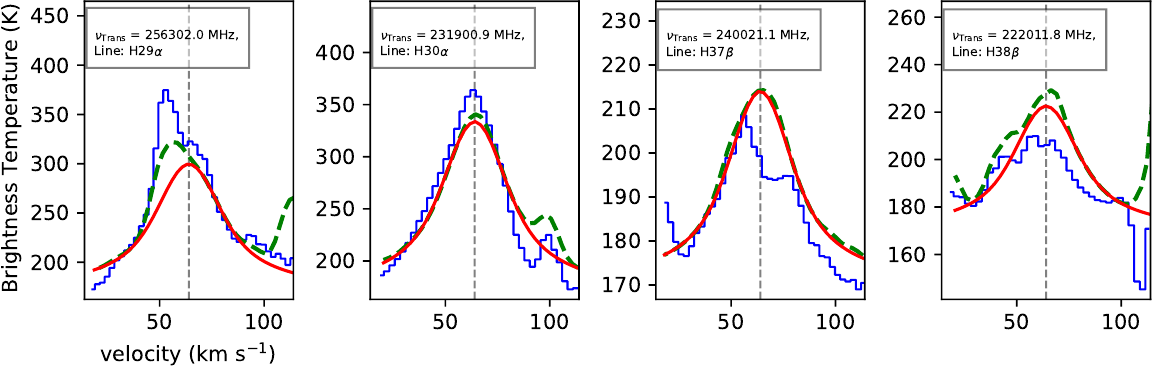}\\
        \label{fig:RRLCoreA01}
    \end{subfigure}
\quad
    \begin{subfigure}[t]{0.88\textwidth}
       \subcaption{Sgr~B2(M), A07 (core):}
       \includegraphics[width=0.88\columnwidth]{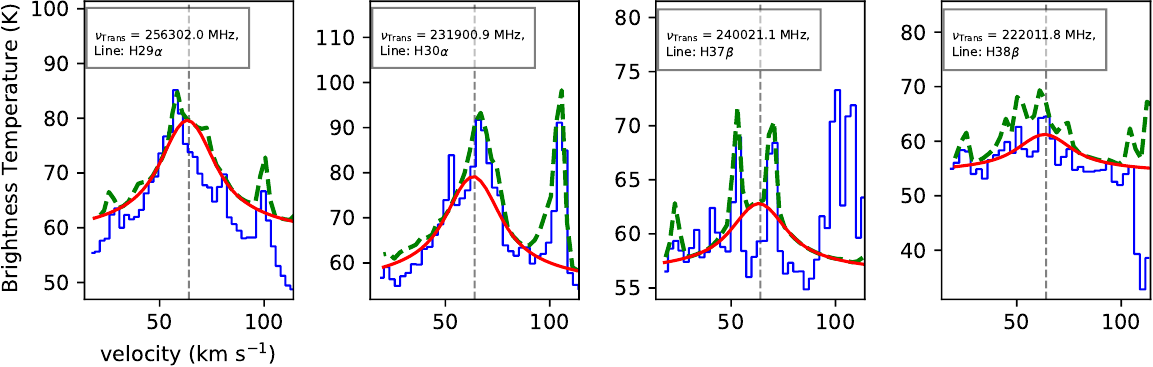}\\
        \label{fig:RRLCoreA07}
    \end{subfigure}
\quad
    \begin{subfigure}[t]{0.88\textwidth}
       \subcaption{Sgr~B2(M), A16 (core):}
       \includegraphics[width=0.88\columnwidth]{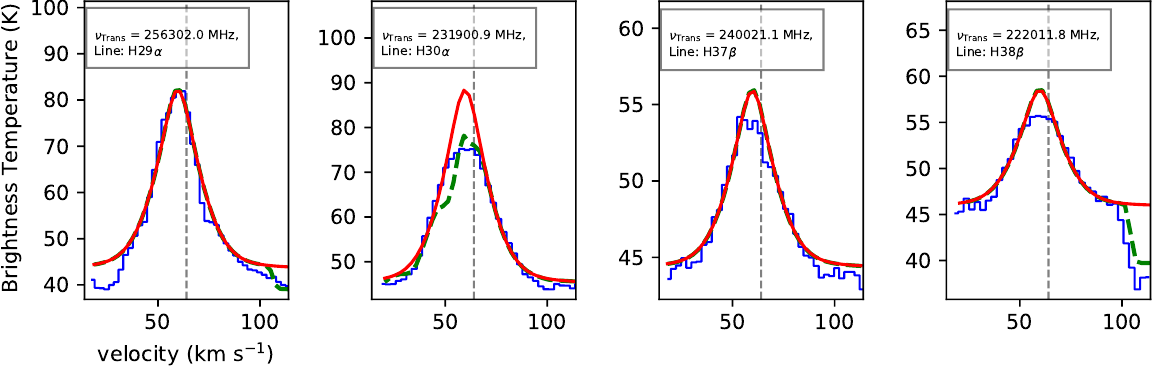}\\
        \label{fig:RRLCoreA16}
    \end{subfigure}
\quad
    \begin{subfigure}[t]{0.88\textwidth}
       \subcaption{Sgr~B2(N), A16 (core):}
       \includegraphics[width=0.88\columnwidth]{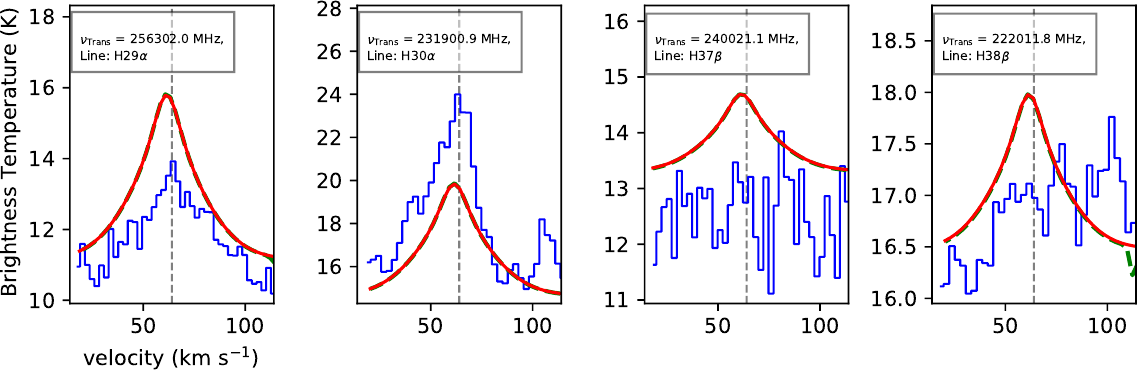}\\
        \label{fig:RRLCoreA24}
    \end{subfigure}
   \caption{Selected RRLs in core spectra of A01~(a), A07~(b), and A16~(c) in Sgr~B2(M) and A16~(c) in Sgr~B2(N). The observational data is described by the blue line, the pure RRL contribution by the red line, and the model spectrum taking all identified molecules into account by the dashed green line. The vertical grey dashed line indicates the source velocity of Sgr~B2(M).}
   \ContinuedFloat
   \label{fig:RRLCoreSgrB2M}
\end{figure*}

\begin{center}
\begin{table*}[!tb]
\caption{Dust and free-free parameters for each envelope around sources in Sgr~B2(M).}
\label{Tab:SgrB2MContEnv}
\centering
\tiny
\begin{tabular}{lccccccc}
    \hline
    \hline
Source  &                  T$_{\rm dust}$ &                     N$_{\rm H, dust}$ &                   $\beta_{\rm dust}$ &             $\gamma_{\rm dust}$ &                                          T$_{\rm e}$ &                                                 EM &               $\gamma_{\rm ff}$ \\
        &                             (K) &                           (cm$^{-2}$) &                                      &                            (\%) &                                                  (K) &                                     (pc cm$^{-6}$) &                            (\%) \\
\hline
\hline
\multicolumn{8}{l}{Envelope} \\
\hline
A01     &                     61 $\pm$ 19 &    1.1(+25)$_{-3.1(+00)}^{+2.2(+00)}$ &        0.9$_{-6.9(-01)}^{+9.9(-01)}$ &  11.4$_{-4.1(+00)}^{+4.5(+00)}$ &                                                    - &                                                  - &                               - \\
A02     &                     77 $\pm$ 38 &    3.1(+24)$_{-1.7(+00)}^{+2.3(+01)}$ &        0.8$_{-6.4(-01)}^{+1.1(+00)}$ &   5.8$_{-3.2(+00)}^{+3.6(+00)}$ &                                                    - &                                                  - &                               - \\
A03     &                     55 $\pm$ 10 &    3.1(+24)$_{-1.9(+00)}^{+1.2(+00)}$ &        2.0$_{-4.0(-01)}^{+9.7(-01)}$ &   5.4$_{-1.2(+00)}^{+1.5(+00)}$ &                                                    - &                                                  - &                               - \\
A04     &                     71 $\pm$ 53 &    3.4(+24)$_{-1.4(+00)}^{+1.3(+00)}$ &        0.7$_{-7.1(-02)}^{+5.5(-02)}$ &   9.7$_{-7.9(+00)}^{+7.9(+00)}$ &                                                    - &                                                  - &                               - \\
A05     &                     71 $\pm$ 47 &    2.1(+24)$_{-1.2(+00)}^{+1.3(+00)}$ &        0.1$_{-2.1(-01)}^{+8.8(-01)}$ &   5.9$_{-4.2(+00)}^{+4.6(+00)}$ &                                                    - &                                                  - &                               - \\
A06     &                     81 $\pm$ 30 &    1.7(+24)$_{-1.1(+00)}^{+3.1(+00)}$ &        1.4$_{-4.7(-01)}^{+4.2(-01)}$ &  12.2$_{-5.1(+00)}^{+5.4(+00)}$ &                                                    - &                                                  - &                               - \\
A07     &                    96 $\pm$ 128 &    9.9(+23)$_{-1.1(+00)}^{+1.0(+00)}$ &        0.2$_{-8.6(-02)}^{+2.6(-01)}$ &   5.8$_{-5.8(+00)}^{+8.3(+00)}$ &                       5817$_{-1.4(+03)}^{+5.5(+03)}$ &                 8.8(+06)$_{-1.9(+00)}^{+1.7(+00)}$ &   0.5$_{-3.6(-02)}^{+8.2(-02)}$ \\
A08     &                     88 $\pm$ 43 &    2.1(+24)$_{-2.9(+00)}^{+4.9(+00)}$ &        0.1$_{-3.8(-01)}^{+5.0(-01)}$ &  15.0$_{-8.0(+00)}^{+8.5(+00)}$ &                                                    - &                                                  - &                               - \\
A09     &                    114 $\pm$ 84 &    1.2(+24)$_{-1.1(+00)}^{+1.1(+00)}$ &        0.2$_{-7.1(-02)}^{+2.1(-01)}$ &  12.2$_{-9.5(+00)}^{+9.7(+00)}$ &                       2781$_{-1.2(+03)}^{+1.2(+03)}$ &                 1.2(+07)$_{-2.4(+00)}^{+1.3(+00)}$ &   1.3$_{-1.8(-01)}^{+1.1(-01)}$ \\
A10$^*$ &                     54 $\pm$ 41 &    1.8(+24)$_{-1.0(+00)}^{+1.0(+00)}$ &        0.1$_{-4.0(-03)}^{+5.5(-01)}$ &                               - &                                                    - &                                                  - &                               - \\
A11     &                     51 $\pm$ 17 &    1.7(+24)$_{-2.1(+00)}^{+2.4(+00)}$ &        0.1$_{-7.9(-01)}^{+1.6(+00)}$ &  13.7$_{-5.3(+00)}^{+6.4(+00)}$ &                                                    - &                                                  - &                               - \\
A12     &                     71 $\pm$ 37 &    2.1(+24)$_{-1.6(+00)}^{+1.6(+00)}$ &        0.5$_{-2.9(-01)}^{+1.5(+00)}$ &  16.3$_{-9.4(+00)}^{+1.1(+01)}$ &                                                    - &                                                  - &                               - \\
A13     &                     72 $\pm$ 38 &    1.7(+24)$_{-1.0(+00)}^{+1.0(+00)}$ &        0.1$_{-2.1(-02)}^{+2.6(-02)}$ &  15.1$_{-8.5(+00)}^{+8.5(+00)}$ &                                                    - &                                                  - &                               - \\
A14     &                     72 $\pm$ 36 &    3.5(+24)$_{-2.8(+00)}^{+2.7(+00)}$ &        0.1$_{-5.9(-02)}^{+7.8(-01)}$ &  32.4$_{-1.8(+01)}^{+1.9(+01)}$ &                                                    - &                                                  - &                               - \\
A15     &                     67 $\pm$ 17 &    1.8(+24)$_{-1.7(+00)}^{+1.7(+00)}$ &        0.7$_{-3.5(-01)}^{+5.5(-01)}$ &  24.6$_{-7.0(+00)}^{+7.5(+00)}$ &                                                    - &                                                  - &                               - \\
A16     &                     65 $\pm$ 15 &    2.2(+24)$_{-1.0(+00)}^{+1.0(+00)}$ &        1.0$_{-4.2(-02)}^{+2.8(-02)}$ &  10.8$_{-2.7(+00)}^{+2.7(+00)}$ &                                                    - &                                                  - &                               - \\
A17     &                     55 $\pm$ 23 &    1.1(+24)$_{-1.0(+00)}^{+1.0(+00)}$ &        0.1$_{-4.9(-02)}^{+5.0(-02)}$ &   8.0$_{-3.8(+00)}^{+3.8(+00)}$ &                                                    - &                                                  - &                               - \\
A18$^*$ &                      83 $\pm$ 6 &    2.3(+24)$_{-1.0(+00)}^{+1.0(+00)}$ &        0.1$_{-3.1(-01)}^{+1.8(-01)}$ &                               - &                                                    - &                                                  - &                               - \\
A19     &                      46 $\pm$ 0 &    1.6(+24)$_{-7.4(+00)}^{+1.4(+01)}$ &        0.1$_{-6.0(-01)}^{+1.1(+00)}$ &  13.5$_{-3.1(-01)}^{+6.7(-01)}$ &                                                    - &                                                  - &                               - \\
A20$^*$ &                     64 $\pm$ 19 &    2.3(+24)$_{-1.1(+00)}^{+1.0(+00)}$ &        0.2$_{-3.1(-01)}^{+6.9(-01)}$ &                               - &                                                    - &                                                  - &                               - \\
A21     &                     86 $\pm$ 10 &    1.9(+24)$_{-2.1(+00)}^{+8.3(+00)}$ &        0.1$_{-4.7(-01)}^{+1.2(+00)}$ &  26.2$_{-3.9(+00)}^{+5.0(+00)}$ &                                                    - &                                                  - &                               - \\
A22     &                      60 $\pm$ 7 &    3.2(+24)$_{-4.1(+00)}^{+2.8(+00)}$ &        0.1$_{-5.9(-01)}^{+9.5(-01)}$ &  27.3$_{-4.1(+00)}^{+4.9(+00)}$ &                                                    - &                                                  - &                               - \\
A23     &                     70 $\pm$ 10 &    1.7(+24)$_{-1.7(+00)}^{+3.9(+00)}$ &        0.1$_{-2.2(-01)}^{+2.2(+00)}$ &  12.9$_{-2.3(+00)}^{+3.8(+00)}$ &                                                    - &                                                  - &                               - \\
A24     &                     50 $\pm$ 87 &    2.5(+24)$_{-1.1(+00)}^{+1.0(+00)}$ &        0.4$_{-9.4(-02)}^{+1.7(-01)}$ &   9.2$_{-9.2(+00)}^{+1.8(+01)}$ &                       4414$_{-1.7(+03)}^{+7.2(+02)}$ &                 2.2(+07)$_{-2.0(+00)}^{+1.4(+00)}$ &   1.8$_{-2.1(-01)}^{+6.5(-02)}$ \\
A25$^*$ &                     53 $\pm$ 51 &    4.1(+24)$_{-1.3(+00)}^{+3.6(+00)}$ &        0.1$_{-4.7(-01)}^{+8.3(-01)}$ &                               - &                                                    - &                                                  - &                               - \\
A26     &                    162 $\pm$ 45 &    1.4(+24)$_{-1.0(+00)}^{+1.0(+00)}$ &        0.2$_{-1.7(-02)}^{+3.0(-02)}$ &  21.1$_{-6.1(+00)}^{+6.2(+00)}$ &                       3369$_{-1.2(+02)}^{+4.7(+02)}$ &                 2.1(+07)$_{-1.1(+00)}^{+1.1(+00)}$ &   2.2$_{-1.9(-02)}^{+6.7(-02)}$ \\
A27$^*$ &                     70 $\pm$ 14 &    1.7(+24)$_{-1.2(+00)}^{+2.3(+00)}$ &        0.4$_{-2.2(-01)}^{+6.0(-01)}$ &                               - &                                                    - &                                                  - &                               - \\
    \hline
    \hline
\end{tabular}
\begin{tablenotes}
   \item Exponents are described by round brackets, e.g., $1.4(16) = 1.4 \cdot 10^{16}$. The errors of the continuum parameters are indicated by subscript (left) and superscript (right) values, e.g. 1.1(+25)$_{-3.1(+00)}^{+2.2(+00)}$ represents the lower limit of $1.1 \cdot 10^{+25} -3.1 \cdot 10^{+00}$ and $1.1 \cdot 10^{+25} + 2.2 \cdot 10^{+00}$ is the upper limit of the corresponding hydrogen column density of $1.1 \cdot 10^{+25}$. For sources marked with an "$^*$", we could not derive a self-consistent description of the continuum of the corresponding core spectrum. Here, T$_{\rm dust}$ indicates the dust temperature in~K, N$_{\rm H, dust}$ the hydrogen column density in~cm$^{-2}$, and $\beta_{\rm dust}$ the power-law index of the dust emissivity. Note that we use temperature units for the fitting. Therefore, the spectral indices for flux units~$\alpha$ is given by $\alpha = \beta + 2$. Additionally, $\gamma_{\rm dust}$ describes the contribution of dust to the total continuum of the corresponding core spectrum in~\%, T$_{\rm e}$ the electron temperature in~K, EM the emission measure in~pc~cm$^{-6}$, and $\gamma_{\rm ff}$ the free-free contribution to the total continuum in~\%.
\end{tablenotes}
\end{table*}
\end{center}

\begin{center}
\begin{table*}[!tb]
\caption{Dust and free-free parameters for each core in Sgr~B2(M).}
\label{Tab:SgrB2MContCore}
\centering
\tiny
\begin{tabular}{lccccccc}
    \hline
    \hline
Source  &                  T$_{\rm dust}$ &                     N$_{\rm H, dust}$ &                   $\beta_{\rm dust}$ &             $\gamma_{\rm dust}$ &                                          T$_{\rm e}$ &                                                 EM &               $\gamma_{\rm ff}$ \\
        &                             (K) &                           (cm$^{-2}$) &                                      &                            (\%) &                                                  (K) &                                     (pc cm$^{-6}$) &                            (\%) \\
    \hline
    \hline
    \multicolumn{8}{l}{Core} \\
    \hline
A01     &                   342 $\pm$ 168 &    2.3(+25)$_{-1.0(+00)}^{+1.0(+00)}$ &        2.6$_{-1.5(-02)}^{+1.4(-01)}$ & 102.7$_{-5.1(+01)}^{+5.2(+01)}$ &                       3808$_{-1.8(+02)}^{+4.4(+01)}$ &                 2.9(+09)$_{-1.0(+00)}^{+1.0(+00)}$ &  52.9$_{-5.5(-01)}^{+1.2(-01)}$ \\
A02     &                   286 $\pm$ 115 &    2.4(+25)$_{-1.0(+00)}^{+1.0(+00)}$ &        0.4$_{-2.9(-03)}^{+3.4(-03)}$ & 106.0$_{-4.4(+01)}^{+4.4(+01)}$ &                                                    - &                                                  - &                               - \\
A03     &                   254 $\pm$ 107 &    2.6(+25)$_{-1.0(+00)}^{+1.0(+00)}$ &        2.3$_{-6.6(-02)}^{+3.3(-02)}$ & 107.6$_{-4.6(+01)}^{+4.7(+01)}$ &                                                    - &                                                  - &                               - \\
A04     &                    243 $\pm$ 90 &    1.2(+25)$_{-1.0(+00)}^{+1.1(+00)}$ &        2.4$_{-3.6(-01)}^{+9.6(-02)}$ &  84.4$_{-3.3(+01)}^{+3.2(+01)}$ &                       9365$_{-1.0(+03)}^{+1.6(+03)}$ &                 8.7(+08)$_{-1.4(+00)}^{+1.1(+00)}$ &  23.4$_{-6.9(-01)}^{+9.6(-01)}$ \\
A05     &                    250 $\pm$ 85 &    1.6(+25)$_{-1.0(+00)}^{+1.2(+00)}$ &        0.7$_{-3.8(-02)}^{+9.1(-02)}$ & 101.7$_{-3.5(+01)}^{+3.6(+01)}$ &                                                    - &                                                  - &                               - \\
A06     &                    256 $\pm$ 86 &    5.5(+24)$_{-6.7(+00)}^{+4.6(+00)}$ &        1.0$_{-3.8(-01)}^{+2.6(+00)}$ &  93.9$_{-3.4(+01)}^{+4.9(+01)}$ &                                                    - &                                                  - &                               - \\
A07     &                    217 $\pm$ 70 &    9.0(+24)$_{-1.0(+00)}^{+1.0(+00)}$ &        1.4$_{-1.9(-02)}^{+3.2(-02)}$ &  82.6$_{-2.7(+01)}^{+2.8(+01)}$ &                       5823$_{-3.1(+00)}^{+9.5(+01)}$ &                 3.9(+08)$_{-1.0(+00)}^{+1.0(+00)}$ &  17.4$_{-0.0(+00)}^{+7.2(-02)}$ \\
A08     &                    211 $\pm$ 44 &    1.2(+24)$_{-1.0(+00)}^{+1.8(+00)}$ &        1.5$_{-1.4(+00)}^{+1.6(-01)}$ &  16.0$_{-4.2(+00)}^{+3.7(+00)}$ &                       8052$_{-7.7(+02)}^{+2.3(+01)}$ &                 1.5(+09)$_{-1.2(+00)}^{+1.0(+00)}$ &  77.6$_{-2.0(+00)}^{+7.2(-02)}$ \\
A09     &                    207 $\pm$ 68 &    6.1(+24)$_{-1.0(+00)}^{+1.0(+00)}$ &        0.8$_{-2.4(-03)}^{+3.0(-03)}$ &  90.4$_{-3.0(+01)}^{+3.0(+01)}$ &                                                    - &                                                  - &                               - \\
A10$^*$ &                             236 &                              7.8(+24) &                                  1.4 &                               - &                                                    - &                                                  - &                               - \\
A11     &                    231 $\pm$ 41 &    2.4(+24)$_{-1.0(+00)}^{+1.0(+00)}$ &        2.6$_{-1.9(-02)}^{+2.5(-02)}$ &  91.8$_{-1.7(+01)}^{+1.7(+01)}$ &                                                    - &                                                  - &                               - \\
A12     &                    225 $\pm$ 55 &    4.4(+24)$_{-1.0(+00)}^{+1.0(+00)}$ &        1.2$_{-1.0(-02)}^{+2.6(-02)}$ &  90.6$_{-2.3(+01)}^{+2.3(+01)}$ &                                                    - &                                                  - &                               - \\
A13     &                    220 $\pm$ 59 &    3.8(+24)$_{-1.2(+00)}^{+1.1(+00)}$ &        2.5$_{-1.1(+00)}^{+5.4(-01)}$ &  90.6$_{-2.8(+01)}^{+2.8(+01)}$ &                                                    - &                                                  - &                               - \\
A14     &                    205 $\pm$ 77 &    3.7(+24)$_{-1.2(+00)}^{+1.6(+00)}$ &        1.7$_{-7.4(-02)}^{+2.0(-01)}$ &  77.0$_{-3.0(+01)}^{+3.1(+01)}$ &                                                    - &                                                  - &                               - \\
A15     &                    219 $\pm$ 95 &    1.1(+23)$_{-2.8(+00)}^{+1.3(+00)}$ &        0.0$_{-7.4(-01)}^{+2.8(+00)}$ &   3.4$_{-1.6(+00)}^{+2.2(+00)}$ &                       8850$_{-4.0(+02)}^{+1.7(+02)}$ &                 6.5(+08)$_{-1.0(+00)}^{+1.0(+00)}$ &  77.4$_{-9.2(-01)}^{+3.8(-01)}$ \\
A16     &                    224 $\pm$ 54 &    4.3(+24)$_{-1.0(+00)}^{+1.0(+00)}$ &        0.6$_{-8.0(-02)}^{+6.5(-02)}$ &  62.6$_{-1.6(+01)}^{+1.6(+01)}$ &                       9445$_{-3.6(+02)}^{+2.9(+02)}$ &                 6.5(+08)$_{-1.1(+00)}^{+1.0(+00)}$ &  37.2$_{-3.7(-01)}^{+2.9(-01)}$ \\
A17     &                    295 $\pm$ 76 &    1.6(+24)$_{-1.0(+00)}^{+1.0(+00)}$ &        0.0$_{-1.0(-03)}^{+5.0(-02)}$ &  58.6$_{-1.5(+01)}^{+1.5(+01)}$ &                      23779$_{-2.0(+03)}^{+1.9(+03)}$ &                 4.6(+08)$_{-1.0(+00)}^{+1.0(+00)}$ &  38.2$_{-9.8(-01)}^{+8.5(-01)}$ \\
A18$^*$ &                             236 &                              7.8(+24) &                                  1.4 &                               - &                                                    - &                                                  - &                               - \\
A19     &                    225 $\pm$ 54 &    2.6(+24)$_{-2.4(+00)}^{+5.1(+00)}$ &        2.8$_{-8.5(-01)}^{+5.2(-01)}$ &  92.0$_{-2.6(+01)}^{+2.6(+01)}$ &                                                    - &                                                  - &                               - \\
A20$^*$ &                             236 &                              7.8(+24) &                                  1.4 &                               - &                                                    - &                                                  - &                               - \\
A21     &                    230 $\pm$ 30 &    4.1(+24)$_{-5.5(+00)}^{+1.3(+01)}$ &        0.6$_{-3.8(-01)}^{+1.9(+00)}$ &  79.1$_{-1.2(+01)}^{+1.9(+01)}$ &                                                    - &                                                  - &                               - \\
A22     &                    218 $\pm$ 34 &    4.1(+24)$_{-1.0(+00)}^{+1.0(+00)}$ &        2.6$_{-1.2(-01)}^{+7.1(-02)}$ &  81.9$_{-1.4(+01)}^{+1.3(+01)}$ &                                                    - &                                                  - &                               - \\
A23     &                    219 $\pm$ 42 &    5.6(+24)$_{-1.1(+00)}^{+1.0(+00)}$ &        0.9$_{-6.7(-02)}^{+8.5(-02)}$ &  92.9$_{-1.9(+01)}^{+1.9(+01)}$ &                                                    - &                                                  - &                               - \\
A24     &                    209 $\pm$ 50 &    4.5(+24)$_{-1.3(+00)}^{+1.0(+00)}$ &        0.0$_{-8.4(-04)}^{+6.4(-01)}$ &  26.7$_{-6.7(+00)}^{+7.4(+00)}$ &                      15214$_{-1.9(+02)}^{+5.8(+02)}$ &                 3.4(+09)$_{-1.0(+00)}^{+1.1(+00)}$ &  77.4$_{-2.8(-01)}^{+8.2(-01)}$ \\
A25$^*$ &                             236 &                              7.8(+24) &                                  1.4 &                               - &                                                    - &                                                  - &                               - \\
A26     &                    195 $\pm$ 63 &    6.6(+24)$_{-1.0(+00)}^{+1.0(+00)}$ &        1.3$_{-3.5(-03)}^{+2.5(-03)}$ &  80.7$_{-2.7(+01)}^{+2.7(+01)}$ &                                                    - &                                                  - &                               - \\
A27$^*$ &                             236 &                              7.8(+24) &                                  1.4 &                               - &                                                    - &                                                  - &                               - \\
    \hline
    \hline
\end{tabular}
\begin{tablenotes}
   \item Exponents are described by round brackets, e.g., $1.4(16) = 1.4 \cdot 10^{16}$. The errors of the continuum parameters are indicated by sub-(left) and superscript (right) values, e.g. 3.7(+24)$_{-1.2(+00)}^{+1.6(+00)}$ represents the lower limit of $3.7 \cdot 10^{+24} - 1.2 \cdot 10^{+00}$ and $3.7 \cdot 10^{+24} + 1.6 \cdot 10^{+00}$ is the upper limit of the corresponding hydrogen column density of $3.7 \cdot 10^{+24}$. For sources marked with an "$^*$", we could not derive a self-consistent description of the continuum of the corresponding core spectrum. Here, T$_{\rm dust}$ indicates the dust temperature in~K, N$_{\rm H, dust}$ the hydrogen column density in~cm$^{-2}$, and $\beta_{\rm dust}$ the power-law index of the dust emissivity. Note that we use temperature units for the fitting. Therefore, the spectral indices for flux units~$\alpha$ is given by $\alpha = \beta + 2$. Additionally, $\gamma_{\rm dust}$ describes the contribution of dust to the total continuum of the corresponding core spectrum in~\%, T$_{\rm e}$ the electron temperature in~K, EM the emission measure in~pc~cm$^{-6}$, and $\gamma_{\rm ff}$ the free-free contribution to the total continuum in~\%.
\end{tablenotes}
\end{table*}
\end{center}

\begin{center}
\begin{table*}[!tb]
\caption{Dust and free-free parameters for each source in Sgr~B2(N).}
\label{Tab:SgrB2NDustPara}
\centering
\tiny
\begin{tabular}{lccccccc}
    \hline
    \hline
Source  &                  T$_{\rm dust}$ &                     N$_{\rm H, dust}$ &                   $\beta_{\rm dust}$ &             $\gamma_{\rm dust}$ &                                          T$_{\rm e}$ &                                                 EM &               $\gamma_{\rm ff}$ \\
        &                             (K) &                           (cm$^{-2}$) &                                      &                            (\%) &                                                  (K) &                                     (pc cm$^{-6}$) &                            (\%) \\
\hline
\hline
    \multicolumn{8}{l}{Envelope} \\
    \hline
A01     &                     39 $\pm$ 21 &    1.0(+25)$_{-4.7(+00)}^{+3.0(+00)}$ &        1.5$_{-4.0(-01)}^{+4.5(-01)}$ &  10.5$_{-6.7(+00)}^{+7.0(+00)}$ &                                                    - &                                                  - &                               - \\
A02     &                      35 $\pm$ 0 &    5.3(+24)$_{-1.0(+00)}^{+1.0(+00)}$ &        0.2$_{-1.0(-01)}^{+2.5(-01)}$ &   9.3$_{-3.5(-02)}^{+9.0(-02)}$ &                                                    - &                                                  - &                               - \\
A03     &                     48 $\pm$ 23 &    2.3(+24)$_{-1.7(+00)}^{+1.6(+00)}$ &        0.9$_{-4.4(-01)}^{+8.0(-01)}$ &   6.6$_{-3.7(+00)}^{+4.0(+00)}$ &                                                    - &                                                  - &                               - \\
A04     &                     59 $\pm$ 20 &    2.7(+24)$_{-1.0(+00)}^{+1.0(+00)}$ &        0.1$_{-3.8(-02)}^{+6.0(-02)}$ &   9.9$_{-3.8(+00)}^{+3.8(+00)}$ &                                                    - &                                                  - &                               - \\
A05     &                      58 $\pm$ 3 &    5.4(+24)$_{-1.0(+00)}^{+1.0(+00)}$ &        2.1$_{-2.4(-02)}^{+7.2(-02)}$ &  27.3$_{-1.9(+00)}^{+2.0(+00)}$ &                                                    - &                                                  - &                               - \\
A06     &                     63 $\pm$ 12 &    2.8(+24)$_{-1.0(+00)}^{+1.0(+00)}$ &        0.8$_{-4.0(-02)}^{+1.8(-02)}$ &  18.6$_{-4.0(+00)}^{+4.0(+00)}$ &                                                    - &                                                  - &                               - \\
A07     &                     58 $\pm$ 16 &    2.6(+24)$_{-3.3(+00)}^{+2.0(+01)}$ &        2.5$_{-1.6(+00)}^{+2.0(-01)}$ &  19.5$_{-7.0(+00)}^{+6.3(+00)}$ &                                                    - &                                                  - &                               - \\
A08$^*$ &                     62 $\pm$ 29 &    9.7(+23)$_{-1.8(+00)}^{+9.9(+01)}$ &        0.1$_{-8.8(-01)}^{+2.9(+00)}$ &                               - &                                                    - &                                                  - &                               - \\
A09     &                     43 $\pm$ 31 &    1.5(+24)$_{-8.0(+00)}^{+1.5(+01)}$ &        0.1$_{-6.9(-01)}^{+7.2(-01)}$ &  12.4$_{-1.0(+01)}^{+1.1(+01)}$ &                                                    - &                                                  - &                               - \\
A10$^*$ &                     68 $\pm$ 20 &    1.6(+24)$_{-1.5(+00)}^{+1.3(+01)}$ &        0.1$_{-1.7(+00)}^{+1.5(+00)}$ &                               - &                                                    - &                                                  - &                               - \\
A11$^*$ &                     68 $\pm$ 13 &    2.1(+24)$_{-5.2(+00)}^{+2.0(+00)}$ &        1.6$_{-7.7(-01)}^{+9.5(-01)}$ &                               - &                                                    - &                                                  - &                               - \\
A12$^*$ &                      77 $\pm$ 0 &    5.1(+23)$_{-1.1(+00)}^{+1.1(+00)}$ &        0.1$_{-9.0(-01)}^{+7.0(-01)}$ &                               - &                                                    - &                                                  - &                               - \\
A13$^*$ &                     53 $\pm$ 22 &    5.5(+23)$_{-1.0(+00)}^{+1.0(+00)}$ &        0.1$_{-8.1(-02)}^{+4.1(-01)}$ &                               - &                                                    - &                                                  - &                               - \\
A14     &                     70 $\pm$ 24 &    5.5(+23)$_{-2.4(+00)}^{+4.7(+01)}$ &        0.1$_{-4.5(-01)}^{+2.6(+00)}$ &   7.6$_{-3.0(+00)}^{+4.3(+00)}$ &                                                    - &                                                  - &                               - \\
A15     &                      86 $\pm$ 5 &    1.4(+24)$_{-1.2(+00)}^{+1.3(+00)}$ &        0.1$_{-7.2(-01)}^{+1.0(+00)}$ &  14.9$_{-1.5(+00)}^{+1.8(+00)}$ &                                                    - &                                                  - &                               - \\
A16$^*$ &                     58 $\pm$ 22 &    1.2(+24)$_{-1.0(+00)}^{+1.0(+00)}$ &        0.3$_{-2.2(-01)}^{+4.9(-01)}$ &                               - &                                                    - &                                                  - &                               - \\
A17     &                     60 $\pm$ 17 &    5.1(+24)$_{-3.1(+01)}^{+3.3(+00)}$ &        1.2$_{-2.6(-01)}^{+1.6(+00)}$ &  71.2$_{-2.3(+01)}^{+3.0(+01)}$ &                                                    - &                                                  - &                               - \\
A18     &                      77 $\pm$ 0 &    7.4(+23)$_{-5.8(+00)}^{+1.6(+01)}$ &        0.1$_{-7.6(-01)}^{+1.5(+00)}$ &  14.3$_{-4.2(-01)}^{+1.0(+00)}$ &                                                    - &                                                  - &                               - \\
A19     &                     56 $\pm$ 24 &    1.6(+24)$_{-1.0(+00)}^{+1.0(+00)}$ &        0.1$_{-3.3(-02)}^{+3.4(-02)}$ &  18.0$_{-8.8(+00)}^{+8.9(+00)}$ &                                                    - &                                                  - &                               - \\
A20$^*$ &                     45 $\pm$ 23 &    3.3(+24)$_{-3.7(+00)}^{+4.9(+00)}$ &        0.1$_{-1.7(+00)}^{+9.0(-01)}$ &                               - &                                                    - &                                                  - &                               - \\
    \hline
    \hline
    \multicolumn{8}{l}{Core} \\
    \hline
A01     &                    248 $\pm$ 64 &    2.1(+25)$_{-1.4(+00)}^{+1.0(+00)}$ &        2.5$_{-0.0(+00)}^{+1.3(+00)}$ & 127.3$_{-3.4(+01)}^{+3.9(+01)}$ &                       9882$_{-7.0(+03)}^{+1.0(+04)}$ &                 3.7(+08)$_{-9.5(+03)}^{+5.0(+00)}$ &  10.1$_{-3.5(+00)}^{+1.8(+00)}$ \\
A02     &                    221 $\pm$ 69 &    1.1(+25)$_{-1.0(+00)}^{+1.0(+00)}$ &        1.4$_{-4.7(-03)}^{+4.4(-03)}$ & 110.6$_{-3.6(+01)}^{+3.5(+01)}$ &                                                    - &                                                  - &                               - \\
A03     &                    240 $\pm$ 74 &    8.4(+24)$_{-2.0(+01)}^{+1.4(+00)}$ &        1.5$_{-1.0(+00)}^{+1.6(-01)}$ & 102.0$_{-3.5(+01)}^{+3.4(+01)}$ &                                                    - &                                                  - &                               - \\
A04     &                    210 $\pm$ 53 &    9.2(+24)$_{-1.0(+00)}^{+1.0(+00)}$ &        0.6$_{-1.4(-01)}^{+7.5(-03)}$ &  99.4$_{-2.7(+01)}^{+2.6(+01)}$ &                                                    - &                                                  - &                               - \\
A05     &                   223 $\pm$ 109 &    5.7(+24)$_{-1.5(+00)}^{+1.6(+00)}$ &        0.8$_{-7.4(-02)}^{+9.2(-01)}$ &  90.9$_{-4.6(+01)}^{+5.2(+01)}$ &                                                    - &                                                  - &                               - \\
A06     &                    196 $\pm$ 49 &    5.0(+24)$_{-1.1(+00)}^{+1.4(+00)}$ &        0.5$_{-4.3(-02)}^{+6.3(-02)}$ &  90.4$_{-2.4(+01)}^{+2.4(+01)}$ &                                                    - &                                                  - &                               - \\
A07     &                    208 $\pm$ 54 &    5.2(+24)$_{-1.0(+00)}^{+1.0(+00)}$ &       -0.1$_{-1.4(-01)}^{+1.2(-01)}$ &  89.6$_{-2.4(+01)}^{+2.4(+01)}$ &                                                    - &                                                  - &                               - \\
A08$^*$ &                             225 &                              6.1(+24) &                                  0.7 &                               - &                                                    - &                                                  - &                               - \\
A09     &                    190 $\pm$ 26 &    2.8(+24)$_{-1.0(+00)}^{+1.0(+00)}$ &        0.0$_{-1.2(-02)}^{+1.3(-02)}$ &  92.6$_{-1.3(+01)}^{+1.4(+01)}$ &                                                    - &                                                  - &                               - \\
A10$^*$ &                             225 &                              6.1(+24) &                                  0.7 &                               - &                                                    - &                                                  - &                               - \\
A11$^*$ &                             225 &                              6.1(+24) &                                  0.7 &                               - &                                                    - &                                                  - &                               - \\
A12$^*$ &                             225 &                              6.1(+24) &                                  0.7 &                               - &                                                    - &                                                  - &                               - \\
A13$^*$ &                             225 &                              6.1(+24) &                                  0.7 &                               - &                                                    - &                                                  - &                               - \\
A14     &                    282 $\pm$ 45 &    1.8(+24)$_{-1.1(+00)}^{+1.0(+00)}$ &        1.4$_{-6.3(-02)}^{+5.2(-02)}$ &  94.3$_{-1.6(+01)}^{+1.6(+01)}$ &                                                    - &                                                  - &                               - \\
A15     &                    231 $\pm$ 39 &    4.7(+24)$_{-5.0(+00)}^{+2.8(+00)}$ &        0.2$_{-1.5(-01)}^{+9.2(-01)}$ &  89.7$_{-1.6(+01)}^{+2.0(+01)}$ &                                                    - &                                                  - &                               - \\
A16$^*$ &                             225 &                              6.1(+24) &                                  0.7 &                               - &                      10320$_{-9.2(+02)}^{+1.7(+03)}$ &                 1.0(+08)$_{-1.0(+00)}^{+1.0(+00)}$ &                               - \\
A17     &                    203 $\pm$ 51 &    1.3(+24)$_{-1.0(+00)}^{+1.0(+00)}$ &        0.1$_{-2.1(-07)}^{+1.2(-03)}$ &  35.1$_{-9.2(+00)}^{+9.2(+00)}$ &                                                    - &                                                  - &                               - \\
A18     &                   267 $\pm$ 132 &    1.5(+24)$_{-1.0(+00)}^{+1.0(+00)}$ &        0.2$_{-1.6(-02)}^{+1.5(-02)}$ &  88.1$_{-4.5(+01)}^{+4.5(+01)}$ &                                                    - &                                                  - &                               - \\
A19     &                    206 $\pm$ 77 &    3.1(+24)$_{-1.0(+00)}^{+1.0(+00)}$ &        0.0$_{-2.8(-03)}^{+1.5(-02)}$ &  86.9$_{-3.4(+01)}^{+3.4(+01)}$ &                                                    - &                                                  - &                               - \\
A20$^*$ &                             225 &                              6.1(+24) &                                  0.7 &                               - &                                                    - &                                                  - &                               - \\
    \hline
    \hline
\end{tabular}
\begin{tablenotes}
   \item Exponents are described by round brackets, e.g., $1.4(16) = 1.4 \cdot 10^{16}$. The errors of the continuum parameters are indicated by sub-(left) and superscript (right) values, e.g. 1.0(+25)$_{-4.7(+00)}^{+3.0(+00)}$ represents the lower limit of $1.0 \cdot 10^{+25} - 4.7 \cdot 10^{+00}$ and $1.0 \cdot 10^{+25} + 3.0 \cdot 10^{+00}$ is the upper limit of the corresponding hydrogen column density of $1.0 \cdot 10^{+25}$. For sources marked with an "$^*$", we could not derive a self-consistent description of the continuum of the corresponding core spectrum. Here, T$_{\rm dust}$ indicates the dust temperature in~K, N$_{\rm H, dust}$ the hydrogen column density in~cm$^{-2}$, and $\beta_{\rm dust}$ the power-law index of the dust emissivity. Note that we use temperature units for the fitting. Therefore, the spectral indices for flux units~$\alpha$ is given by $\alpha = \beta + 2$. Additionally, $\gamma_{\rm dust}$ describes the contribution of dust to the total continuum of the corresponding core spectrum in~\%, T$_{\rm e}$ the electron temperature in~K, EM the emission measure in~pc~cm$^{-6}$, and $\gamma_{\rm ff}$ the free-free contribution to the total continuum in~\%.
\end{tablenotes}
\end{table*}
\end{center}

\subsubsection{Free-free parameters of the envelope}\label{subsusbsec:freefreeenv}

Some envelope spectra (A07, A09, A24, and A26 in Sgr~B2(M)) contain RRLs, which we used to determine the free-free contribution to the corresponding continuum levels as well. The frequency ranges covered by our observation, contain two H$\alpha$ (H29$\alpha$ and H30$\alpha$), three H$\beta$ (H36$\beta$, H37$\beta$, H38$\beta$), three H$\gamma$ (H41$\gamma$, H42$\gamma$, H43$\gamma$), four H$\delta$ (H44$\delta$, H46$\delta$, H47$\delta$, H48$\delta$), five H$\epsilon$ (H47$\epsilon$, H48$\epsilon$, H49$\epsilon$, H50$\epsilon$, H51$\epsilon$), and four H$\zeta$ (H50$\zeta$, H52$\zeta$, H53$\zeta$, H54$\zeta$) transitions. Although the contributions of transitions with $\Delta n \ge 4$ ($\delta$-, $\epsilon$-, and $\zeta$-transitions) are very small, they can still help in determining the model parameters (electron temperatures, T$_{\rm e}$, and the emission measures, EM) because even very weak transitions contain useful informations and provide additional constraints. In contrast to the aforementioned procedure we have performed a full line survey analysis of each envelope spectrum containing RRLs taking local-overlap into account, see Sect.~\ref{subsec:localoverlap}, because the RRLs there contain non-negligible admixtures of other molecules, see Fig.~\ref{fig:RRLEnvSgrB2M}. In this analysis, we start with modelling all molecules and RRLs using the new XCLASS-GUI included in the extended XCLASS package, which offers the possibility to interactively model observational data and to describes molecules and recombination lines in LTE. The GUI can be used to generate synthetic spectra from input physical parameters, which can be overlaid on the observed spectra and/or fitted to the observations to obtain the best fit to the physical parameters. For all RRLs, we used a single emission component covering the full beam. Furthermore, for species where only one transition is included in the survey, e.g.\ CO and CS, a reliable quantitative description of their contribution is not possible. Since line overlap plays a major role for many sources, it is necessary to model the contributions of these molecules as well. Therefore, we fix the excitation temperatures for components describing emission features to a value of 200~K\footnote{A temperature of 200~K might be to high for molecules located in the envelope, but for molecules with only one transition a reliable temperature estimation is not possible. Here we have used the indicated value only for a phenomenological description of the line shape of the corresponding molecule. The exact value has no further meaning for our analysis.}. For components describing absorption features, we assume an excitation temperature of 2.7~K. In the next step we compute the average gas temperature from all identified molecules with more than one transition, where we take all components into account which describe emission features. After that, we perform one final fit, where all molecule and RRL parameters are fitted together with the hydrogen column density~$N_{\rm H}$ and spectral index~$\beta$ to achieve a self-consistent description of the molecular and recombination lines and the continuum level. Subsequently, we re-compute again the average gas temperature to get the dust temperature of the envelope.

\subsubsection{Continuum parameters of core}\label{subsusbsec:contcore}

In the next step, we estimate the continuum parameters for each core spectrum. Similar to the procedure described above, we obtain the dust temperature for each core from the averaged gas temperature. Here, we make use of the results of the analysis of the full molecular line survey. Here, we modeled all RRLs with a single emission component, whose source size $\theta_{\rm core}$ is given by the diameter of a circle that has the same area $A_{\rm core}$ as the corresponding polygon describing the source\footnote{Here, we assume a common source size for all molecules and RRLs within a source, since calculating each individual source size would be too computationally expensive due to the calculation of the local overlap. Therefore, our derived column densities describe lower limits only.}, see Figs.~\ref{fig:CorePosSgrB2M}~-~\ref{fig:CorePosSgrB2N}, i.e.\ we determine the source size $\theta_{\rm core}$, see Tab.~\ref{Tab:PhysParameters}, using
\begin{align}\label{myxclass:SourceSize}
  A^{\rm core} &= \pi \, \left(\frac{\theta_{\rm core}}{2} \right)^2 \nonumber\\
  \Rightarrow \theta_{\rm core} &= 2 \, \sqrt{\frac{A_{\rm core}}{\pi}}.
\end{align}
For the temperature estimation we consider all components of molecules describing emission features in the corresponding core spectrum and which have more than one transition within the frequency ranges covered by the observation. Afterwards, we use again XCLASS to derive the corresponding hydrogen column density $N_{\rm H}$ and spectral index~$\beta$, taking into account both the continuum contributions from the envelope layer and a possibly existing free-free contribution from the core layer, see Fig.~\ref{fig:RRLCoreSgrB2M}. The obtained dust and free-free parameters for each layer and source are described in Tabs.~\ref{Tab:SgrB2MContEnv}~-~\ref{Tab:SgrB2NDustPara}. Additionally, we calculated the contribution $\gamma$ of each portion to the total continuum of the corresponding core spectrum by determining the ratio of the integrated intensity of each contribution and the total continuum. Here, each contribution is calculated without taking the interaction with other contributions into account, which is why the ratios described in Tabs.~\ref{Tab:SgrB2MContEnv}~-~\ref{Tab:SgrB2NDustPara} should be regarded as upper limits.

For some sources we were not able to derive a self-consistent description of the continuum of the corresponding core spectrum. For sources A10, A20, A25, and A27 in Sgr~B2(M) and A07, A08, A10, A12, A13, and A20 in Sgr~B2(N), we could not find RRLs despite negative slopes of the continuum levels. Additionally, the derived free-free parameters for source A16 in Sgr~B2(N) can not describe the observed slope. In addition, sources A18 in Sgr~B2(M) and A11 in Sgr~B2(N) show positive slopes that cannot be explained by optically thin dust emissions. As mentioned by \citetads{2017A&A...604A...6S}, for some faint sources the slope of the continuum levels might be falsified by calibration issues and the frequency-dependent filtering out of extended emission. For all sources where a self-consistent description of the core continuum was not possible, we apply a phenomenological description of the continuum and use averaged dust parameters for the core layers in sources in Sgr~B2(M) and Sgr~B2(N), respectively.

\begin{figure*}[!tb]
   \centering
   \includegraphics[width=1.0\textwidth]{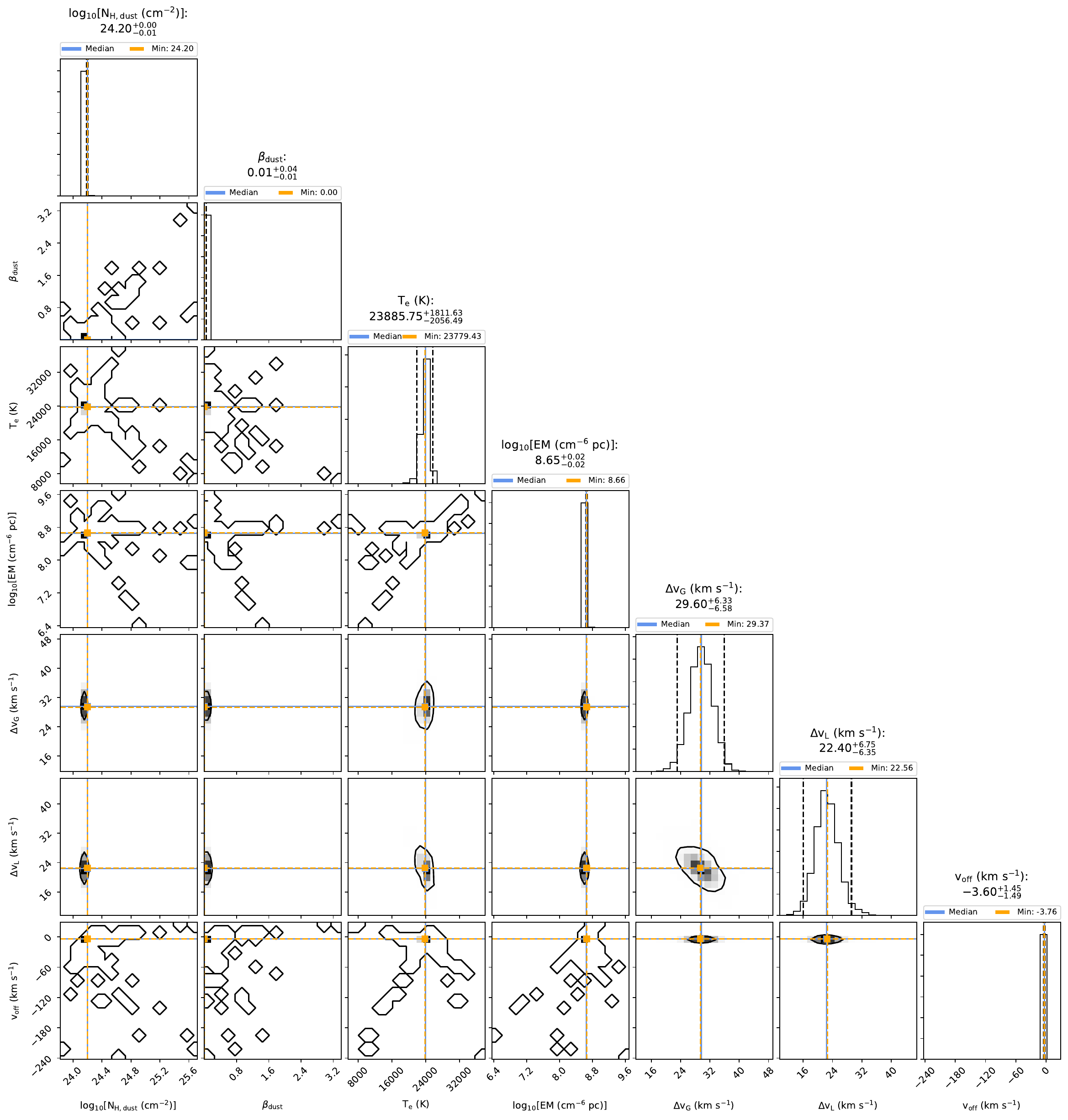}\\
   \caption{A corner plot \citepads{corner} showing the one and two dimensional projections of the posterior probability distributions of the continuum parameters of core A17, in Sgr~B2(M). On top of each column the probability distribution for each free parameter is shown together with the 50~\% quantile (median) and the corresponding left and right errors. The left and right dashed lines indicate the lower and upper limits of the corresponding HPD interval, respectively. The blue line indicates the median of the distribution. The dashed orange lines indicate the parameter values of the best fit. The plots in the lower left corner describe the projected 2D histograms of two parameters. In order to get a better estimation of the errors, we determine the error of the hydrogen column density and the emission measure on log10 scale and use the velocity offset (v$_{\rm off}$) related to the source velocity of v$_{\rm LSR}$ = 64~km~s$^{-1}$.}
   \label{fig:ErrorEstim}
\end{figure*}

\begin{figure}[!tb]
   \centering
   \includegraphics[width=0.5\textwidth]{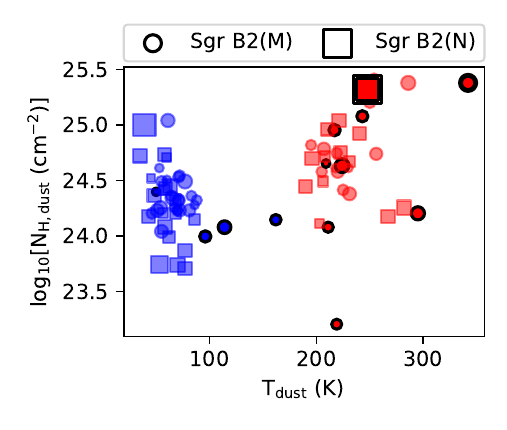}\\
   \caption{Derived dust temperatures and hydrogen column densities for each envelope (blue) and core (red) layer in Sgr~B2(M) (circles) and N (squares). Layers containing RRLs are indicated by a black border. The size of each scatter point corresponds to the corresponding source size.}
   \label{fig:AllContTdustNH}
\end{figure}

\begin{figure}[!tb]
   \centering
   \includegraphics[width=0.5\textwidth]{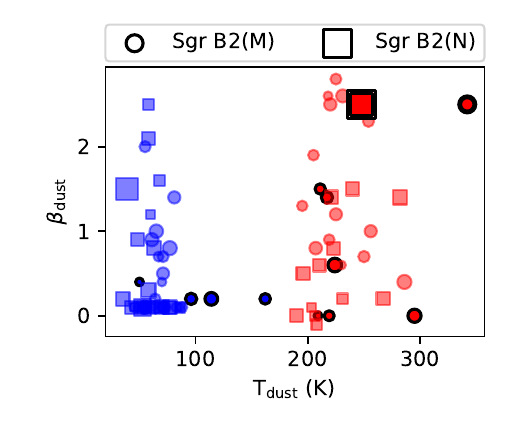}\\
   \caption{Derived dust temperatures and spectral indices for each envelope (blue) and core (red) layer in Sgr~B2(M) (circles) and N (squares). Layers containing RRLs are indicated by a black border. The size of each scatter point corresponds to the corresponding source size.}
   \label{fig:AllContTdustBeta}
\end{figure}

\subsubsection{Errors of continuum parameters}\label{subsusbsec:errorscont}

The errors of the continuum parameters described in Tabs.~\ref{Tab:SgrB2MContEnv}~-~\ref{Tab:SgrB2NDustPara}, are derived by two different methods: The errors of the dust temperatures for spectra without RRLs describe the standard errors of the corresponding means, while the errors of the other parameters were estimated using the \texttt{emcee}\footnote{\url{https://emcee.readthedocs.io/en/stable/}} package \citepads{2013PASP..125..306F}, which implements the affine-invariant ensemble sampler of \citetads{2010CAMCS...5...65G}, to perform a Markov chain Monte Carlo (MCMC) algorithm approximating the posterior distribution of the model parameters by random sampling in a probabilistic space. Here, the MCMC algorithm starts at the estimated maximum of the likelihood function, that is the continuum model parameters described in Tabs.~\ref{Tab:SgrB2MContEnv}~-~\ref{Tab:SgrB2NDustPara}, and draws 30 samples (walkers) of model parameters from the likelihood function in a small ball around the a priori preferred position. For each parameter we used 500 steps to sample the posterior. The probability distribution and the corresponding highest posterior density (HPD) interval of each continuum parameter are calculated afterwards. Details of the HPD interval are described in \citetads{2021A&A...651A...9M}. In order to get a more reliable error estimation, the errors for the hydrogen column densities and emission measures are calculated on log scale, i.e.\ these parameters are converted to their log10 values before applying the MCMC algorithm and converted back to linear scale after finishing the error estimation procedure. For most sources, the log10-errors of the hydrogen column densities and spectral indices are tiny, so the linear values are usually on the order of one. Finally the errors of the ratios $\gamma$ are calculated using the continuum parameters, where each parameter is reduced (enhanced) by the corresponding left (right) error value.

The posterior distributions of the continuum parameters of core A17 in Sgr~B2(M) are shown in Fig.~\ref{fig:ErrorEstim}, the distributions for the other cores are shown in the appendix~\ref{app:ErrorSgrB2}. For all histograms we find a more or less unimodal distribution, that is only one best fit within the given parameter ranges.

\begin{figure}[!tb]
   \centering
   \includegraphics[width=0.5\textwidth]{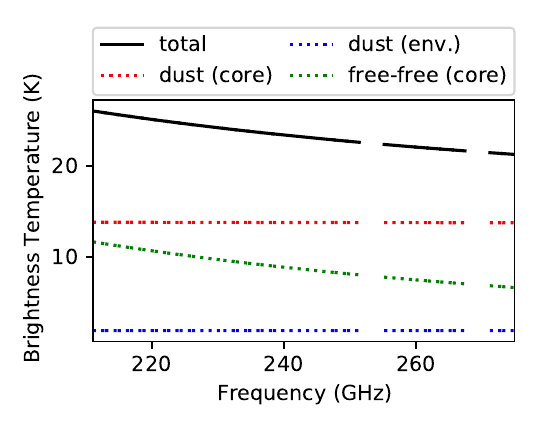}\\
   \caption{Continuum contributions to core A17 in Sgr~B2(M). Each contribution is computed without taking the interaction with other contributions into account.}
   \label{fig:ContContrSgrB2MA17}
\end{figure}

\begin{center}
\begin{table*}[!htb]
\caption{Physical parameters for each source in Sgr~B2(M) and Sgr~B2(N).}
\label{Tab:PhysParameters}
\centering
\tiny
\begin{tabular}{l@{\hspace{0.2cm}}c@{\hspace{0.2cm}}c@{\hspace{0.2cm}}c@{\hspace{0.2cm}}c@{\hspace{0.2cm}}c@{\hspace{0.2cm}}c@{\hspace{0.2cm}}c@{\hspace{0.2cm}}c@{\hspace{0.2cm}}c@{\hspace{0.2cm}}c@{\hspace{0.2cm}}}
    \hline
    \hline
Source  &        $\theta^{\rm core}$ &   M$_{\rm d+g}^{\rm core}$ & n$_{{\rm H}_{2}}^{\rm core}$ &   n$_{\rm e}^{\rm core}$ &   M$_{\rm i}^{\rm core}$ & $\Dot{\rm N}_{\rm i}^{\rm core}$ &  R$_{\rm St}^{\rm core}$ &   c$_{\rm s}^{\rm core}$ & t$_{\rm exp}^{\rm core}$ &        $\frac{P_{\rm e}^{\rm core}}{P_{\rm M}^{\rm core}}$ \\
        &                     ($''$) &              (M$_{\odot}$) &           ($10^7$ cm$^{-3}$) &       ($10^4$ cm$^{-3}$) &  ($10^{-3}$ M$_{\odot}$) &             ($10^{47}$ s$^{-1}$) &           ($10^{-3}$ pc) &            (km s$^{-1}$) &           ($10^{4}$ yrs) &                                                         () \\
    \hline
    \hline
    \multicolumn{11}{l}{Sgr~B2(M)} \\
    \hline
A01     &                 1.66 (0.6) &             197.36 (727.0) &                11.10 (111.1) &             20.68 (44.0) &            810.80 (72.0) &                   1116.48 (98.0) &                        - &                        - &                        - &                                                          - \\
A02     &                 1.87 (0.6) &             297.32 (815.0) &                10.27 (103.2) &                 - (30.0) &                 - (64.0) &                         - (59.0) &                        - &                        - &                        - &                                                          - \\
A03     &                 1.18 (0.7) &             107.25 (286.0) &                 17.52 (29.8) &                        - &                        - &                                - &                        - &                        - &                        - &                                                          - \\
A04     &                 1.16 (0.6) &              70.82 (183.0) &                  8.51 (25.5) &              13.67 (9.4) &            180.70 (18.0) &                      80.08 (5.2) &                        - &                        - &                        - &                                                          - \\
A05     &                 1.23 (0.6) &              78.85 (191.0) &                 10.47 (27.7) &                        - &                        - &                                - &                        - &                        - &                        - &                                                          - \\
A06     &                 1.40 (0.6) &              59.44 (150.0) &                  3.15 (21.7) &                        - &                        - &                                - &                        - &                        - &                        - &                                                          - \\
A07     &                 1.20 (0.5) &              58.19 (126.0) &                  5.99 (23.2) &               8.96 (5.2) &             132.80 (6.5) &                      56.41 (1.1) &                        - &                        - &                        - &                                                          - \\
A08     &                 0.98 (0.3) &               21.97 (46.0) &                  1.01 (19.6) &             19.65 (30.0) &            159.67 (11.0) &                    114.75 (10.0) &                        - &                        - &                        - &                                                          - \\
A09     &                 1.49 (0.6) &              54.16 (109.0) &                  3.27 (15.8) &                        - &                        - &                                - &                        - &                        - &                        - &                                                          - \\
A10$^*$ &                 1.67 (1.1) &                  - (160.0) &                      - (6.9) &                        - &                        - &                                - &                        - &                        - &                        - &                                                          - \\
A11     &                 1.59 (0.6) &               34.10 (84.0) &                  1.24 (12.1) &                        - &                        - &                                - &                        - &                        - &                        - &                                                          - \\
A12     &                 1.38 (0.8) &               38.89 (87.0) &                   2.54 (6.8) &                        - &                        - &                                - &                        - &                        - &                        - &                                                          - \\
A13     &                 1.45 (0.9) &               37.40 (88.0) &                   2.09 (5.3) &                        - &                        - &                                - &                        - &                        - &                        - &                                                          - \\
A14     &                 1.08 (0.6) &               22.69 (48.0) &                   2.73 (6.1) &                        - &                        - &                                - &                        - &                        - &                        - &                                                          - \\
A15     &                 0.94 (0.6) &                 2.65 (5.0) &                   0.09 (0.8) &             13.16 (16.0) &             92.01 (23.0) &                     41.06 (11.0) &                     3.19 &                    12.23 &                     0.31 &                                                      11.17  \\
A16     &                 1.72 (1.2) &               17.93 (39.0) &                   1.99 (1.3) &               9.63 (6.6) &           419.22 (100.0) &                    129.97 (21.0) &                        - &                        - &                        - &                                                          - \\
A17     &                 1.60 (1.2) &               16.06 (45.0) &                   0.79 (1.5) &               8.45 (4.0) &            295.55 (67.0) &                      38.41 (8.2) &                        - &                        - &                        - &                                                          - \\
A18$^*$ &                 0.98 (0.7) &                   - (21.0) &                      - (2.1) &                        - &                        - &                                - &                        - &                        - &                        - &                                                          - \\
A19     &                 1.13 (0.9) &               12.12 (29.0) &                   1.86 (1.9) &                        - &                        - &                                - &                        - &                        - &                        - &                                                          - \\
A20$^*$ &                 1.08 (1.0) &                   - (26.0) &                      - (1.3) &                        - &                        - &                                - &                        - &                        - &                        - &                                                          - \\
A21     &                 0.71 (0.6) &                4.61 (10.0) &                   4.55 (1.5) &                        - &                        - &                                - &                        - &                        - &                        - &                                                          - \\
A22     &                 0.76 (0.7) &                5.00 (12.0) &                   4.31 (1.4) &                        - &                        - &                                - &                        - &                        - &                        - &                                                          - \\
A23     &                 1.09 (1.0) &               10.47 (22.0) &                   4.11 (1.1) &                        - &                        - &                                - &                        - &                        - &                        - &                                                          - \\
A24     &                 0.55 (0.4) &                   1.16 (-) &                     6.59 (-) &             39.47 (19.0) &              55.26 (8.3) &                      47.96 (4.8) &                     0.23 &                    16.03 &                     0.70 &                                                       0.87 \\
A25$^*$ &                 0.73 (0.6) &                    - (9.0) &                      - (1.2) &                        - &                        - &                                - &                        - &                        - &                        - &                                                          - \\
A26     &                 1.03 (0.9) &                9.44 (18.0) &                   5.10 (1.0) &                        - &                        - &                                - &                        - &                        - &                        - &                                                          - \\
A27$^*$ &                 0.74 (0.6) &                    - (8.0) &                      - (1.0) &                        - &                        - &                                - &                        - &                        - &                        - &                                                          - \\
    \hline
    \hline
    \multicolumn{7}{l}{Sgr~B2(N)} \\
    \hline
A01     &                 4.55 (1.4) &           1542.64 (4083.0) &                  3.63 (98.6) &               4.51 (4.8) &          3641.60 (130.0) &                    509.88 (19.0) &                        - &                        - &                        - &                                                          - \\
A02     &                 1.74 (0.6) &             132.48 (294.0) &                  4.98 (36.6) &                        - &                        - &                                - &                        - &                        - &                        - &                                                          - \\
A03     &                 1.57 (0.7) &              91.54 (221.0) &                  4.29 (22.7) &                  - (4.4) &                 - (13.0) &                          - (1.7) &                        - &                        - &                        - &                                                          - \\
A04     &                 1.53 (0.8) &              97.46 (198.0) &                  4.85 (16.9) &                        - &                        - &                                - &                        - &                        - &                        - &                                                          - \\
A05     &                 1.54 (0.8) &              75.23 (163.0) &                  2.94 (11.6) &                        - &                        - &                                - &                        - &                        - &                        - &                                                          - \\
A06     &                 1.74 (0.9) &              86.90 (163.0) &                  2.28 (10.2) &                        - &                        - &                                - &                        - &                        - &                        - &                                                          - \\
A07     &                 1.06 (0.4) &               25.14 (49.0) &                  3.92 (12.7) &                        - &                        - &                                - &                        - &                        - &                        - &                                                          - \\
A08$^*$ &                 1.30 (0.6) &                   - (49.0) &                      - (7.1) &                        - &                        - &                                - &                        - &                        - &                        - &                                                          - \\
A09     &                 1.58 (1.0) &               37.53 (66.0) &                   1.41 (3.5) &                        - &                        - &                                - &                        - &                        - &                        - &                                                          - \\
A10$^*$ &                 1.03 (0.6) &                   - (27.0) &                      - (3.4) &                  - (7.7) &                 - (17.0) &                          - (4.1) &                        - &                        - &                        - &                                                          - \\
A11$^*$ &                 1.20 (0.8) &                   - (38.0) &                      - (2.7) &                        - &                        - &                                - &                        - &                        - &                        - &                                                          - \\
A12$^*$ &                 1.70 (1.1) &                   - (66.0) &                      - (2.8) &                        - &                        - &                                - &                        - &                        - &                        - &                                                          - \\
A13$^*$ &                 2.71 (2.0) &                  - (166.0) &                      - (2.1) &                        - &                        - &                                - &                        - &                        - &                        - &                                                          - \\
A14     &                 1.87 (1.5) &               27.71 (79.0) &                   0.75 (1.8) &                        - &                        - &                                - &                        - &                        - &                        - &                                                          - \\
A15     &                 1.05 (0.8) &               10.77 (23.0) &                   3.54 (1.9) &                        - &                        - &                                - &                        - &                        - &                        - &                                                          - \\
A16$^*$ &                 1.84 (1.6) &                   - (75.0) &                      - (1.5) &                 3.71 (-) &               198.30 (-) &                        22.04 (-) &                        - &                        - &                        - &                                                          - \\
A17     &                 0.74 (0.5) &                6.37 (12.0) &                   1.39 (2.2) &                        - &                        - &                                - &                        - &                        - &                        - &                                                          - \\
A18     &                 1.59 (1.4) &               22.73 (57.0) &                   0.76 (1.6) &                        - &                        - &                                - &                        - &                        - &                        - &                                                          - \\
A19     &                 0.92 (0.8) &                9.83 (19.0) &                   2.69 (1.4) &                        - &                        - &                                - &                        - &                        - &                        - &                                                          - \\
A20$^*$ &                 0.60 (0.4) &                    - (6.0) &                      - (1.4) &                  - (5.8) &                  - (4.3) &                          - (0.8) &                        - &                        - &                        - &                                                          - \\
    \hline
    \hline
\end{tabular}
\begin{tablenotes}
   \item Here, $\theta^{\rm core}$, indicates the source size, Eq.~\eqref{myxclass:SourceSize}, M$_{\rm d+g}^{\rm core}$ the dust and gas masses, Eq.~\eqref{cont:DustGasMass}, n$_{{\rm H}_{2}}^{\rm core}$ the hydrogen density, Eq.~\eqref{cont:HydrogenDensity}, n$_{\rm e}^{\rm core}$ the electron density, Eq.~\eqref{cont:ElectronDensity}, M$_{\rm i}^{\rm core}$ the ionized gas mass, Eq.~\eqref{cont:IonizedMass}, $\Dot{\rm N}_{\rm i}^{\rm core}$ the number of ionizing photons per second, Eq.~\eqref{cont:NumPhotons}, R$_{\rm St}^{\rm core}$ the initial Strömgren radius, Eq.~\eqref{cont:RStroemgren}, c$_{\rm s}^{\rm core}$ isothermal sound speed, Eq.~\eqref{cont:cs}, t$_{\rm exp}^{\rm core}$ the dynamical age of an \hii~region, Eq.~\eqref{cont:HIIAge}, and $\frac{P_{\rm e}^{\rm core}}{P_{\rm M}^{\rm core}}$ the ratio of the electron and the molecular pressure, Eq.~\eqref{cont:equicond}, respectively. The values in round brackets indicate the values obtained by \citetads{2017A&A...604A...6S}.  For sources marked with an "$^*$", we could not derive a self-consistent description of the continuum of the corresponding core spectrum.
\end{tablenotes}
\end{table*}
\end{center}

\section{Results and discussion}\label{sec:Results}

\subsection{Results}\label{susbsec:results}

In both regions, Sgr~B2(M) and Sgr~B2(N), most of the cores and their local envelopes are dominated by the contribution of thermal dust, where the dense, dust-dominated cores are optically thick toward the center and optically thin in the outer regions, see Figs.~\ref{fig:AllContTdustNH}~-~\ref{fig:AllContTdustBeta}. The greater number of \hii~regions in Sgr~B2(M) cause strong thermal free-free emissions, which for some cores are the dominant continuum contributions to the total continuum levels. However, we also detect ionized gas localized between the sources, as we found in the envelope around cores A09 and A26 in Sgr~B2(M).

\subsubsection{Results of Sgr~B2(M)}\label{subsusbsec:resultsM}

For local envelopes around the hot cores in Sgr~B2(M) we found only a small variation of the dust parameters except for those envelopes containing RRLs (A07, A09, A24, and A26). The dust temperatures vary between 46 K (A19) and 88 K (A08) for envelopes whose spectra do not contain RRLs. For envelopes where free-free ionized gas emission has been found, the dust temperatures are located in the range of 50~K (A24) and 162~K (A26). Additionally, the envelope around the central core A01 shows the highest hydrogen column density of $1.1 \times 10^{25}$~cm$^{-2}$, while the lowest hydrogen column densities are found for envelopes containing RRLs. For the majority of spectra we find spectral dust indices of $\beta = 0.1$ ($\alpha = 2.1$) (A05, A08, A10, A11, A13, A14, A17, A18, A19, A21, A22, A23, A25), which are associated with optically thick dust emission (see e.g.\, \citetads{2014MNRAS.444.2303S}, \citetads{2015ApJ...811..118R}). For core A03, we derive an index of $\beta = 2.0$ ($\alpha = 4.0$), which indicates optically thin dust emission. Envelopes containing free-free emission in their spectra have spectral dust indices between 0.2 and 0.4, i.e.\ moderate optically thick dust emission. The contributions of the dust emission from the envelopes to the total continua observed toward the hot cores alter between 5.4~\% (A03) and 32.4~\% (A14).
%
From the RRLs we obtain electronic temperatures between 2781~(A09) and 5817~K (A07), finding no correlation with dust temperatures. The corresponding emission measures are located within a small range between $8.8 \times 10^{6}$~(pc~cm$^{-6}$) (A07) and $2.2 \times 10^{7}$~(pc~cm$^{-6}$) (A24). The free-free emissions in the envelopes contribute almost nothing to the observed continuum and range from 0.5~\% (A07) and 2.2~\% (A26).
%
Eight hot core spectra in Sgr~B2(M) contain RRLs. For two other sources (A09, A26) we could not detect RRLs in the corresponding core spectra, although RRLs are found in their envelopes, which may be due to the fact that there is ionized gas localized between the sources and that we used averaged core spectra, see e.g.\, Fig.~\ref{fig:MapSgrB2NA14}, where possible contributions from individual pixels with a small RRL contribution were averaged out. For example, the polygon used to calculate the averaged core spectrum of A26 contains a small \hii~region identified by \citetads{2015ApJ...815..123D}, see Fig.~\ref{fig:CorePosSgrB2M}, but we could not detect any RRLs in the corresponding core spectrum. On the other hand, the identification of RRLs in the core spectrum of A17 in Sgr~B2(M) is remarkable, because \citetads{2015ApJ...815..123D} do not describe an \hii~region in the vicinity of this source. The hot cores in Sgr~B2(M) have dust temperatures ranging from 195~K (A26) to 342~K (A01), but unlike the envelope spectra, we do not find a general correlation between cores containing RRLs and high dust temperatures, which again may be due to the use of averaged core spectra. Moreover, we obtain the highest hydrogen column density not for the central core A01, as \citetads{2017A&A...604A...6S}, but for core A03. This discrepancy could be due to the fact that, unlike \citetads{2017A&A...604A...6S}, we decompose the contributions from the envelope and core, where the envelope around core A01 has the highest hydrogen column density of all envelopes in Sgr~B2(M). Furthermore, the continuum of core A01 contains strong free ionized gas emissions. In contrast to the envelopes the majority of core spectra shows spectral dust indices above 0.4. Only three cores have spectral indices of $\beta = 0.0$ ($\alpha = 0.0$). Half of all cores (A01, A04, A07, and A08) containing RRLs have spectral indices above 1.0, which are associated with optically thin dust emission. For cores whose spectra do not contain RRLs and for which we could derive a self-consistent description of the continuum level, we obtain approximately the same spectral indices as \citetads{2017A&A...604A...6S}.

The obtained electronic temperatures T$_{\rm e}$ show a large variation between 3808~K (A01) and 23779~K (A17). Although electronic temperatures above 10000~K are unusual for \hii-regions (\citetads{1967ApJ...149L..61Z}, \citetads{2002ASSL..282.....G}), the temperatures agree quite well with those derived by \citetads{2006ApJ...653.1226Q}, where temperatures between 1850~K and 21810~K were found using high-precision radio recombination line and continuum observations of more than 100~\hii~regions in the Galactic disk. Taking Non-LTE effects into account, \citetads{1995ApJ...442L..29M} derived an electronic temperature of T$_{\rm e}$~=~23700~K towards a source in Sgr~B2(M), which is very close to the LTE temperature we determined for source A17. Additionally, \citetads{2006ApJ...653.1226Q} identified a Galactic \hii~region (G220.508-2.8) with an electronic temperature of T$_{\rm e}$~=~21810~K. Furthermore, the error estimation for A17, see Fig.~\ref{fig:ErrorEstim}, shows that the best description of the free-free contribution occurs at the indicated temperature. Although we cannot rule out the possibility that the slope of the continuum level of core A17 is affected by interferometric filtering, this high temperature seems to actually exist. But it is unclear what mechanism heats the gas to such high temperatures. The electron temperature of an \hii~region in thermal equilibrium is determined by the balance of competing heating and cooling mechanisms. Among others, the electronic temperature T$_{\rm e}$ can be influenced by the effective temperature of the ionizing star or by the electron density, which inhibits cooling and increases T$_{\rm e}$ by collisional excitation in the high electron density \hii~regions. Additionally, T$_{\rm e}$ is affected by the dust grains, which are involved in heating and cooling in complex ways (see e.g.\, \citetads{1986PASP...98..995M}, \citetads{1991ApJ...374..580B}, \citetads{1995ApJ...454..807S}). Photoelectric heating occurs due to the ejection of electrons from the dust grains, while the gas is cooled by collisions of fast particles with the grains. Furthermore, the electron temperature decreases with distance from the star, because the field of ionizing radiation is attenuated by dust grains. However, the electron temperature also increases when coolants are depleted at the dust grains. According to \citepads{1980pim..book.....D}, photoionization is not able to heat the material to such a high temperature unless there is a strong depletion of metals in this region compared to other \hii~regions in Sgr~B2. Since heavy elements cool photoionized gas, the electron temperatures of \hii~regions are directly related to the abundance of the heavy elements: A low electron temperature T$_{\rm e}$ corresponds to a higher heavy element abundance due to the higher cooling rate and vice versa. Another possibility is that the heating is caused by a non-equilibrium situation at the edge of the expanding \hii~region. The electronic temperatures of the other cores correspond quite well to those obtained by \citetads{1993ApJ...412..684M}, suggesting that the metal abundance in most cores of Sgr~B2 is similar to that of Orion.\\

The corresponding emission measures ranges from $4.7 \times 10^{8}$~(pc~cm$^{-6}$) (A17) to $4.4 \times 10^{9}$~(pc~cm$^{-6}$) (A01). The free-free contributions to the total continuum levels vary between 15.7~\% and 75.8~\%, where the high contributions are caused by the fact, that seven of the hot cores in Sgr~B2(M) contain one or more \hii~regions, see Fig.~\ref{fig:CorePosSgrB2M}, while core A17 was not associated with \hii~regions before, which may be caused by the strong dust contribution, see Fig.\ref{fig:ContContrSgrB2MA17}.

\subsubsection{Results of Sgr~B2(N)}\label{subsusbsec:resultsN}

Similar to Sgr~B2(M), we found only a small variation of the dust parameters for the local envelopes around the hot cores in Sgr~B2(N). The dust temperatures are located in a range between 35~K (A02) and 86~K (A15) while the column densities vary between $5.1 \times 10^{23}$~cm$^{-2}$ (A12) and $1.0 \times 10^{25}$~cm$^{-2}$ (A01). Similar to the envelope spectra around sources in Sgr~B2(M), we find spectral dust indices for most of the envelopes (A04, A08, A09, A10, A12, A13, A14, A15, A18, A19, and A20) in Sgr~B2(N) of 0.1. The highest dust index of $\beta = 2.5$ ($\alpha = 4.5$) is found for the envelope around core A07. In contrast to Sgr~B2(M), we do not find RRLs in the envelope spectra in  Sgr~B2(N). The dust emissions from the envelopes contribute between 6.6~\% (A03) and 71.2~\% (A17) to the total continuum level. The strong contribution of the envelope around source A17 shows that the contribution of the envelope is indispensable for a realistic modeling of a sources with low continuum levels.
%
For the hot cores in Sgr~B2(N) we derived dust temperatures between 190~K (A09) and 282~K (A14), where the high core dust temperature of source A19 is noteworthy, because this source is not associated with an \hii~region. However, we cannot exclude an influence of interferometric filtering on molecular lines used for temperature estimation. The column densities alter between $1.3 \times 10^{24}$~cm$^{-2}$ (A17) and $2.1 \times 10^{25}$~cm$^{-2}$ (A01) and the spectral index between  $\beta = -0.1$ ($\alpha = 1.9$) (A07) and $\beta = 2.5$ ($\alpha = 4.5$) (A01). Only two cores contain RRLs with electronic temperatures of 9877~K (A01) and 9921~K (A16). The corresponding emission measures are located in a range between $1.0 \times 10^{8}$~(pc~cm$^{-6}$) (A16) and $3.9 \times 10^{8}$~(pc~cm$^{-6}$) (A01), which is almost an order of magnitude lower than the highest emission measure for core A01 in Sgr~B2(M). For A01 the free-free contribution is 10.5~\%.

In the following we present in more detail the results obtained from the fitting towards the cores of Sgr~B2(M) and (N).

%
%

\subsection{Physical properties of the continuum sources}

In Tab.~\ref{Tab:PhysParameters}, we summarize the physical parameters derived from the continuum parameters described above. Here, we compute the (dust and gas) masses for each source determined from the expression \citepads{1983QJRAS..24..267H}
\begin{equation}\label{cont:DustGasMass}
  M_{\rm d + g} = \frac{S_\nu \cdot D^2}{B_\nu (T_{\rm dust}^{\rm core}) \cdot \kappa_\nu},
\end{equation}
where $S_\nu$ is the flux density at 242 GHz, $D$ is the distance (8.34~kpc for Sgr~B2), $B_\nu (T_{\rm dust}^{\rm core})$ is the Planck function at a core dust temperature $T_{\rm dust}^{\rm core}$, and $\kappa_\nu$ is the absorption coefficient per unit of total mass (gas and dust) density. Assuming a spherical and homogeneous core, we use the following expression to estimate the hydrogen density $n_{H_2}^{\rm core}$ from the hydrogen column density $N_{H_2}^{\rm core}$
\begin{equation}\label{cont:HydrogenDensity}
  n_{H_2}^{\rm core} = \frac{N_{H_2}^{\rm core}}{d^{\rm core}},
\end{equation}
where $d^{\rm core} = D \cdot \theta^{\rm core}$ indicates the diameter and $\theta^{\rm core}$ the corresponding source size of the source. The electron density $n_e^{\rm core}$ is computed in a similar way using the derived emission measure EM$^{\rm core}$
\begin{equation}\label{cont:ElectronDensity}
  n_e^{\rm core} = \sqrt{ \frac{{\rm EM}^{\rm core}}{d^{\rm core}}},
\end{equation}
where we assume spherical and homogeneous \hii~regions.

The ionized gas mass $M_i^{\rm core}$ is estimated using the following expression
\begin{equation}\label{cont:IonizedMass}
  M_i^{\rm core} = n_e^{\rm core} \cdot \frac{4}{3} \, \pi \, \left(\frac{d^{\rm core}}{2}\right)^3 \cdot m_p,
\end{equation}
where $m_p$ indicates the proton mass. Finally we calculate the number of ionizing photons per second $\Dot{N_i}^{\rm core}$ \citepads{2016A&A...588A.143S}
\begin{align}\label{cont:NumPhotons}
  \Dot{N}_i^{\rm core} &= \int n_e^2 \, \left(\tilde{\beta} - \tilde{\beta_1}\right) dV,
\end{align}
where we assume a Strömgren sphere. Here, $\tilde{\beta}$ and $\tilde{\beta_1}$ are the rate coefficients for recombinations to all levels and to the ground state, respectively. The term $\left(\tilde{\beta} - \tilde{\beta_1}\right)$ describes the recombination coefficient to level 2 or higher and can be described using the expression derived by \citetads{1968ApJ...154..391R},
\begin{align}\label{cont:RecombCoeff}
    \left(\frac{\tilde{\beta} - \tilde{\beta_1}}{ {\rm cm}^3 \, {\rm s}^{-1}}\right) = 4.1 \cdot 10^{-10} \, \left( \frac{{\rm T}_e^{\rm core}}{\rm K}\right)^{-0.8},
\end{align}
who approximate the recombination coefficient given by \citetads{1959MNRAS.119...81S} for electron temperatures T$_{\rm e}$.

Unlike other cores containing \hii~regions, cores A15 and A24 in Sgr~B2(M) each match more or less in position and size a single \hii~region identified by \citetads{2015ApJ...815..123D}. In the following, we will estimate the age of these \hii~regions. We start with calculating the initial Strömgren radius ($R_{\rm St}$) of an \hii~region, which is given by \citepads{1968ITPA...28.....S}:
\begin{align}\label{cont:RStroemgren}
    R_{\rm St} = \left( \frac{3}{16 \, \pi \, \left(\tilde{\beta} - \tilde{\beta_1}\right)} \cdot \frac{\Dot{N}_i^{\rm core}}{\left[n_{H_2}^{\rm core}\right]^2}\right)^{1/3},
\end{align}
where $\Dot{N}_i^{\rm core}$ indicates the number of ionizing photons per second,  Eq.\eqref{cont:NumPhotons}, $n_{H_2}^{\rm core}$ the hydrogen density, Eq.\eqref{cont:HydrogenDensity}, and $\left(\tilde{\beta} - \tilde{\beta_1}\right)$ the recombination coefficient, Eq.\eqref{cont:RecombCoeff}, respectively.

Assuming expansion into a homogeneous molecular cloud, the dynamical age of both regions can now be computed by using the following expansion equation (\citetads{1968ITPA...28.....S}, \citetads{1980pim..book.....D})
\begin{align}\label{cont:HIIAge}
    t_{\rm exp} = \frac{4}{7} \, \frac{R_{\rm St}}{c_s} \, \left[ \left(\frac{r^{\rm core}}{R_{\rm St}} \right)^{7/4} - 1 \right],
\end{align}
where $r^{\rm core} = d^{\rm core} / 2$ describes the radius of the \hii~region and $R_{\rm St}$ the Strömgren radius, Eq.\eqref{cont:RStroemgren}. In addition, $c_s$ indicates the isothermal sound speed, which is given by \citepads{2011piim.book.....D}
\begin{align}\label{cont:cs}
    c_{\rm s} = \sqrt{ \frac{2 \, k_B \, T_e^{\rm core}}{m_H} },
\end{align}
where $m_H$ is the hydrogen atomic mass.\\

Additionally, we compute the ratio of the electron and the molecular pressure \citepads{2019PASJ...71..128T},
\begin{align}\label{cont:equicond}
    \frac{P_e}{P_M} = \frac{2 \, n_e^{\rm core} \, k_B \, T_e^{\rm core}}{n_{H_2}^{\rm core} \, k_B \, T_{\rm dust}^{\rm core}} = \frac{2 \, n_e^{\rm core} \, T_e^{\rm core}}{n_{H_2}^{\rm core} \, T_{\rm dust}^{\rm core}},
\end{align}
where $n_e^{\rm core}$ represents the electron density, Eq. \eqref{cont:ElectronDensity}, $n_{H_2}^{\rm core}$ the hydrogen density, Eq.\eqref{cont:HydrogenDensity}, $T_{\rm dust}^{\rm core}$ the dust, and $T_e^{\rm core}$ the electronic temperature, respectively. A ratio greater than one means that the corresponding \hii~region is in the expansion phase, since the pressure of the ionized gas exceeds the pressure of the neutral gas.

Compared to \citetads{2017A&A...604A...6S}, we find much lower dust and gas masses, i.e.\ between 27~\% and 57~\% of their masses, which is due to our elevated dust temperatures. \citetads{2017A&A...604A...6S} assume a dust temperature of 100~K, while our core dust temperatures range from 190~K to 342~K. However, the mass distribution of the most massive cores is almost unchanged for both regions.
%
In addition, we obtain reduced H$_2$ volume densities n$_{{\rm H}_2}$ for most sources in Sgr~B2(M) and N. For the central sources, our densities are in the range of 4~\% and 153~\% of the previous analysis results, while the volume densities for some outlying sources, especially in Sgr~B2(M), exceed the results of \citetads{2017A&A...604A...6S} by a factor of up to five (A26). These discrepancies are among others due to the different sizes of the sources used in our analysis compared to those described in \citetads{2017A&A...604A...6S}. Nevertheless, the highest hydrogen density, which we determine in our analysis is in the order of $10^8$~cm$^{-3}$, which corresponds to $10^6$~M$_\odot$~pc$^{-3}$, still one orders of magnitude larger than the typical stellar densities found in super star clusters \citepads[e.g.\ $\sim10^5$~M$_\odot$~pc$^{-3}$,][]{2010ARA&A..48..431P}.

Finally, we calculated the physical parameters of the ionized gas for the sources in which we identified RRLs. In Sgr~B2(M), we find RRLs in all sources except A02, for which \citetads{2017A&A...604A...6S} had also found contributions from ionized gas. This is quite different for Sgr~B2(N), in which we could find RRLs only for the central source A01. In all other sources which were connected with \hii~regions by \citetads{2017A&A...604A...6S} we could not find RRLs. However, we identify RRLs in A16 that were not associated with ionized gas by \citetads{2017A&A...604A...6S}. The \hii~regions we identified fit very well with the results of \citetads{2015ApJ...815..123D} for both regions. According to \citetads{2015ApJ...815..123D}, only sources A01, A10, and A16 in Sgr~B2(N) contain \hii~regions. The electron densities n$_e$ vary between 47~\% and 211~\% of the values derived from \citetads{2017A&A...604A...6S}, with the differences due to the different analysis techniques. In contrast to \citetads{2017A&A...604A...6S}, which found the highest electron density for core A01 in Sgr~B2(M), we obtain the highest electron density for core A24, which is associated with a single, bright \hii~region, see Fig.~\ref{fig:CorePosSgrB2M}. All other cores containing RRLs show comparable densities. For core A15 and A24 in Sgr~B2(M) we also determined the age of the corresponding \hii~region (A15:~3100~yr, A24:~7000~yr), which are comparable to results obtained by \citetads{2022A&A...666A..31M}. Additionally, the gas pressure ratio of the observed ionized gas and the observed ambient molecular gas is close to unity for A24, which indicates that the pressure-driven expansion for the corresponding \hii~region is coming to a halt, while for core A15, we find a remarkable pressure-driven expansion for this \hii~region.

\section{Conclusions}\label{sec:Conclusions}

Many of the hot cores identified by \citetads{2017A&A...604A...6S} include large amounts of dust. In addition, some cores contain one or more \hii~regions. In this work, which is the first of two papers on the complete analysis of the full spectral line surveys towards these hot cores, we have quantified the dust and, if contained, the free-free contributions to the continuum levels. In contrast to previous analyses, we derived the corresponding parameters here not only for each core but also for their local surrounding envelope and determined their physical properties. Especially for some outlying sources, the contributions of these envelopes are not negligible. In general, the distribution of RRLs we found in the core spectra fits well with the distribution of \hii~regions described by \citetads{2015ApJ...815..123D}. Only for core A02 in Sgr~B2(M) and A10 in Sgr~B2(N) we can not identify RRLs in the corresponding spectra, although \hii~regions are contained in these sources. Additionally, we found RRLs in core A17 of Sgr~B2(M), despite the fact that no \hii~region is known to be nearby. The average dust temperature for envelopes around sources in Sgr~B2(M) is 73~K while in Sgr~B2(N), however, we obtain only 59~K, which may be caused by the enhanced number of \hii~region in Sgr~B2(M) compared to N. For the cores we obtain average dust temperatures around 236~K (Sgr~B2(M)) and 225~K (Sgr~B2(N)) and see no correlations between occurrence of RRLs and enhanced dust temperatures, although one would expect this in the presence of ionized gas. For the average hydrogen column densities we get $2.5 \times 10^{24}$~cm$^{-2}$ ($2.6 \times 10^{24}$~cm$^{-2}$) for the envelopes and $7.8 \times 10^{24}$~cm$^{-2}$ ($6.1 \times 10^{24}$~cm$^{-2}$) for Sgr~B2(M) and N, respectively. The derived electronic temperatures are located in a range between 2781~K and 9921~K, while two cores show electronic temperatures of 15214~K and 23779~K. The highest emission measures in Sgr~B2(M) are found in cores A01 and A24, while the two cores in Sgr~B2(N) containing RRLs have almost the same emission measure. In Sgr~B2(M), the three inner sources are the most massive, whereas in Sgr~B2(N) the innermost core A01 dominates all other sources in mass and size. This analysis of the dust and ionized gas contribution to the continuum emission enables a full detailed analysis of the spectral line content which will be presented in a following paper (Möller et al. in prep.).

%


\begin{acknowledgements}
    This work was supported by the Deutsche Forschungsgemeinschaft (DFG) through grant Collaborative Research Centre 956 (subproject A6 and C3, project ID 184018867) and from BMBF/Verbundforschung through the projects ALMA-ARC 05A14PK1 and ALMA-ARC 05A20PK1. A.S.M. acknowledges support from the RyC2021-032892-I grant funded by MCIN/AEI/10.13039/501100011033 and by the European Union `Next GenerationEU'/PRTR, as well as the program Unidad de Excelencia Mar\'ia de Maeztu CEX2020-001058-M. This paper makes use of the following ALMA data: ADS/JAO.ALMA\#2013.1.00332.S. ALMA is a partnership of ESO (representing its member states), NSF (USA) and NINS (Japan), together with NRC (Canada), MOST and ASIAA (Taiwan), and KASI (Republic of Korea), in cooperation with the Republic of Chile. The Joint ALMA Observatory is operated by ESO, AUI/NRAO and NAOJ.
\end{acknowledgements}


\bibliographystyle{aa}
\bibliography{SgrB2ADS.bib}

    \newpage
    \Online
    \begin{appendix}

\section{Error estimation}\label{app:ErrorSgrB2}

\subsection{Error estimation of continuum parameters for cores in Sgr~B2(M)}\label{app:ErrorSgrB2-M}

Corner plots \citepads{{corner}} showing the one and two dimensional projections of the posterior probability distributions of the continuum parameters of each core in Sgr~B2(M). On top of each column the probability distribution for each free parameter is shown together with the value of the best fit and the corresponding left and right errors. The left and right dashed lines indicate the lower and upper limits of the corresponding highest posterior density (HPD) interval, respectively. The dashed line in the middle indicates the mode of the distribution. The blue lines indicate the parameter values of the best fit. The plots in the lower left corner describe the projected 2D histograms of two parameters and the contours the HPD regions, respectively. In order to get a better estimation of the errors, we determine the error of the hydrogen column density and the emission measure on log scale and use the velocity offset (v$_{\rm off}$) related to the source velocity of v$_{\rm LSR}$ = 64~km~s$^{-1}$.

\begin{figure*}[!htb]
   \centering
   \includegraphics[width=0.50\textwidth]{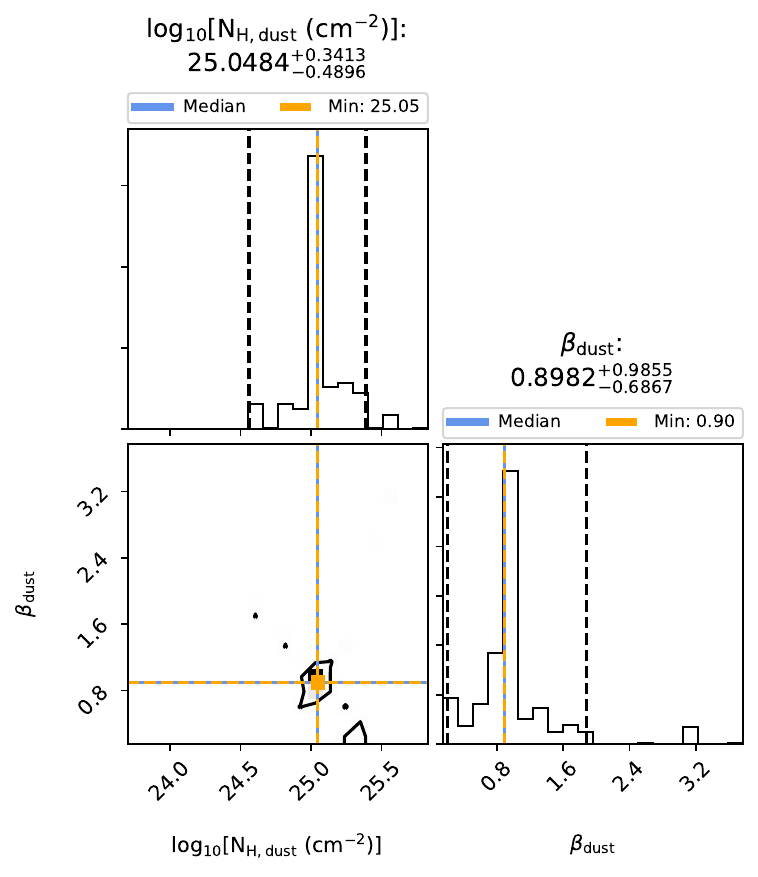}\\
   \caption{Corner plot for source A01 in Sgr~B2(M), envelope layer.}
   \label{fig:SgrB2-MErrorA01env}
\end{figure*}

\begin{figure*}[!htb]
   \centering
   \includegraphics[width=0.99\textwidth]{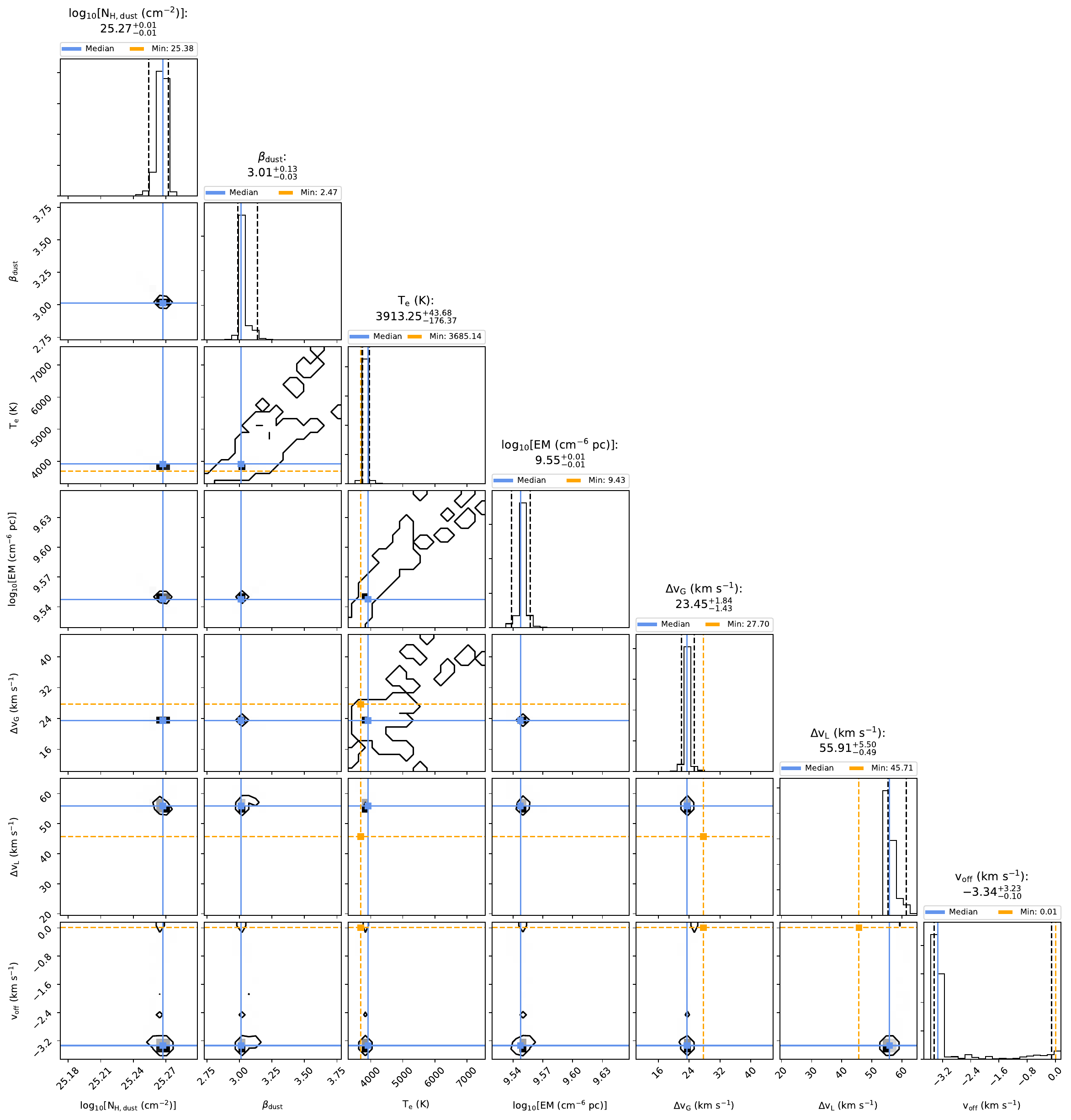}\\
   \caption{Corner plot for source A01 in Sgr~B2(M), core layer.}
   \label{fig:SgrB2-MErrorA01core}
\end{figure*}
\newpage
\clearpage

\begin{figure*}[!htb]
   \centering
   \includegraphics[width=0.50\textwidth]{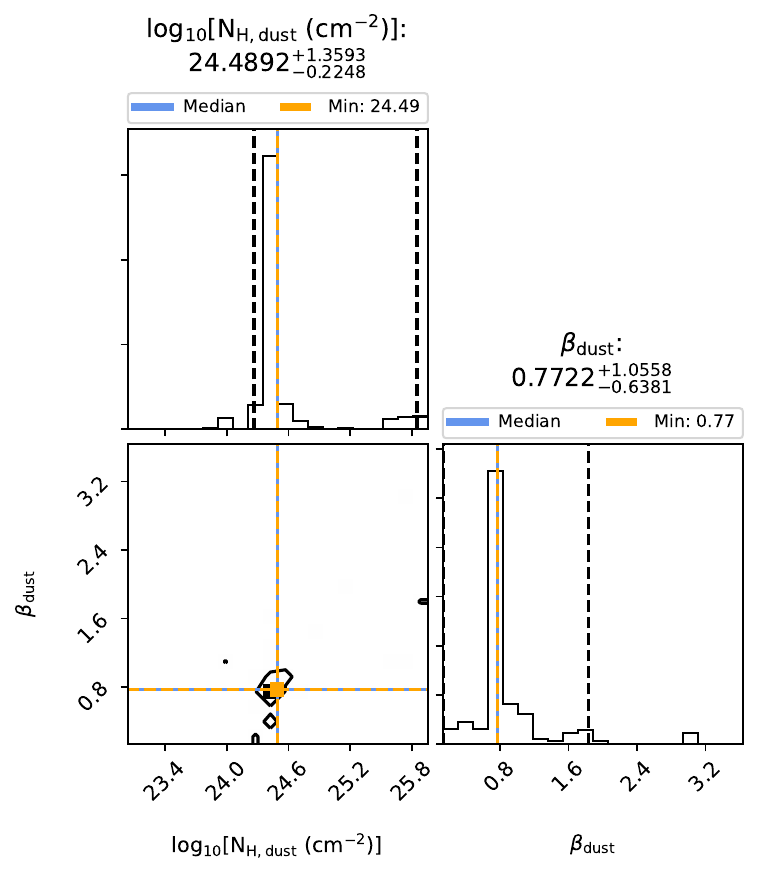}\\
   \caption{Corner plot for source A02 in Sgr~B2(M), envelope layer.}
   \label{fig:SgrB2-MErrorA02env}
\end{figure*}

\begin{figure*}[!htb]
   \centering
   \includegraphics[width=0.50\textwidth]{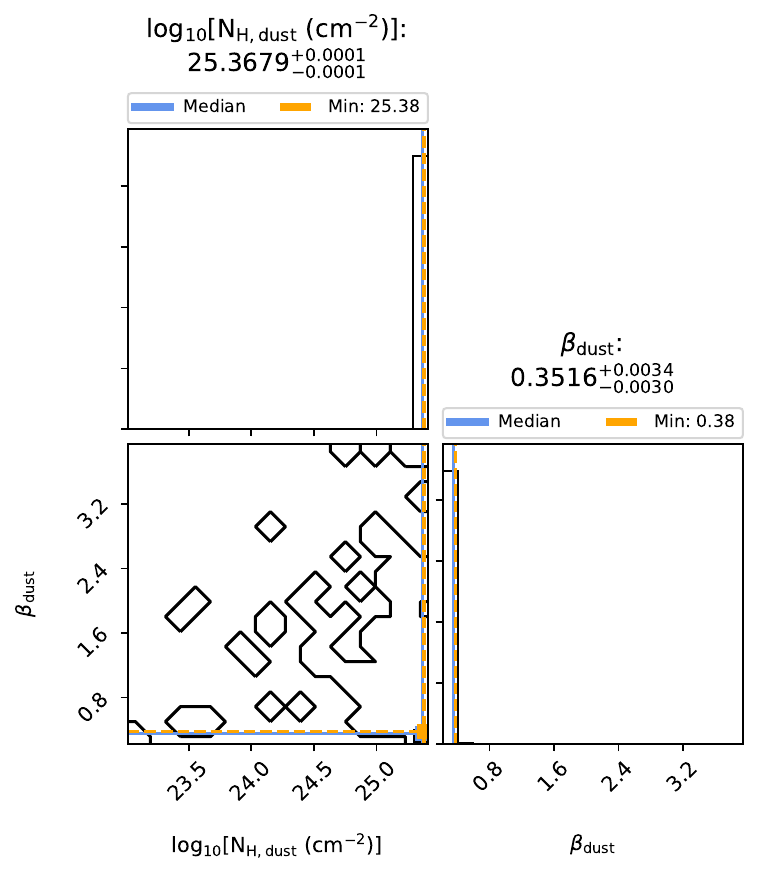}\\
   \caption{Corner plot for source A02 in Sgr~B2(M), core layer.}
   \label{fig:SgrB2-MErrorA02core}
\end{figure*}
\newpage
\clearpage

\begin{figure*}[!htb]
   \centering
   \includegraphics[width=0.50\textwidth]{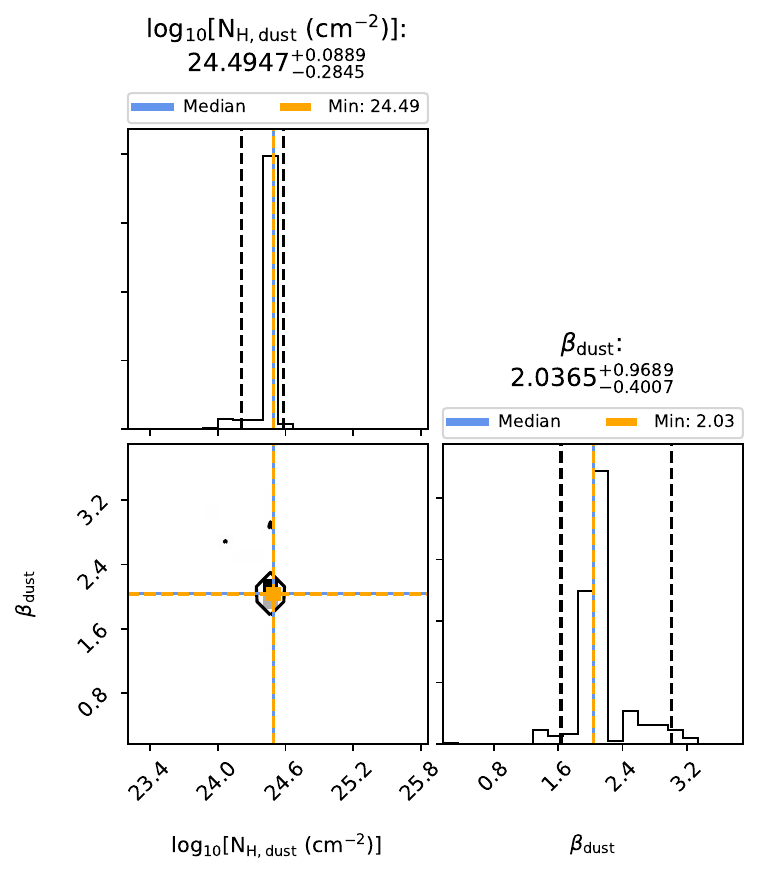}\\
   \caption{Corner plot for source A03 in Sgr~B2(M), envelope layer.}
   \label{fig:SgrB2-MErrorA03env}
\end{figure*}

\begin{figure*}[!htb]
   \centering
   \includegraphics[width=0.50\textwidth]{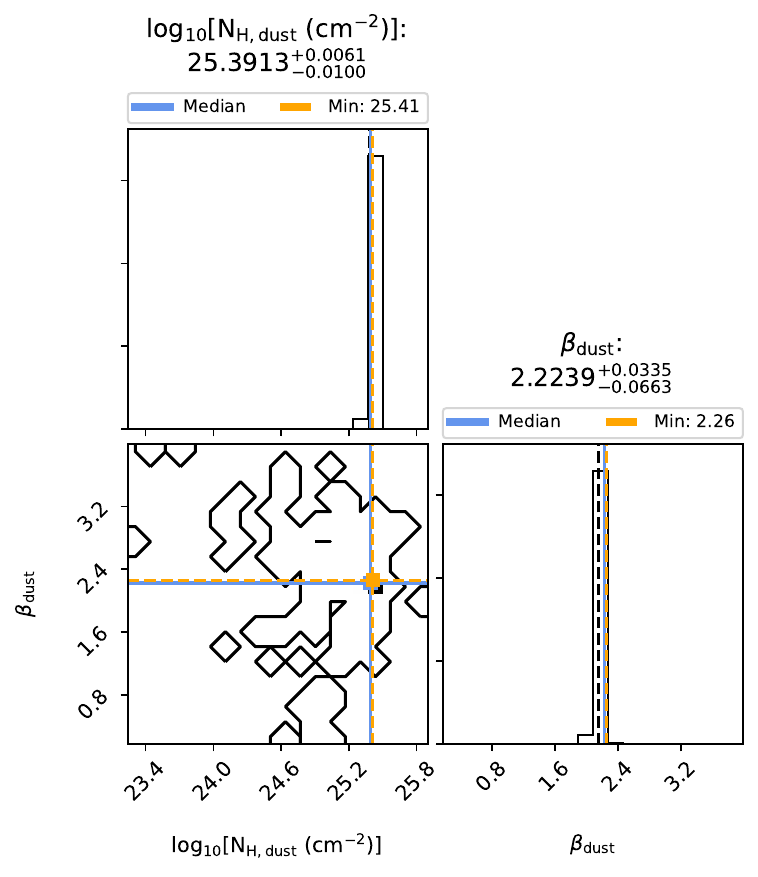}\\
   \caption{Corner plot for source A03 in Sgr~B2(M), core layer.}
   \label{fig:SgrB2-MErrorA03core}
\end{figure*}
\newpage
\clearpage

\begin{figure*}[!htb]
   \centering
   \includegraphics[width=0.50\textwidth]{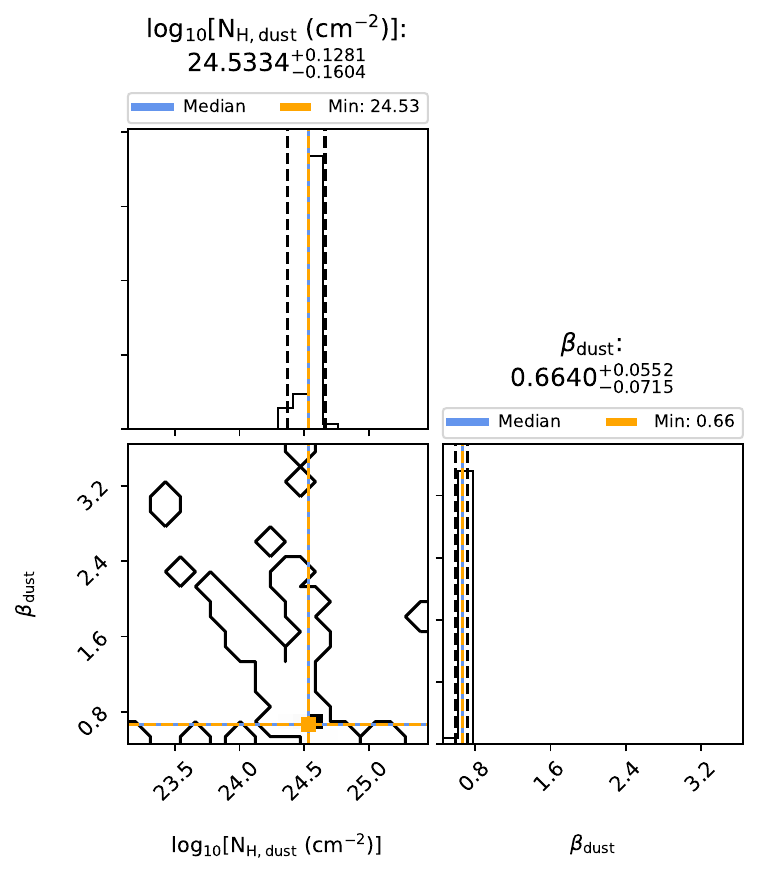}\\
   \caption{Corner plot for source A04 in Sgr~B2(M), envelope layer.}
   \label{fig:SgrB2-MErrorA04env}
\end{figure*}

\begin{figure*}[!htb]
   \centering
   \includegraphics[width=0.99\textwidth]{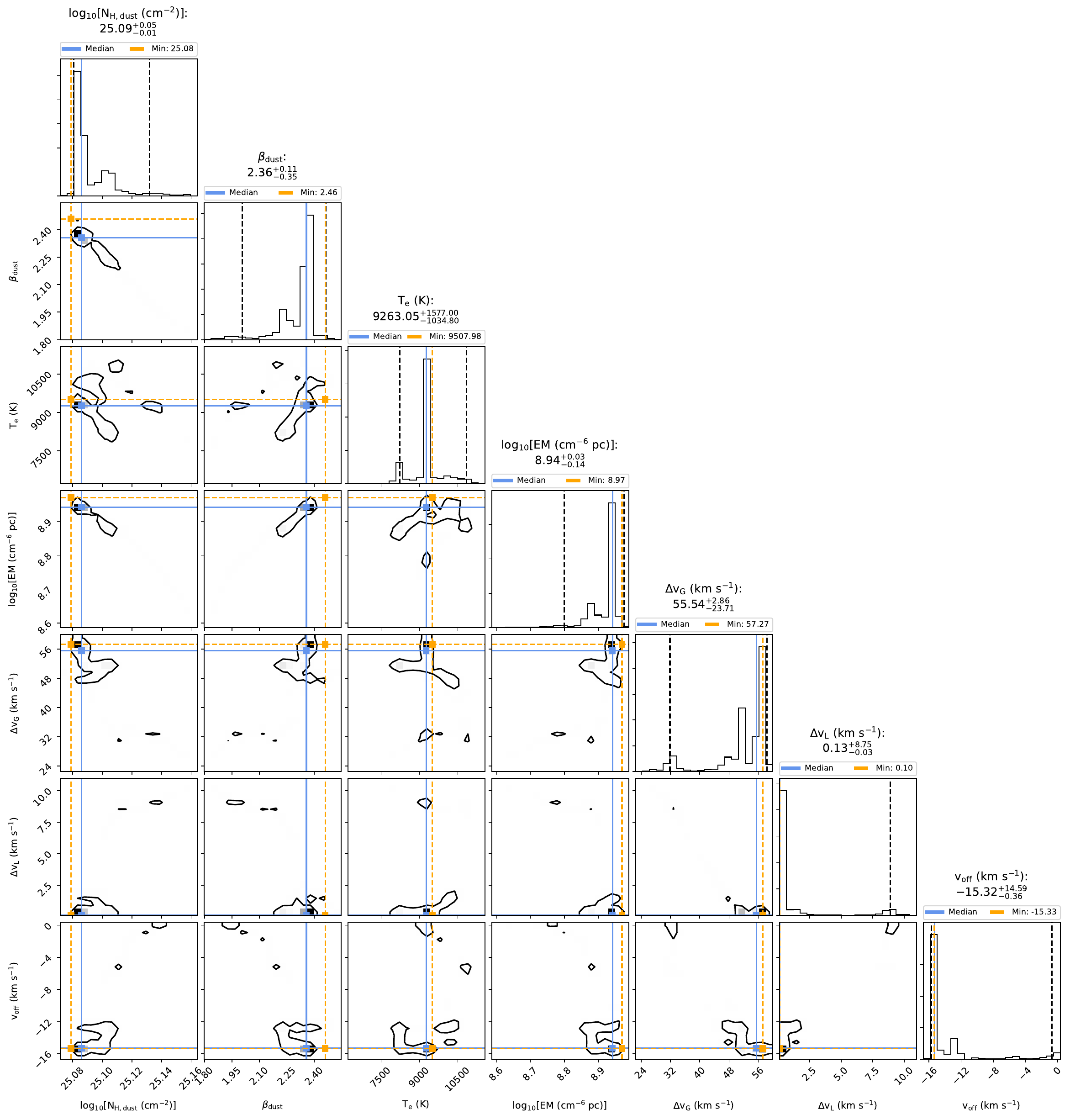}\\
   \caption{Corner plot for source A04 in Sgr~B2(M), core layer.}
   \label{fig:SgrB2-MErrorA04core}
\end{figure*}
\newpage
\clearpage

\begin{figure*}[!htb]
   \centering
   \includegraphics[width=0.50\textwidth]{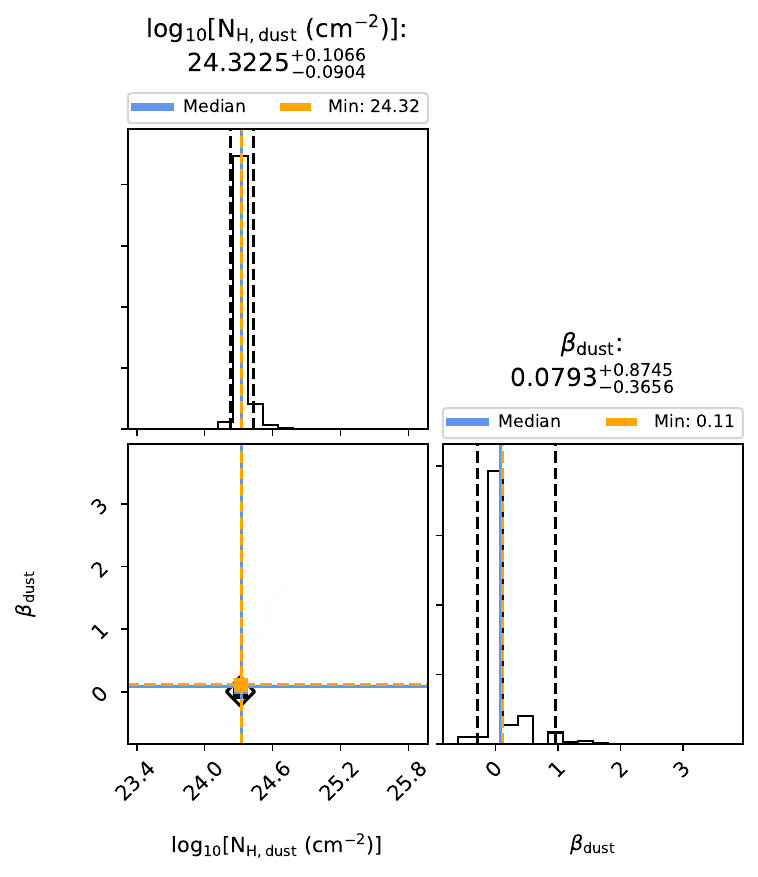}\\
   \caption{Corner plot for source A05 in Sgr~B2(M), envelope layer.}
   \label{fig:SgrB2-MErrorA05env}
\end{figure*}

\begin{figure*}[!htb]
   \centering
   \includegraphics[width=0.50\textwidth]{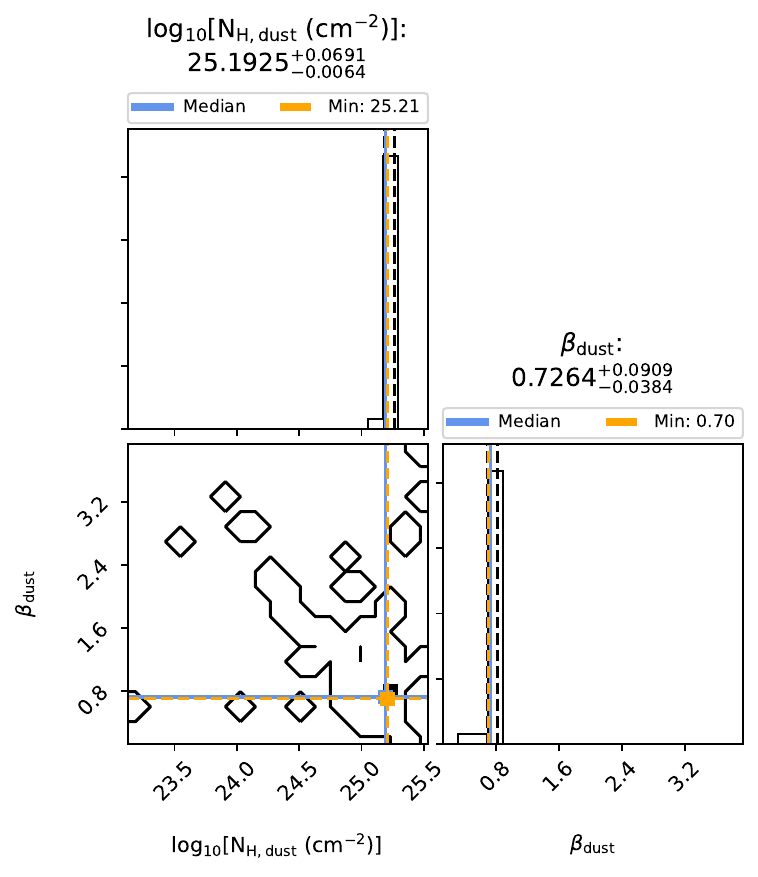}\\
   \caption{Corner plot for source A05 in Sgr~B2(M), core layer.}
   \label{fig:SgrB2-MErrorA05core}
\end{figure*}
\newpage
\clearpage

\begin{figure*}[!htb]
   \centering
   \includegraphics[width=0.50\textwidth]{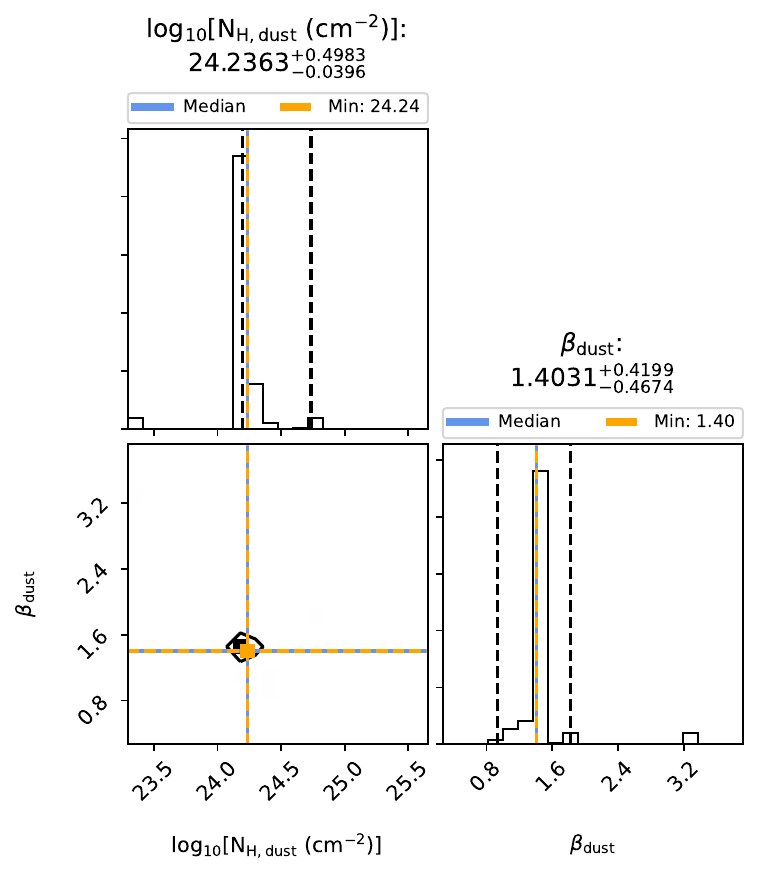}\\
   \caption{Corner plot for source A06 in Sgr~B2(M), envelope layer.}
   \label{fig:SgrB2-MErrorA06env}
\end{figure*}

\begin{figure*}[!htb]
   \centering
   \includegraphics[width=0.50\textwidth]{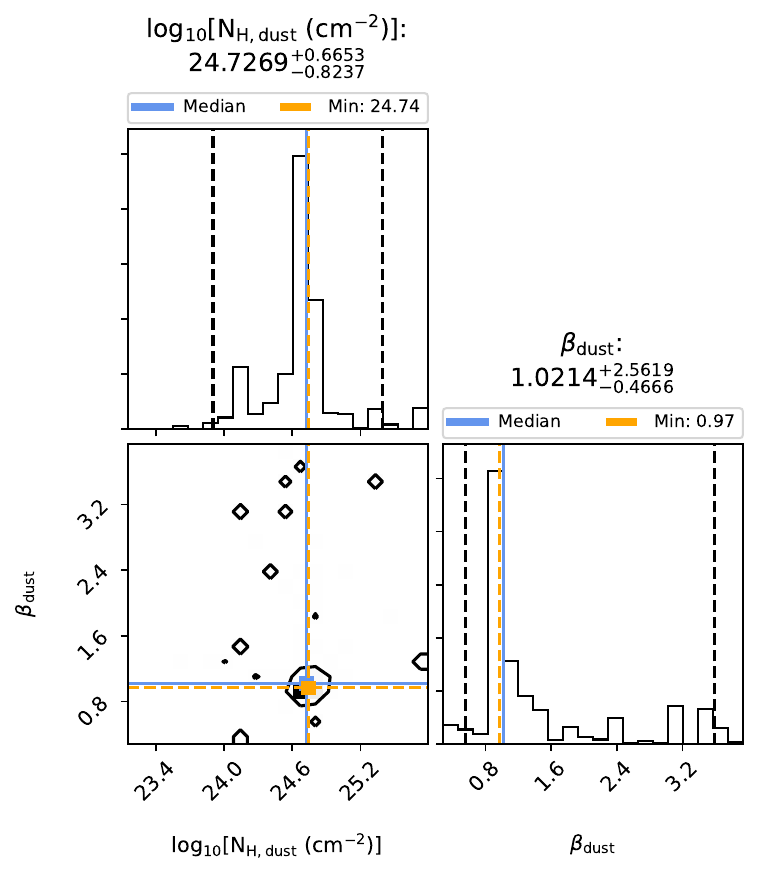}\\
   \caption{Corner plot for source A06 in Sgr~B2(M), core layer.}
   \label{fig:SgrB2-MErrorA06core}
\end{figure*}
\newpage
\clearpage

\begin{figure*}[!htb]
   \centering
   \includegraphics[width=0.99\textwidth]{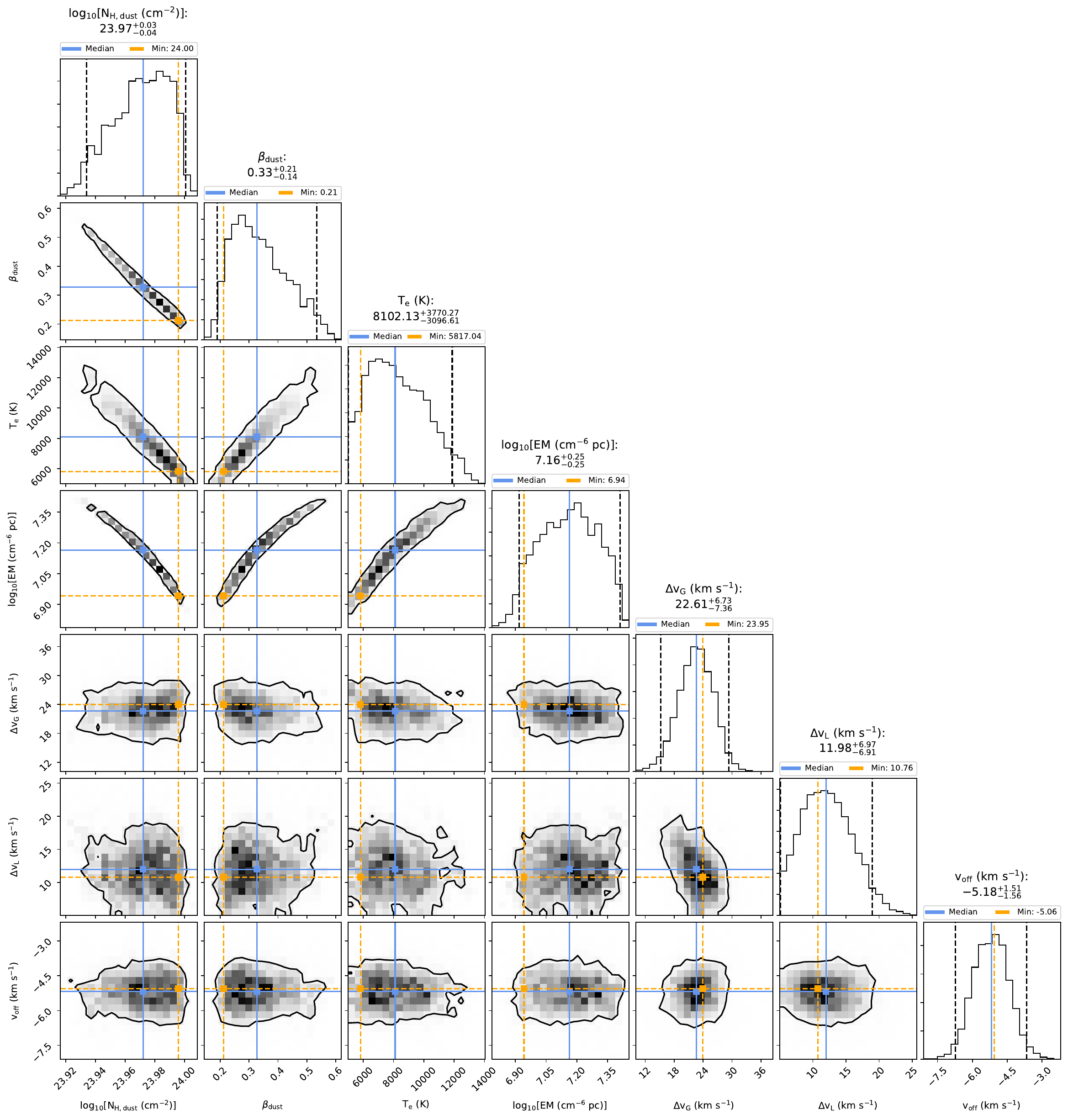}\\
   \caption{Corner plot for source A07 in Sgr~B2(M), envelope layer.}
   \label{fig:SgrB2-MErrorA07env}
\end{figure*}

\begin{figure*}[!htb]
   \centering
   \includegraphics[width=0.99\textwidth]{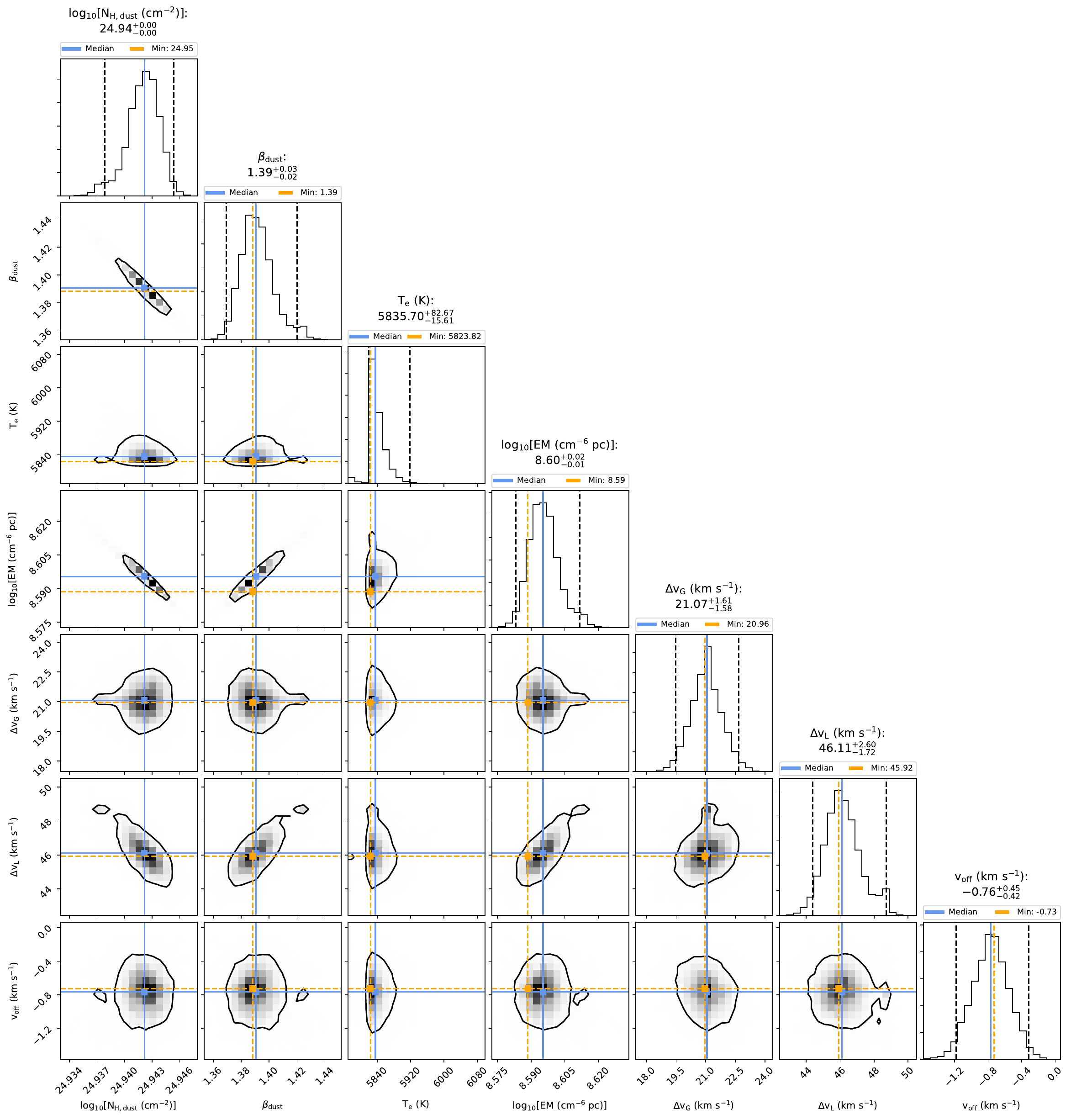}\\
   \caption{Corner plot for source A07 in Sgr~B2(M), core layer.}
   \label{fig:SgrB2-MErrorA07core}
\end{figure*}
\newpage
\clearpage

\begin{figure*}[!htb]
   \centering
   \includegraphics[width=0.50\textwidth]{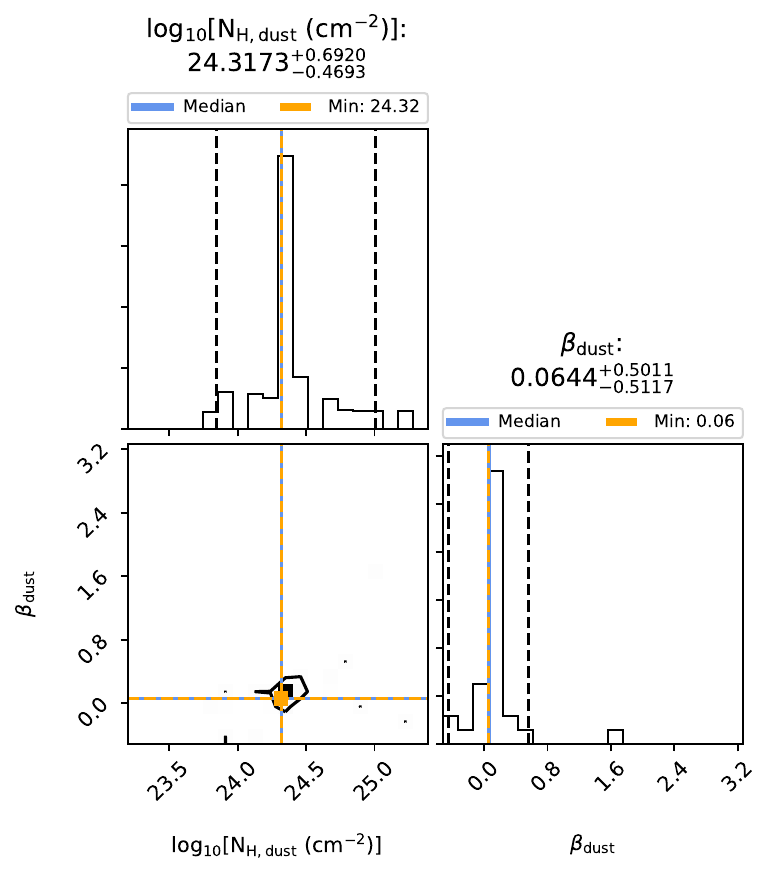}\\
   \caption{Corner plot for source A08 in Sgr~B2(M), envelope layer.}
   \label{fig:SgrB2-MErrorA08env}
\end{figure*}

\begin{figure*}[!htb]
   \centering
   \includegraphics[width=0.99\textwidth]{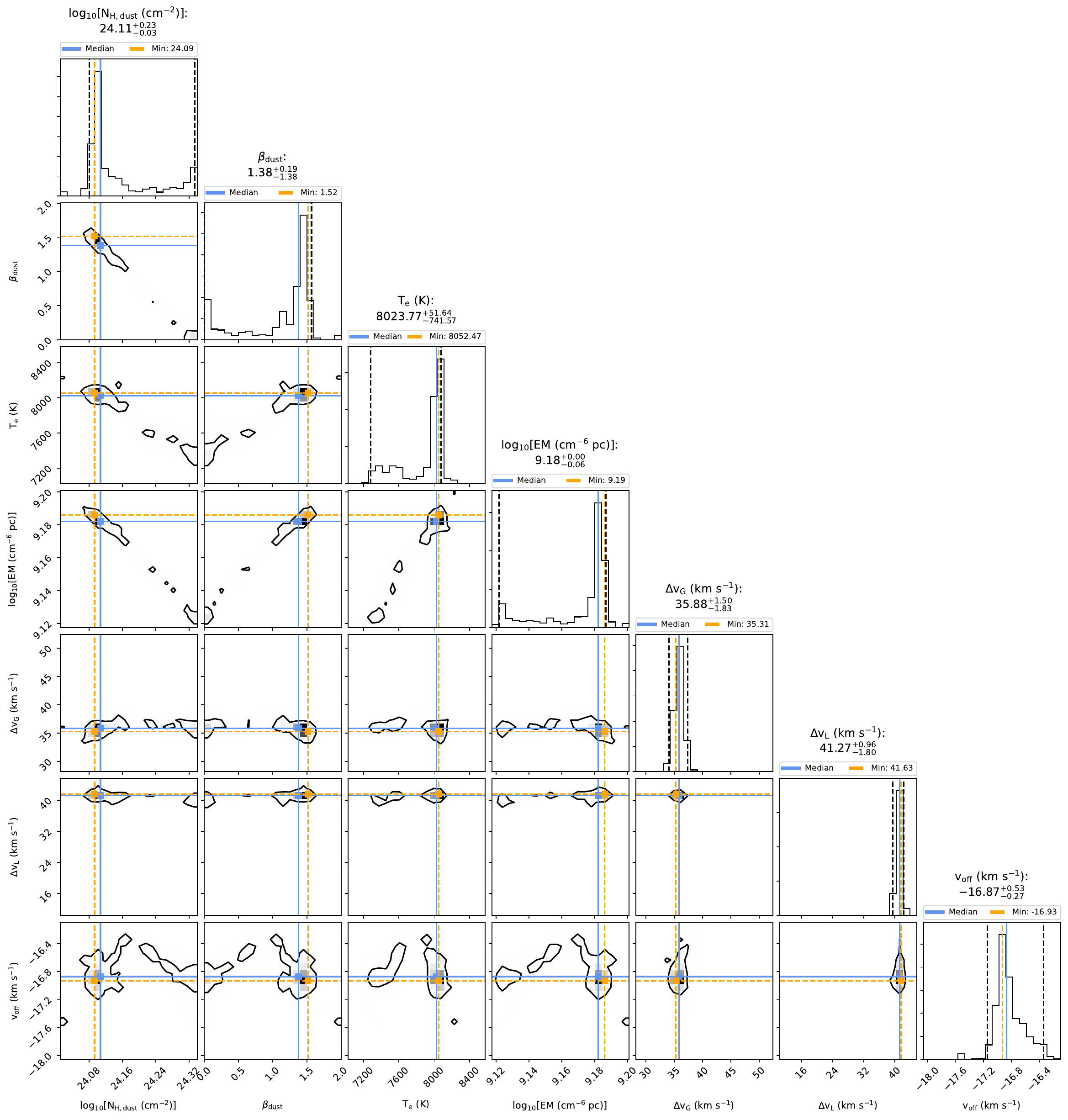}\\
   \caption{Corner plot for source A08 in Sgr~B2(M), core layer.}
   \label{fig:SgrB2-MErrorA08core}
\end{figure*}
\newpage
\clearpage

\begin{figure*}[!htb]
   \centering
   \includegraphics[width=0.99\textwidth]{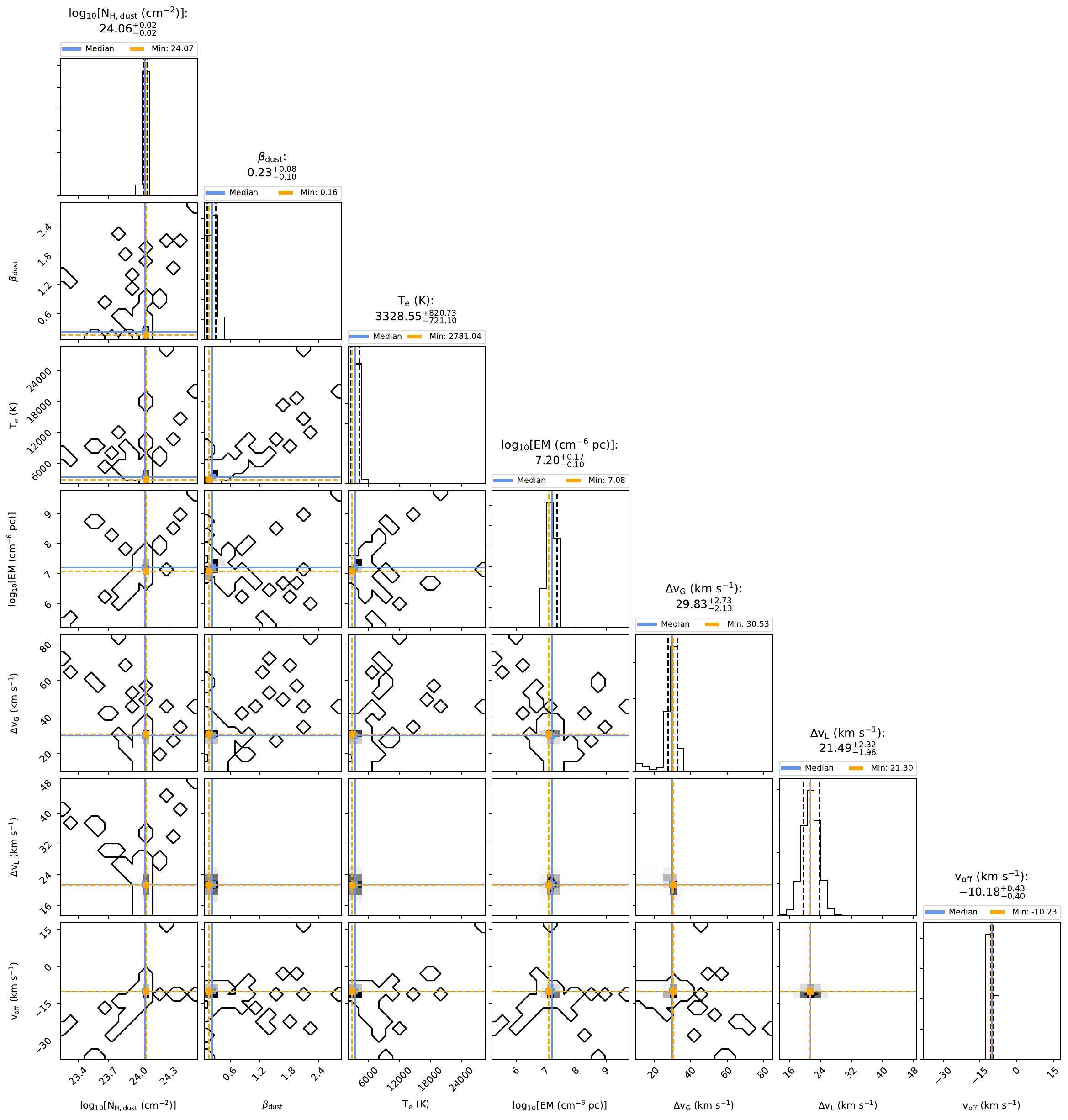}\\
   \caption{Corner plot for source A09 in Sgr~B2(M), envelope layer.}
   \label{fig:SgrB2-MErrorA09env}
\end{figure*}

\begin{figure*}[!htb]
   \centering
   \includegraphics[width=0.50\textwidth]{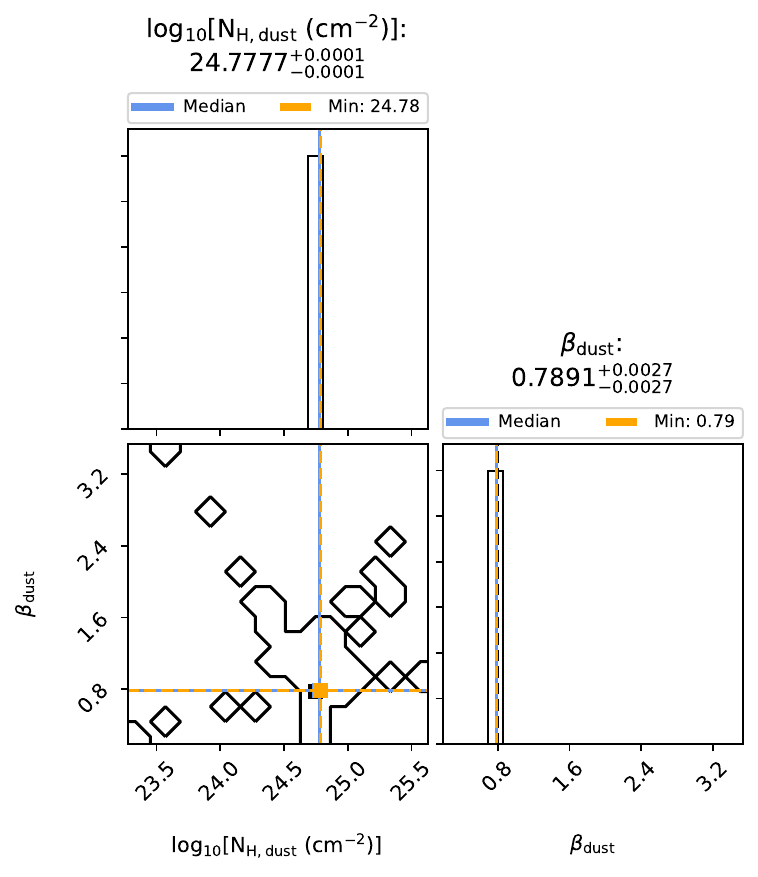}\\
   \caption{Corner plot for source A09 in Sgr~B2(M), core layer.}
   \label{fig:SgrB2-MErrorA09core}
\end{figure*}
\newpage
\clearpage

\begin{figure*}[!htb]
   \centering
   \includegraphics[width=0.50\textwidth]{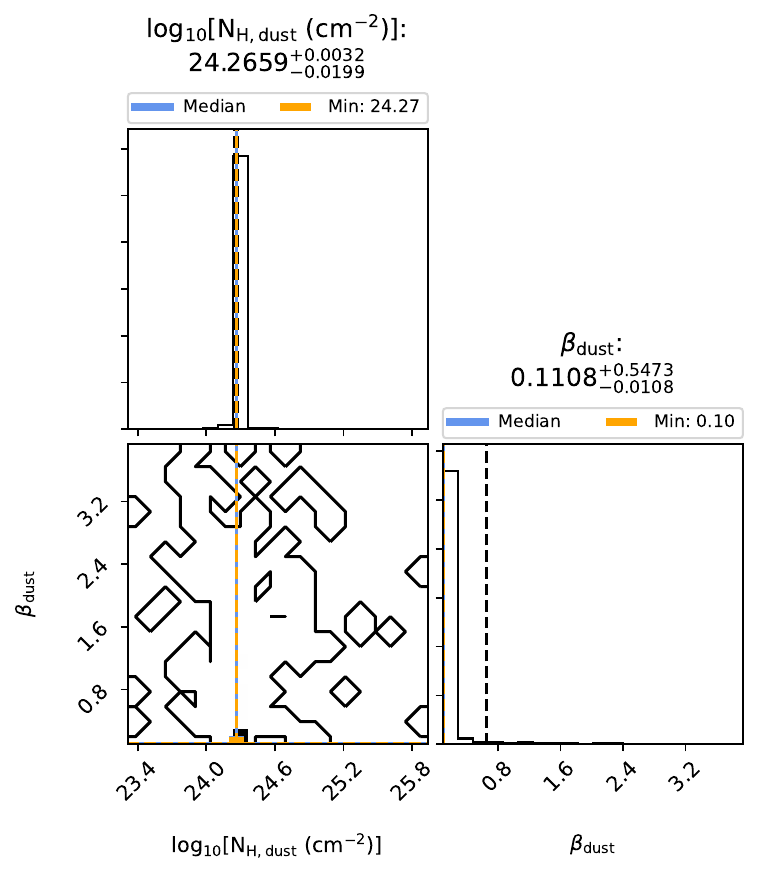}\\
   \caption{Corner plot for source A10 in Sgr~B2(M), envelope layer.}
   \label{fig:SgrB2-MErrorA10env}
\end{figure*}
\newpage
\clearpage

\begin{figure*}[!htb]
   \centering
   \includegraphics[width=0.50\textwidth]{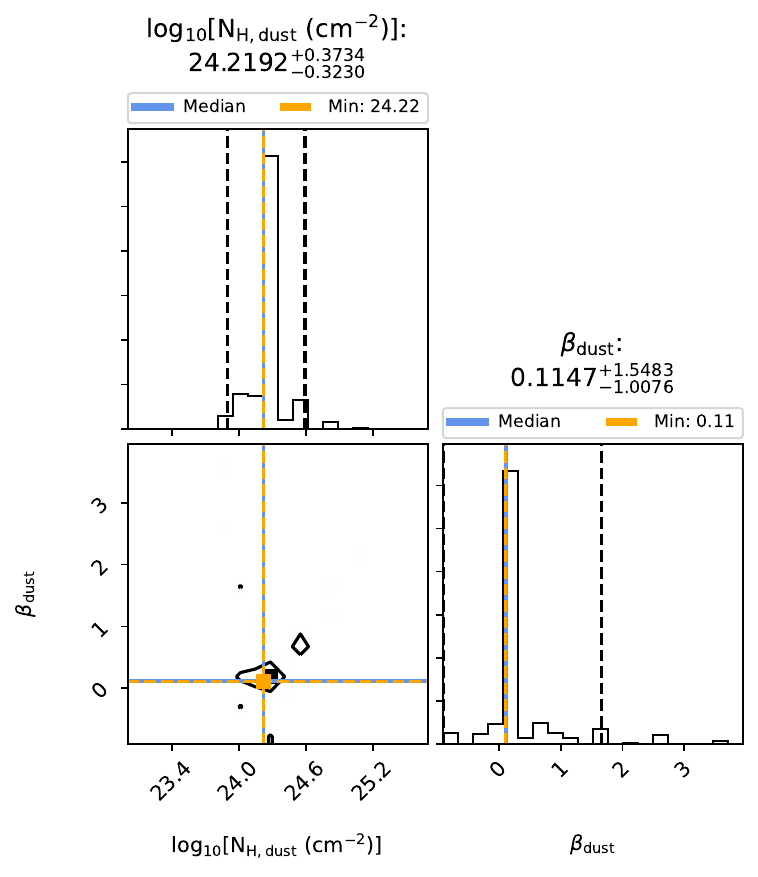}\\
   \caption{Corner plot for source A11 in Sgr~B2(M), envelope layer.}
   \label{fig:SgrB2-MErrorA11env}
\end{figure*}

\begin{figure*}[!htb]
   \centering
   \includegraphics[width=0.50\textwidth]{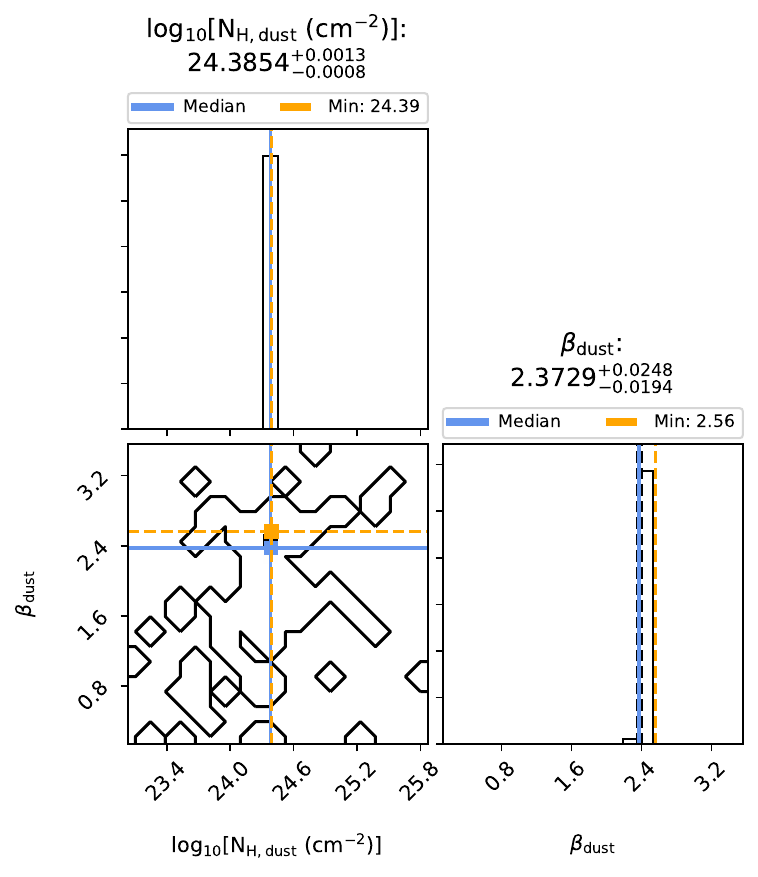}\\
   \caption{Corner plot for source A11 in Sgr~B2(M), core layer.}
   \label{fig:SgrB2-MErrorA11core}
\end{figure*}
\newpage
\clearpage

\begin{figure*}[!htb]
   \centering
   \includegraphics[width=0.50\textwidth]{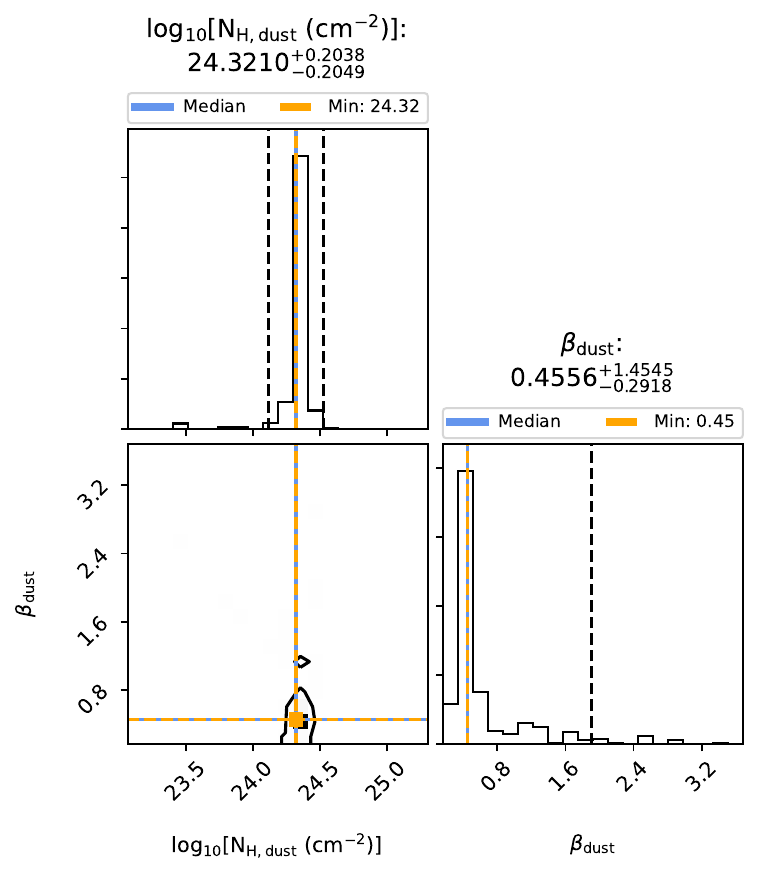}\\
   \caption{Corner plot for source A12 in Sgr~B2(M), envelope layer.}
   \label{fig:SgrB2-MErrorA12env}
\end{figure*}

\begin{figure*}[!htb]
   \centering
   \includegraphics[width=0.50\textwidth]{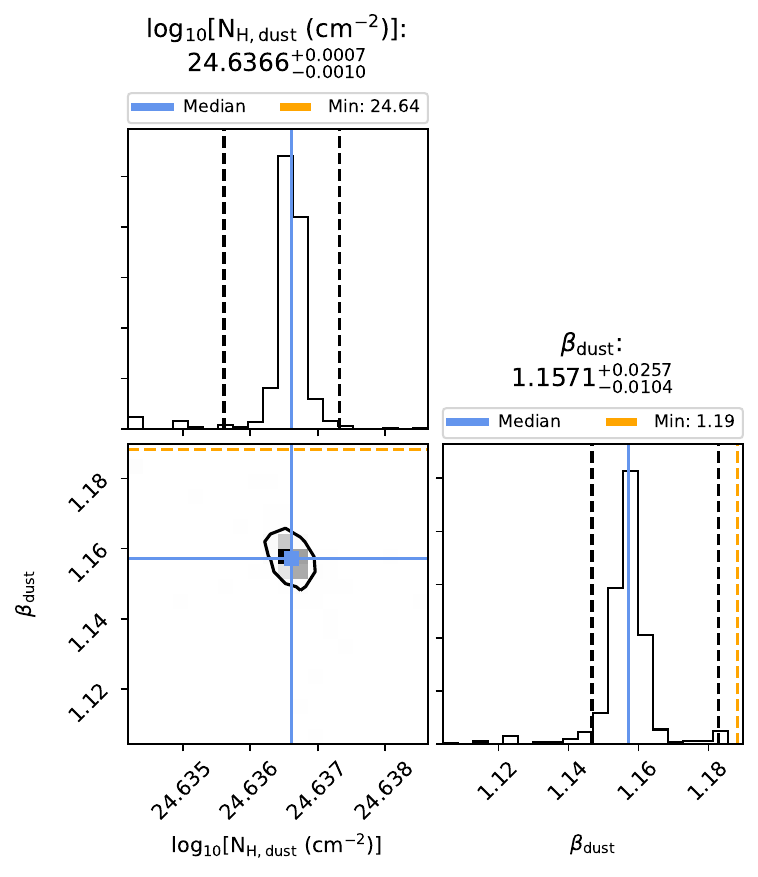}\\
   \caption{Corner plot for source A12 in Sgr~B2(M), core layer.}
   \label{fig:SgrB2-MErrorA12core}
\end{figure*}
\newpage
\clearpage

\begin{figure*}[!htb]
   \centering
   \includegraphics[width=0.50\textwidth]{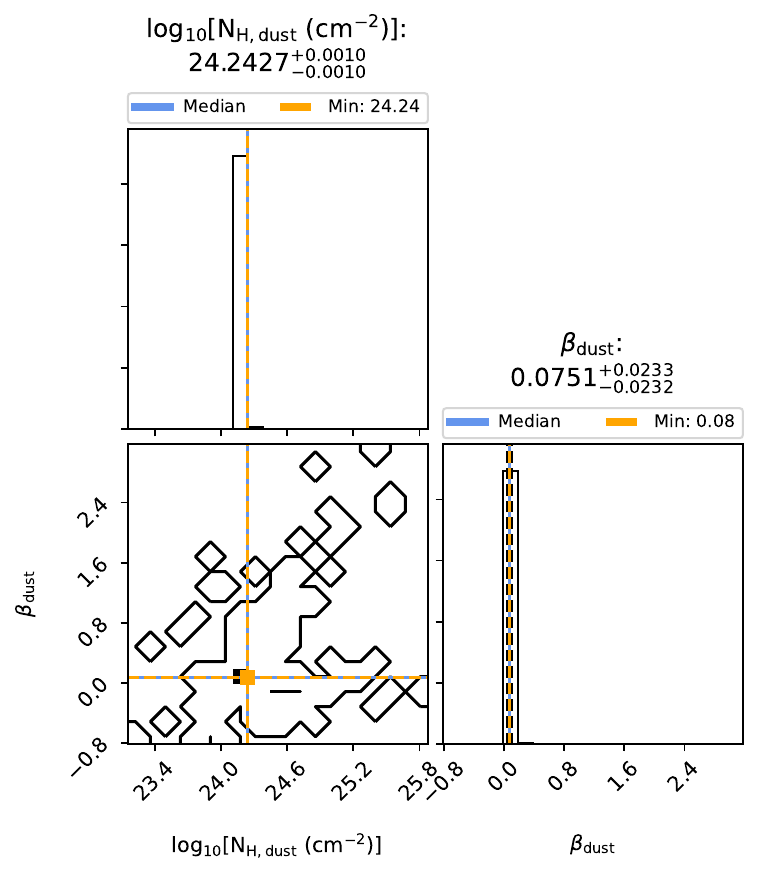}\\
   \caption{Corner plot for source A13 in Sgr~B2(M), envelope layer.}
   \label{fig:SgrB2-MErrorA13env}
\end{figure*}

\begin{figure*}[!htb]
   \centering
   \includegraphics[width=0.50\textwidth]{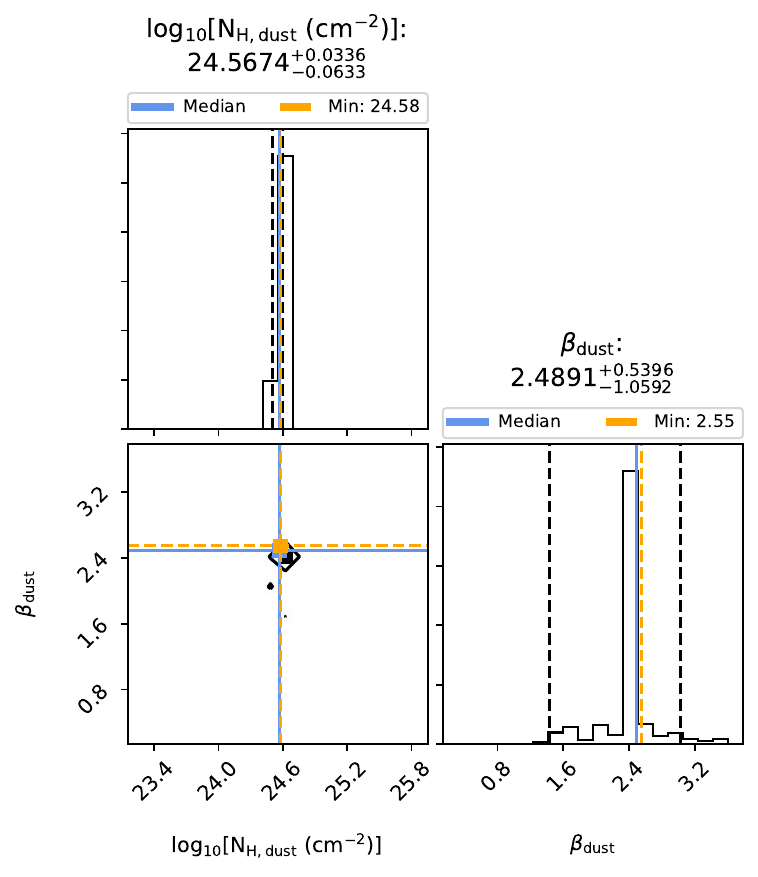}\\
   \caption{Corner plot for source A13 in Sgr~B2(M), core layer.}
   \label{fig:SgrB2-MErrorA13core}
\end{figure*}
\newpage
\clearpage

\begin{figure*}[!htb]
   \centering
   \includegraphics[width=0.50\textwidth]{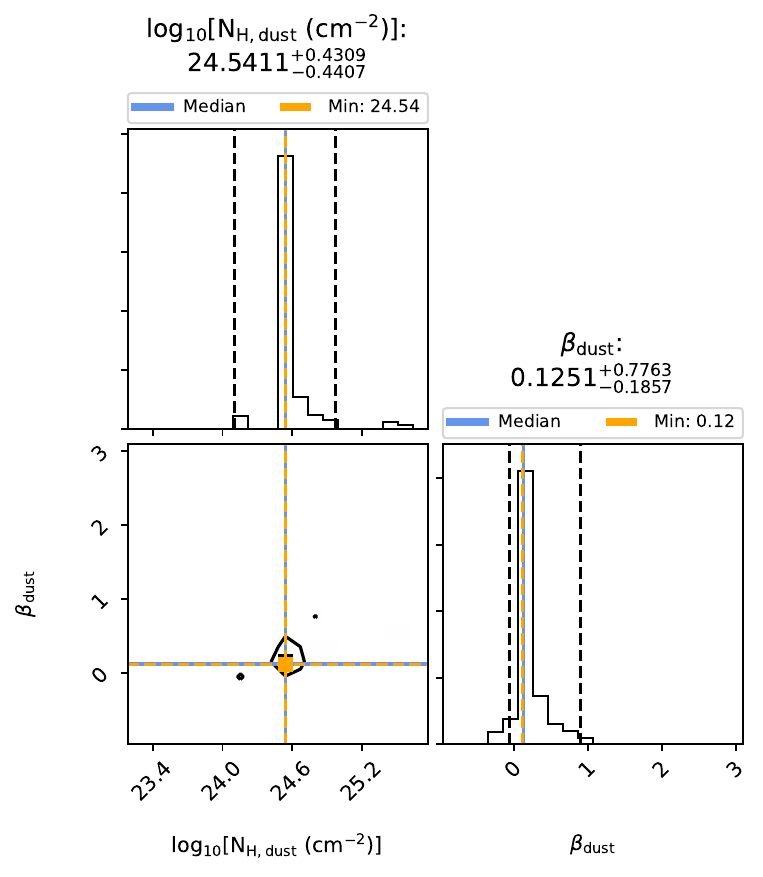}\\
   \caption{Corner plot for source A14 in Sgr~B2(M), envelope layer.}
   \label{fig:SgrB2-MErrorA14env}
\end{figure*}

\begin{figure*}[!htb]
   \centering
   \includegraphics[width=0.50\textwidth]{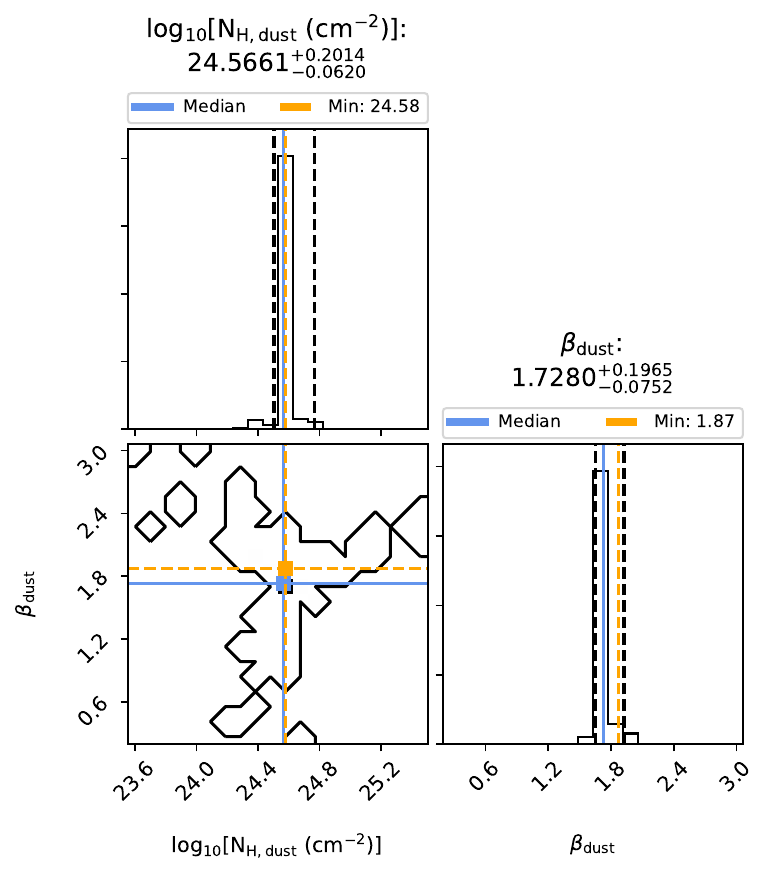}\\
   \caption{Corner plot for source A14 in Sgr~B2(M), core layer.}
   \label{fig:SgrB2-MErrorA14core}
\end{figure*}
\newpage
\clearpage

\begin{figure*}[!htb]
   \centering
   \includegraphics[width=0.50\textwidth]{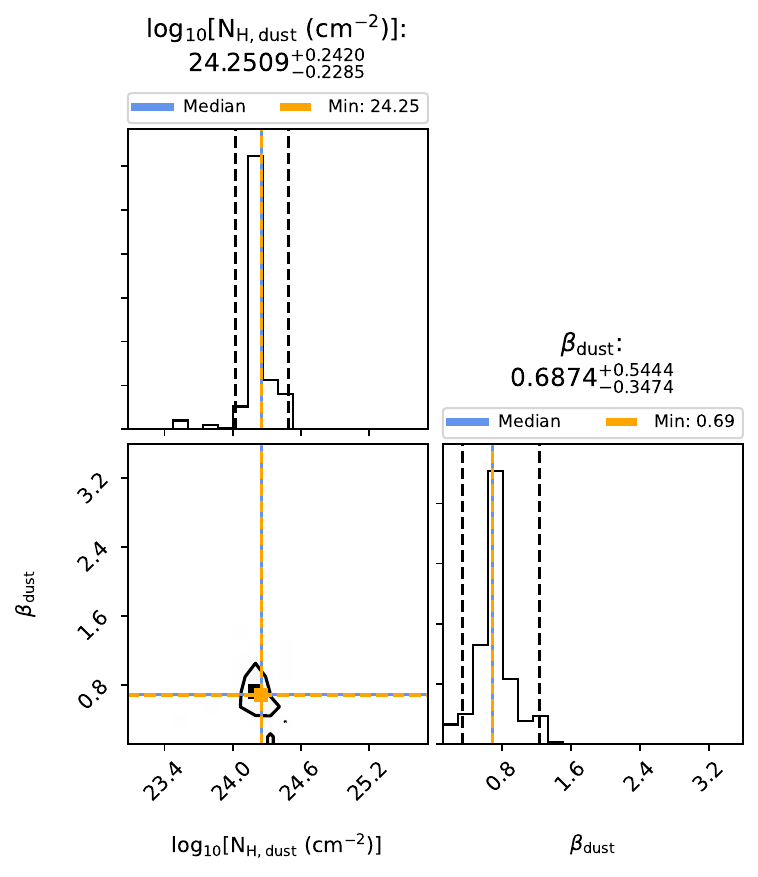}\\
   \caption{Corner plot for source A15 in Sgr~B2(M), envelope layer.}
   \label{fig:SgrB2-MErrorA15env}
\end{figure*}

\begin{figure*}[!htb]
   \centering
   \includegraphics[width=0.99\textwidth]{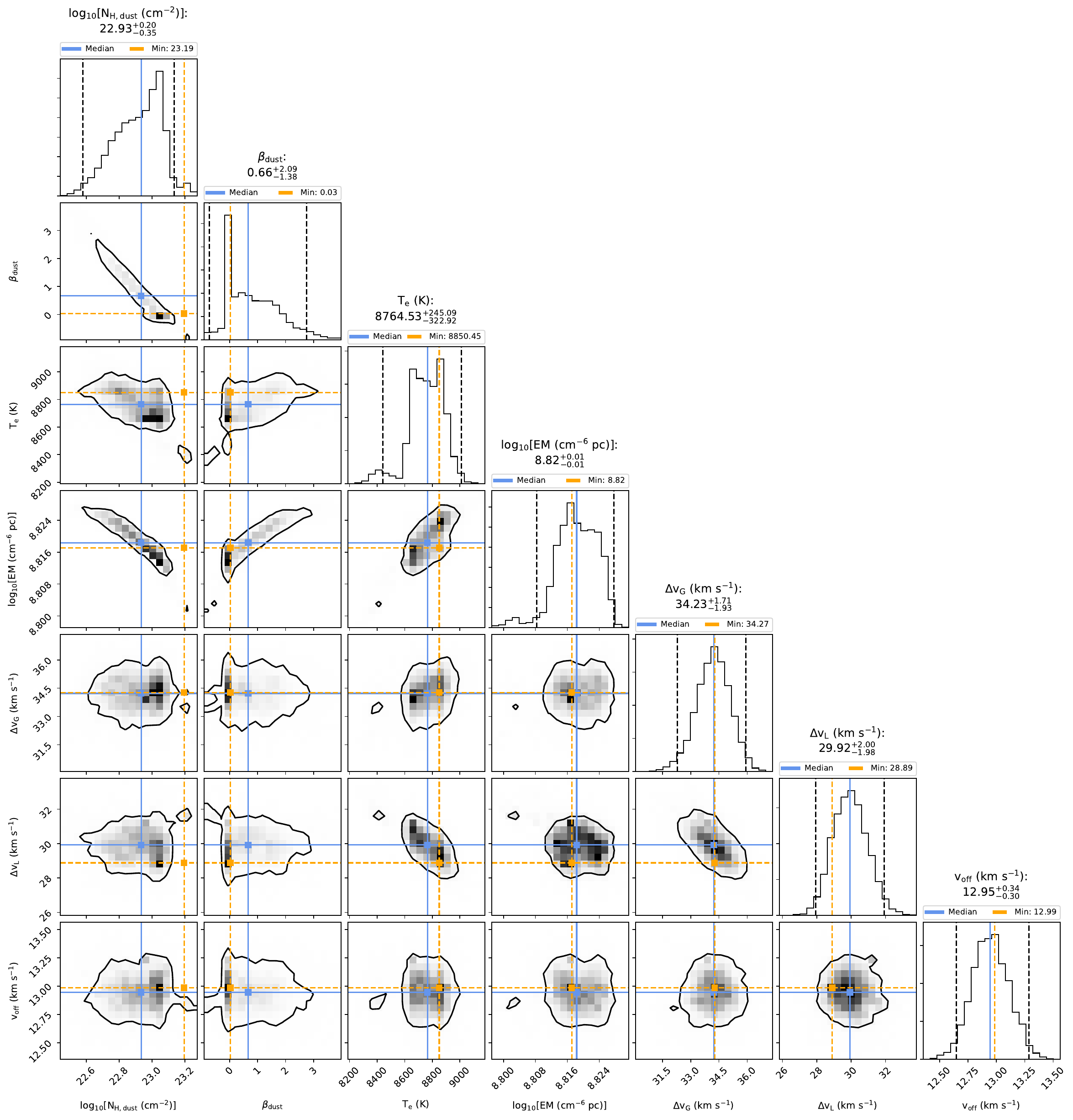}\\
   \caption{Corner plot for source A15 in Sgr~B2(M), core layer.}
   \label{fig:SgrB2-MErrorA15core}
\end{figure*}
\newpage
\clearpage

\begin{figure*}[!htb]
   \centering
   \includegraphics[width=0.50\textwidth]{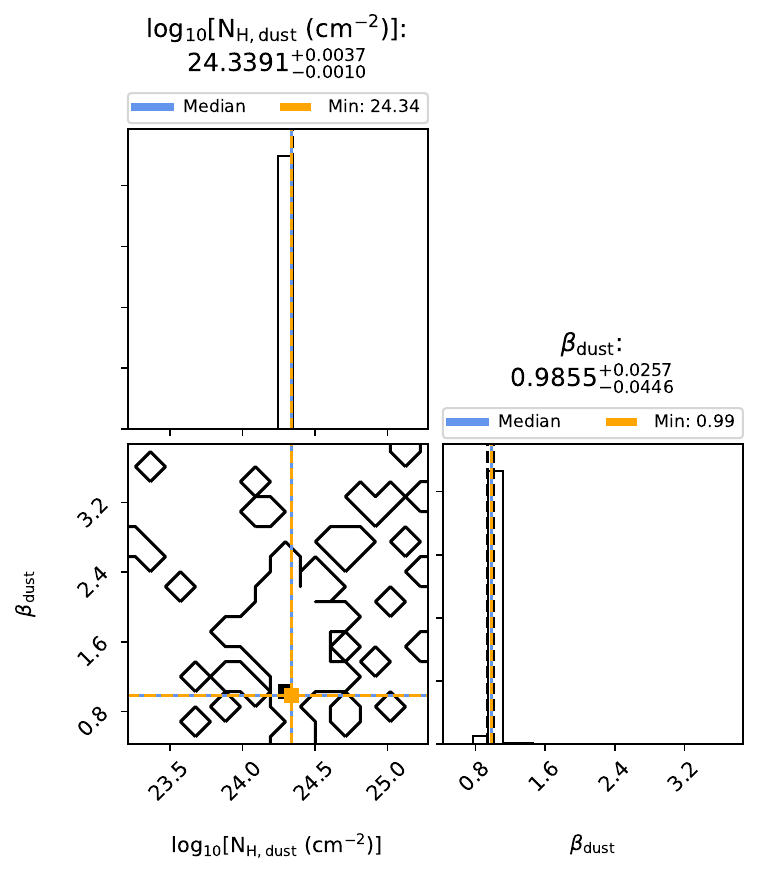}\\
   \caption{Corner plot for source A16 in Sgr~B2(M), envelope layer.}
   \label{fig:SgrB2-MErrorA16env}
\end{figure*}

\begin{figure*}[!htb]
   \centering
   \includegraphics[width=0.99\textwidth]{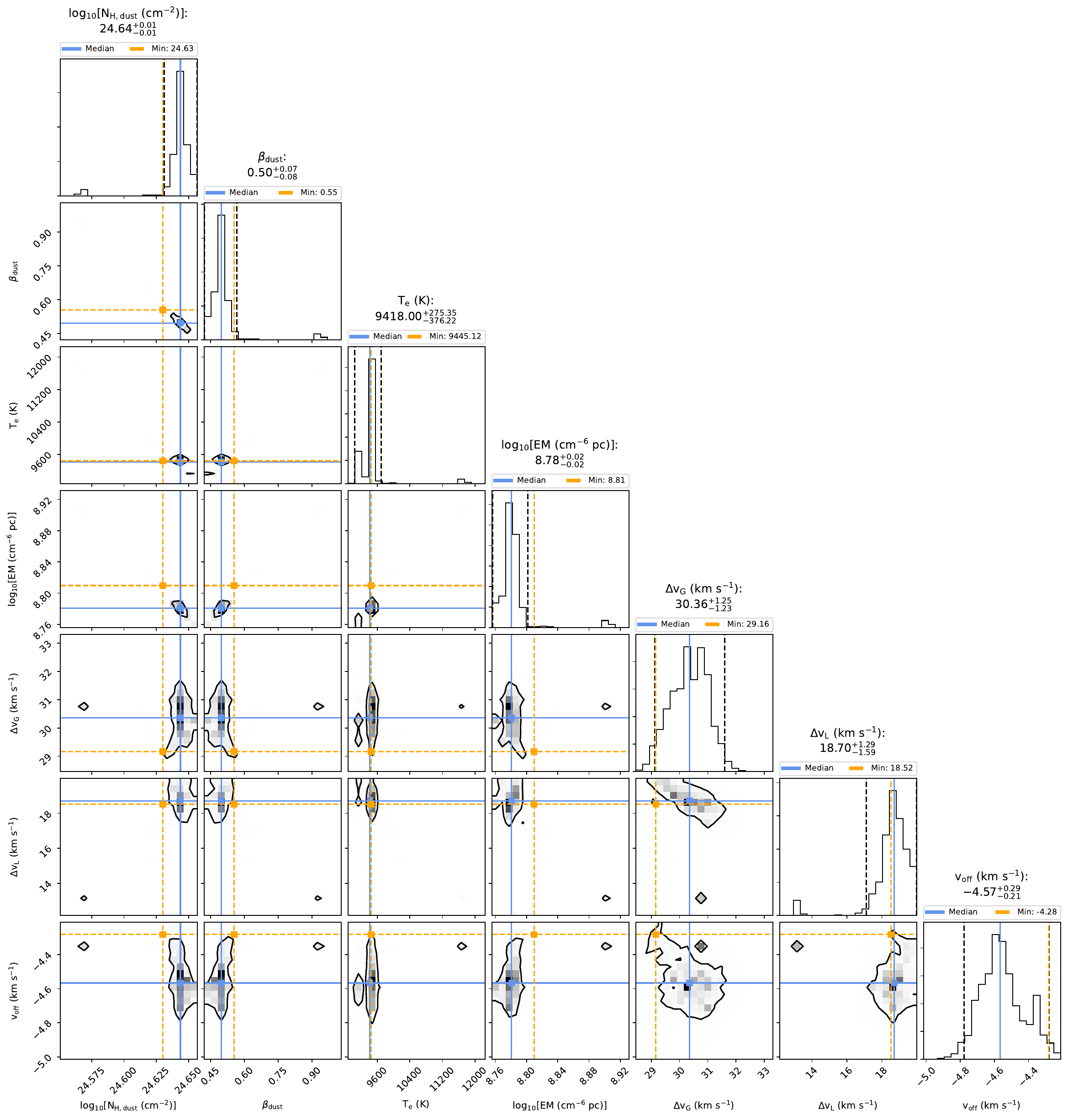}\\
   \caption{Corner plot for source A16 in Sgr~B2(M), core layer.}
   \label{fig:SgrB2-MErrorA16core}
\end{figure*}
\newpage
\clearpage

\begin{figure*}[!htb]
   \centering
   \includegraphics[width=0.50\textwidth]{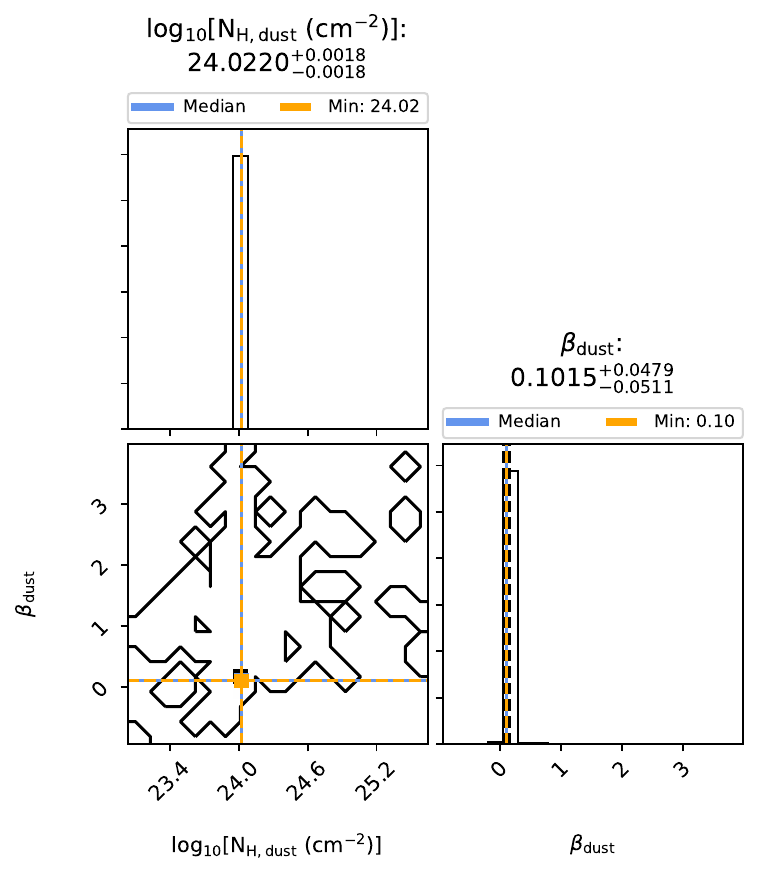}\\
   \caption{Corner plot for source A17 in Sgr~B2(M), envelope layer.}
   \label{fig:SgrB2-MErrorA17env}
\end{figure*}

\begin{figure*}[!htb]
   \centering
   \includegraphics[width=0.99\textwidth]{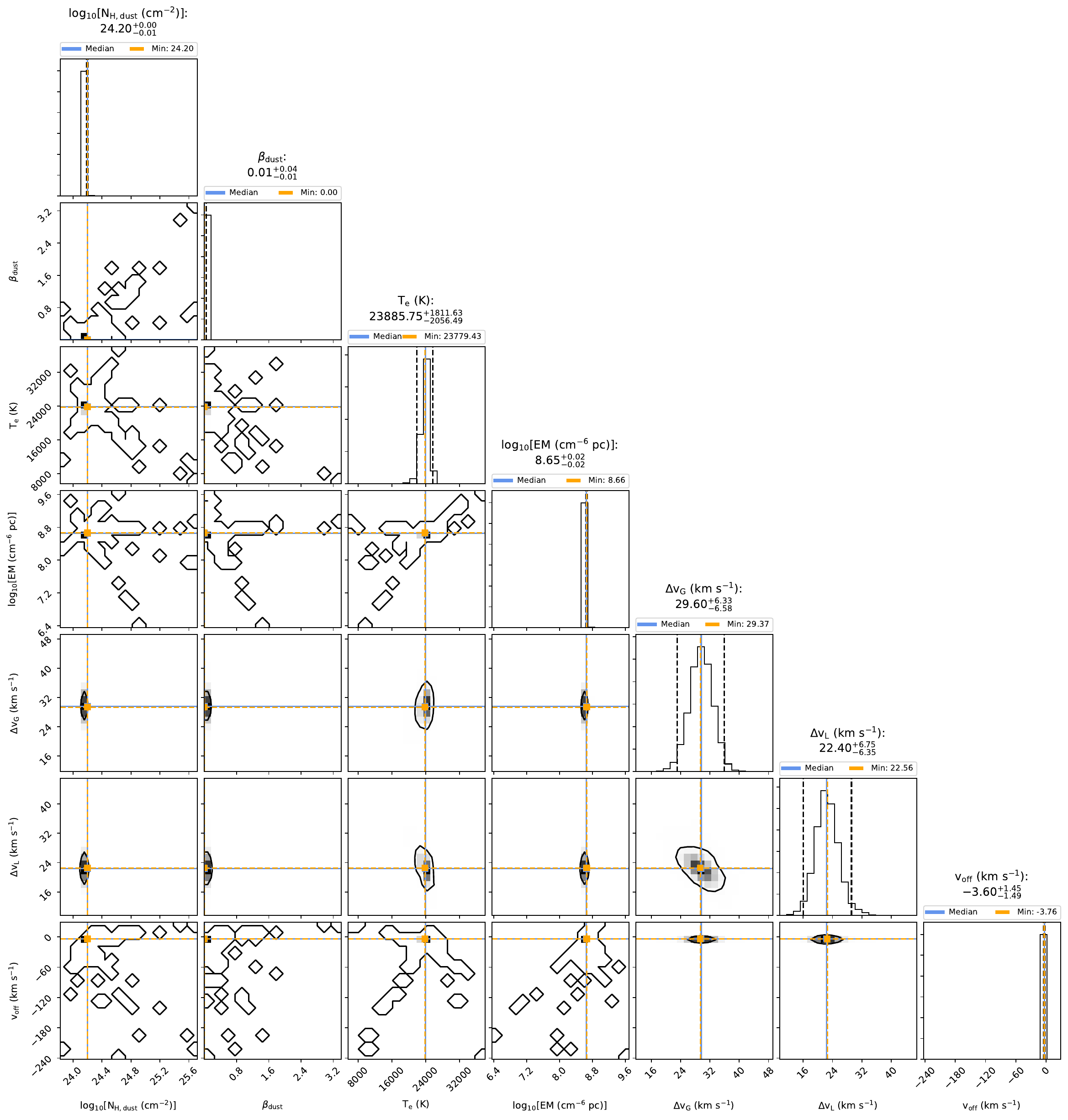}\\
   \caption{Corner plot for source A17 in Sgr~B2(M), core layer.}
   \label{fig:SgrB2-MErrorA17core}
\end{figure*}
\newpage
\clearpage

\begin{figure*}[!htb]
   \centering
   \includegraphics[width=0.50\textwidth]{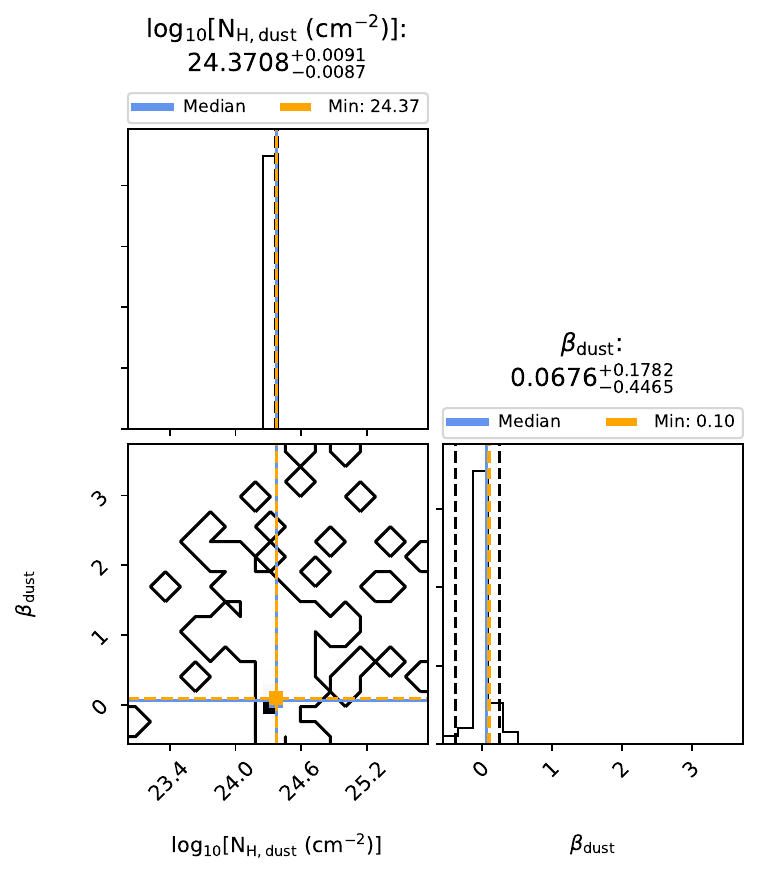}\\
   \caption{Corner plot for source A18 in Sgr~B2(M), envelope layer.}
   \label{fig:SgrB2-MErrorA18env}
\end{figure*}
\newpage
\clearpage

\begin{figure*}[!htb]
   \centering
   \includegraphics[width=0.50\textwidth]{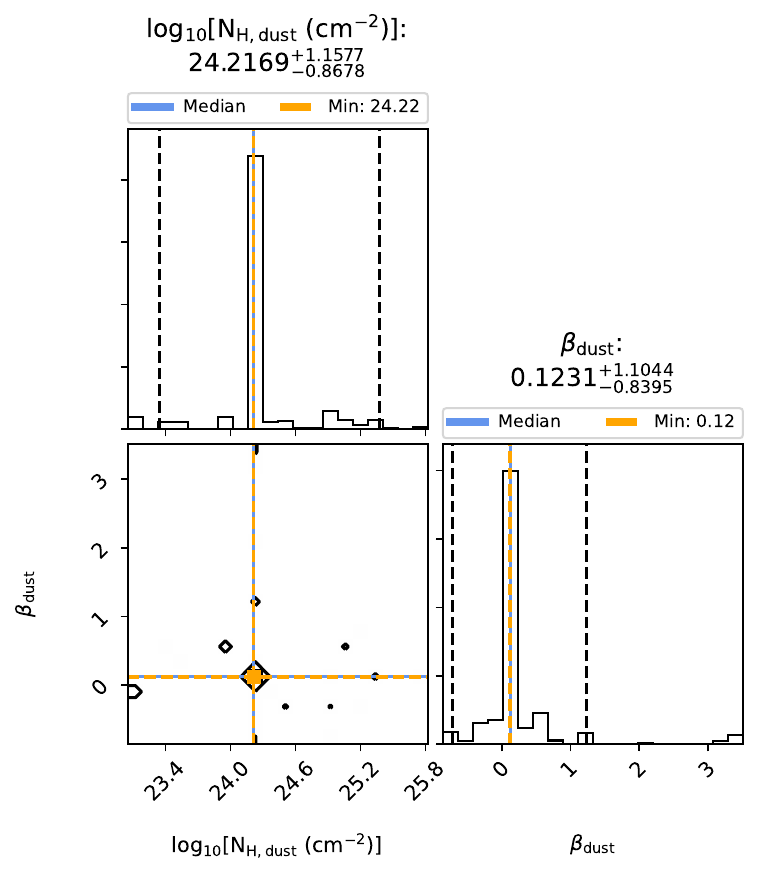}\\
   \caption{Corner plot for source A19 in Sgr~B2(M), envelope layer.}
   \label{fig:SgrB2-MErrorA19env}
\end{figure*}

\begin{figure*}[!htb]
   \centering
   \includegraphics[width=0.50\textwidth]{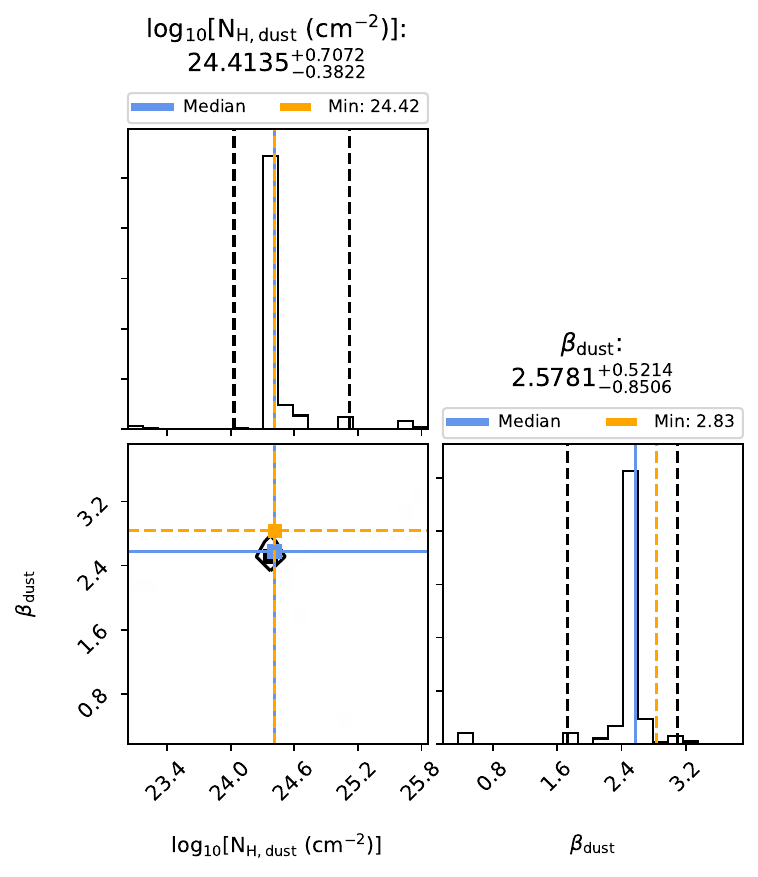}\\
   \caption{Corner plot for source A19 in Sgr~B2(M), core layer.}
   \label{fig:SgrB2-MErrorA19core}
\end{figure*}
\newpage
\clearpage

\begin{figure*}[!htb]
   \centering
   \includegraphics[width=0.50\textwidth]{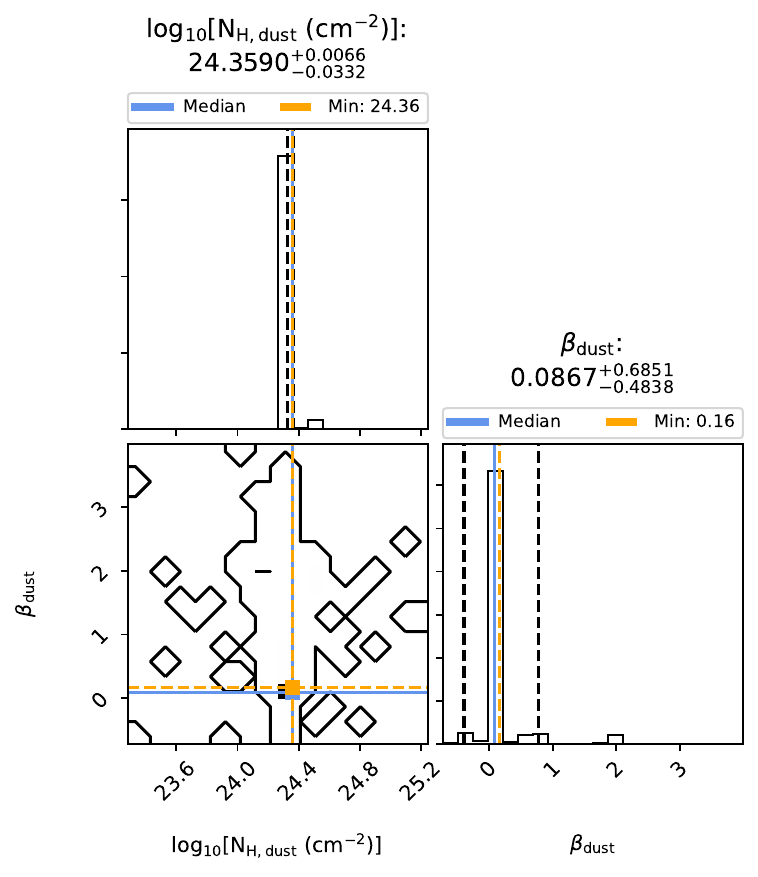}\\
   \caption{Corner plot for source A20 in Sgr~B2(M), envelope layer.}
   \label{fig:SgrB2-MErrorA20env}
\end{figure*}
\newpage
\clearpage

\begin{figure*}[!htb]
   \centering
   \includegraphics[width=0.50\textwidth]{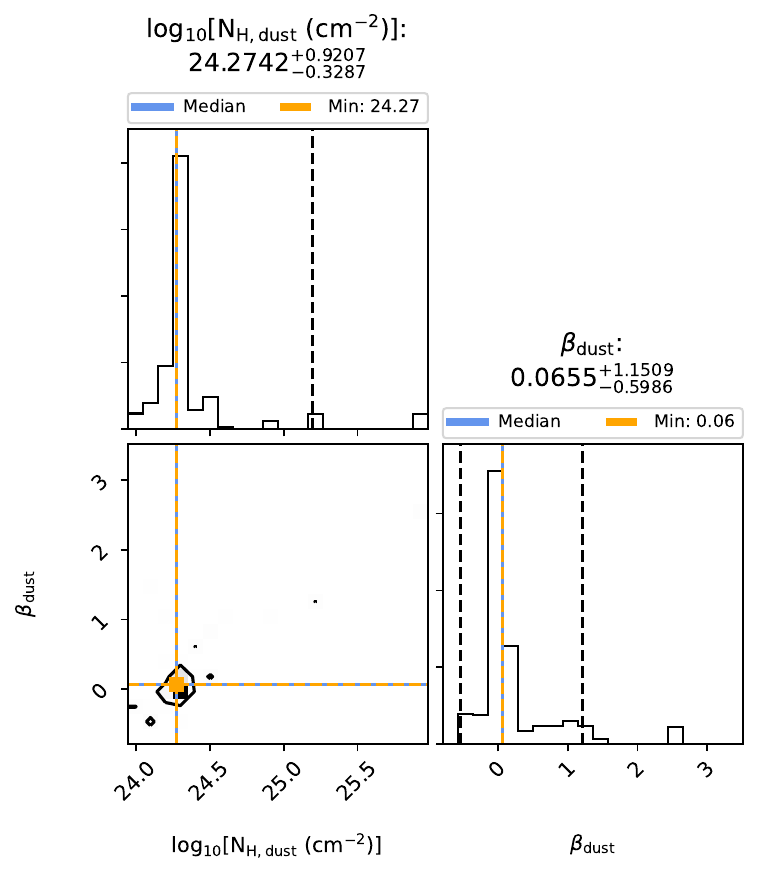}\\
   \caption{Corner plot for source A21 in Sgr~B2(M), envelope layer.}
   \label{fig:SgrB2-MErrorA21env}
\end{figure*}

\begin{figure*}[!htb]
   \centering
   \includegraphics[width=0.50\textwidth]{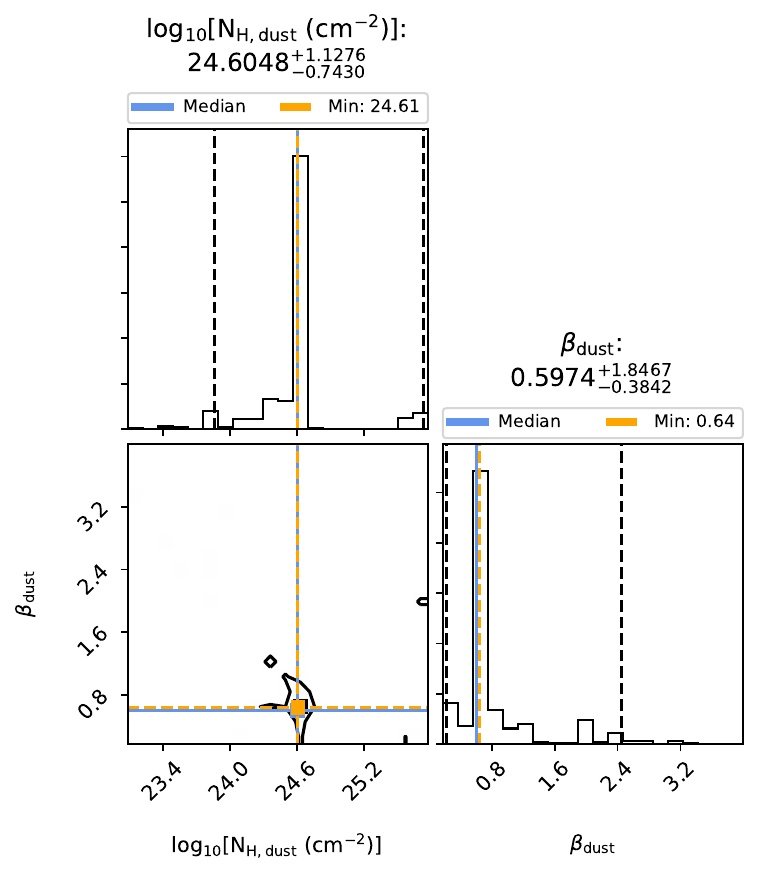}\\
   \caption{Corner plot for source A21 in Sgr~B2(M), core layer.}
   \label{fig:SgrB2-MErrorA21core}
\end{figure*}
\newpage
\clearpage

\begin{figure*}[!htb]
   \centering
   \includegraphics[width=0.50\textwidth]{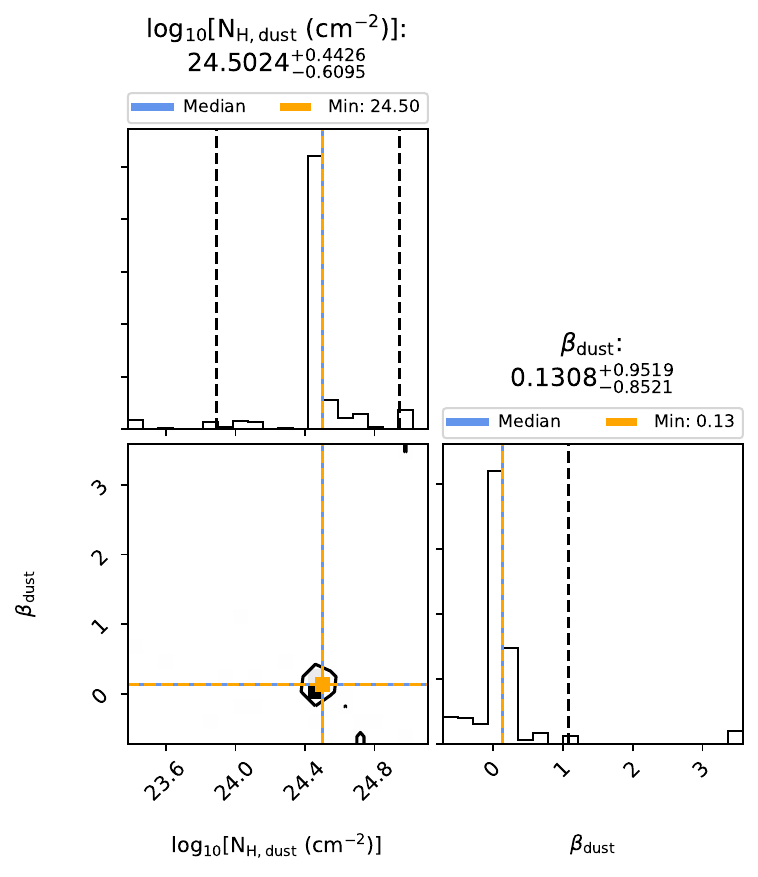}\\
   \caption{Corner plot for source A22 in Sgr~B2(M), envelope layer.}
   \label{fig:SgrB2-MErrorA22env}
\end{figure*}

\begin{figure*}[!htb]
   \centering
   \includegraphics[width=0.50\textwidth]{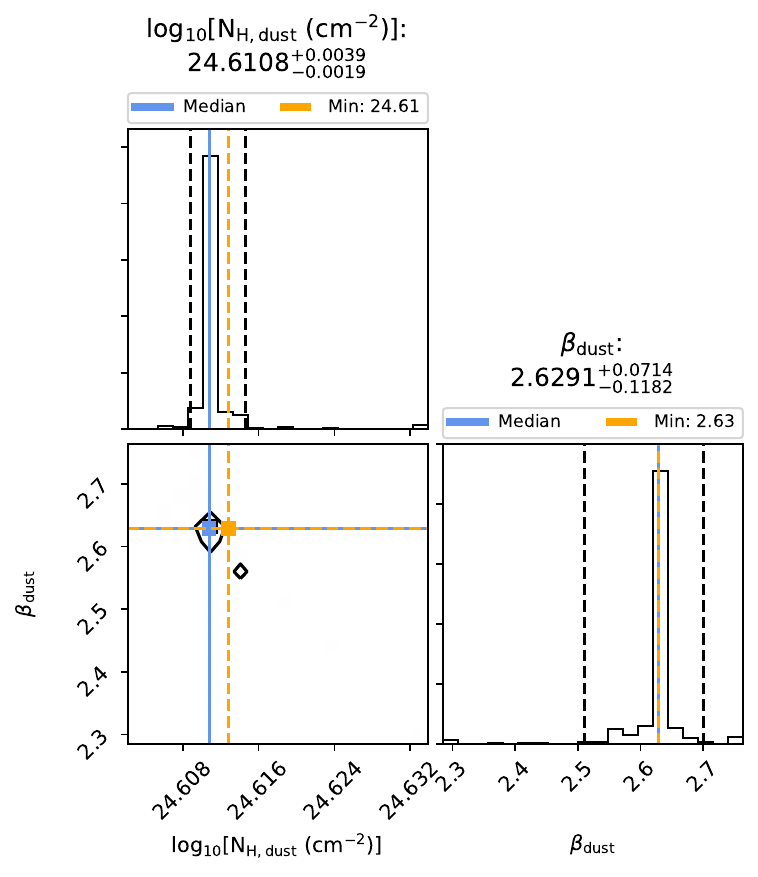}\\
   \caption{Corner plot for source A22 in Sgr~B2(M), core layer.}
   \label{fig:SgrB2-MErrorA22core}
\end{figure*}
\newpage
\clearpage

\begin{figure*}[!htb]
   \centering
   \includegraphics[width=0.50\textwidth]{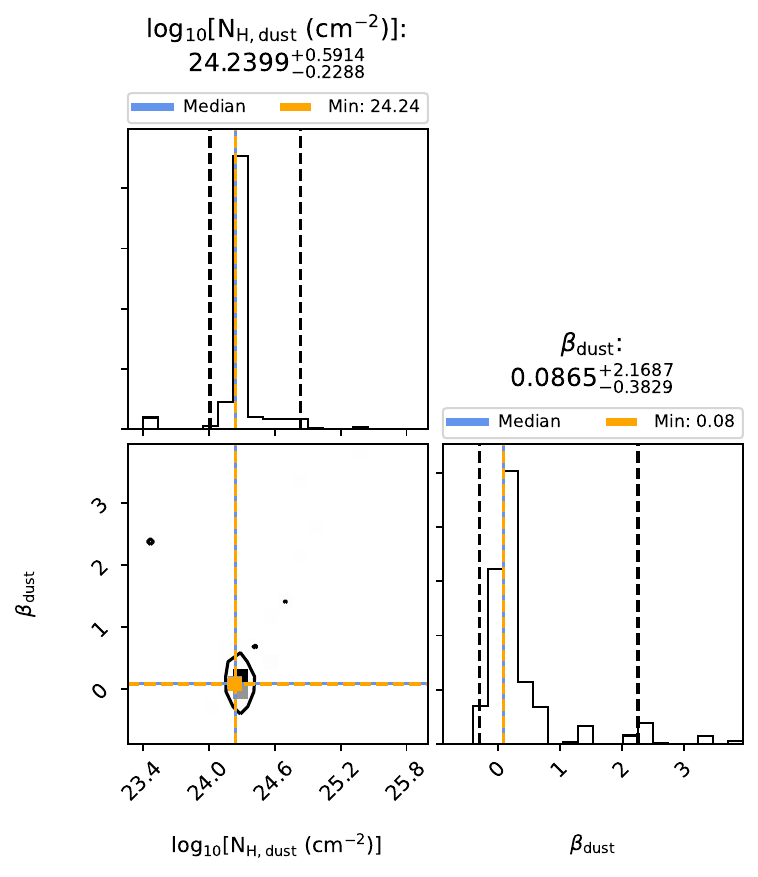}\\
   \caption{Corner plot for source A23 in Sgr~B2(M), envelope layer.}
   \label{fig:SgrB2-MErrorA23env}
\end{figure*}

\begin{figure*}[!htb]
   \centering
   \includegraphics[width=0.50\textwidth]{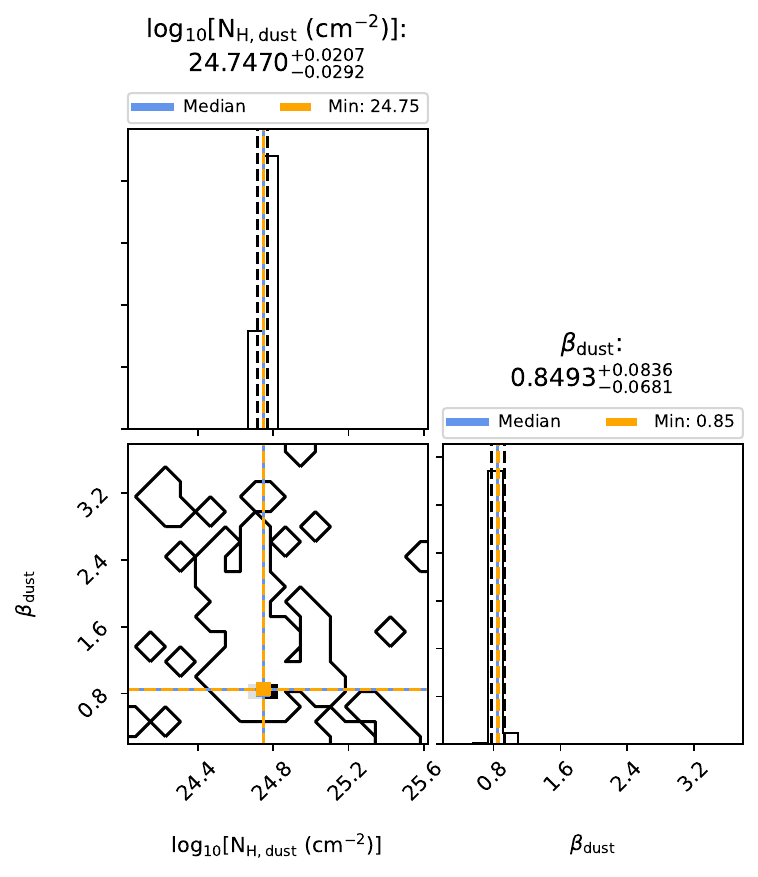}\\
   \caption{Corner plot for source A23 in Sgr~B2(M), core layer.}
   \label{fig:SgrB2-MErrorA23core}
\end{figure*}
\newpage
\clearpage

\begin{figure*}[!htb]
   \centering
   \includegraphics[width=0.99\textwidth]{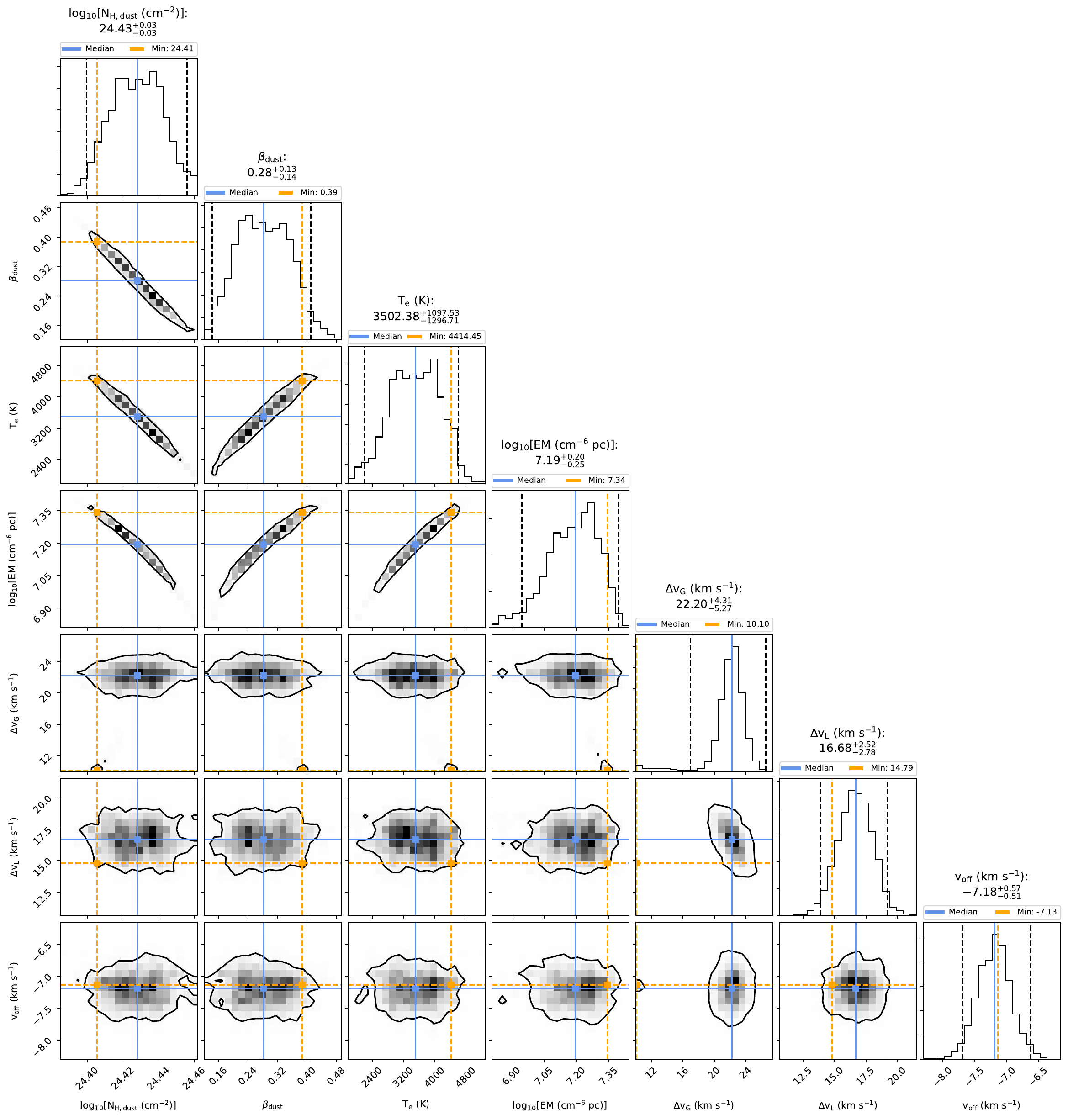}\\
   \caption{Corner plot for source A24 in Sgr~B2(M), envelope layer.}
   \label{fig:SgrB2-MErrorA24env}
\end{figure*}

\begin{figure*}[!htb]
   \centering
   \includegraphics[width=0.99\textwidth]{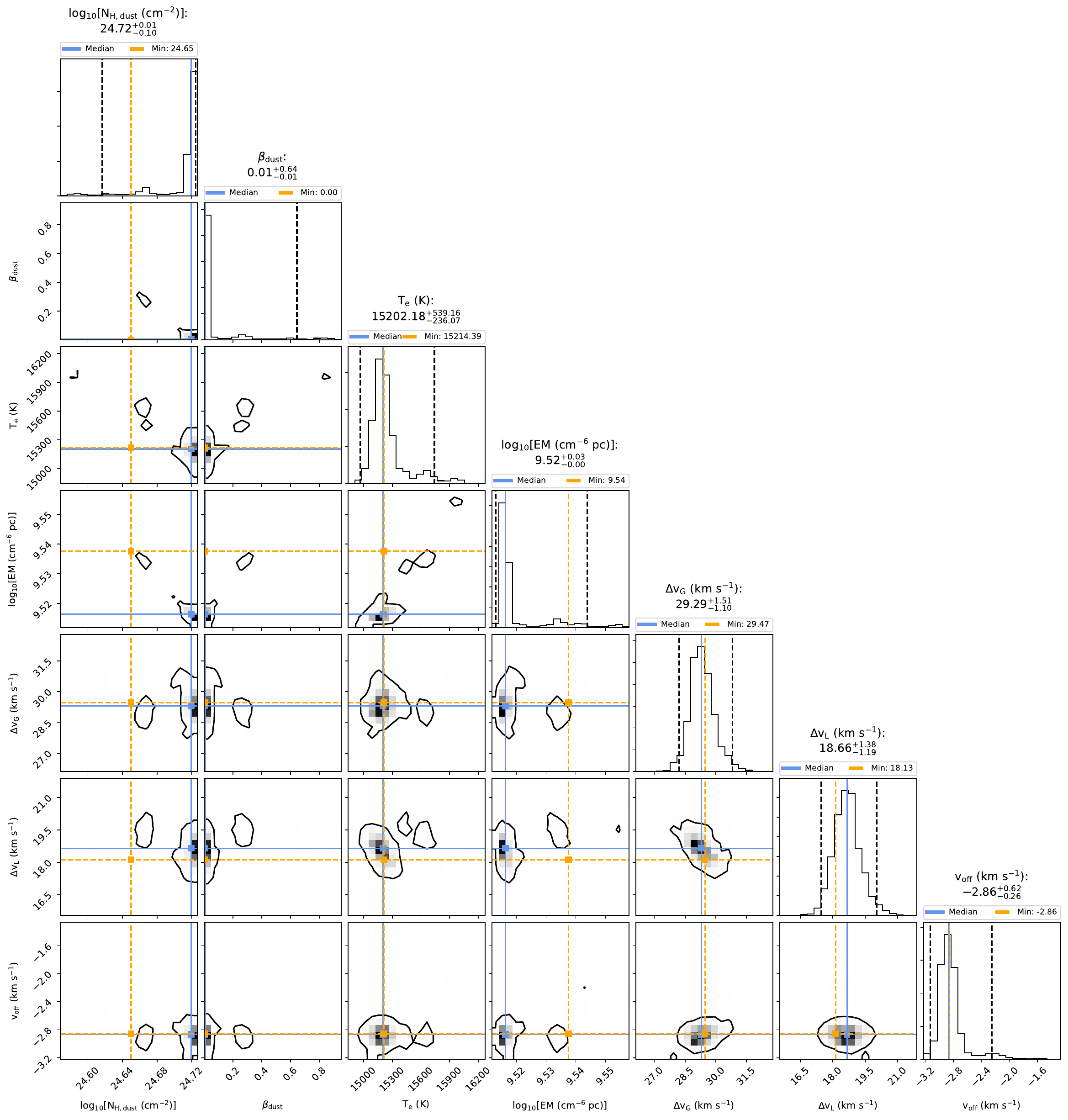}\\
   \caption{Corner plot for source A24 in Sgr~B2(M), core layer.}
   \label{fig:SgrB2-MErrorA24core}
\end{figure*}
\newpage
\clearpage

\begin{figure*}[!htb]
   \centering
   \includegraphics[width=0.50\textwidth]{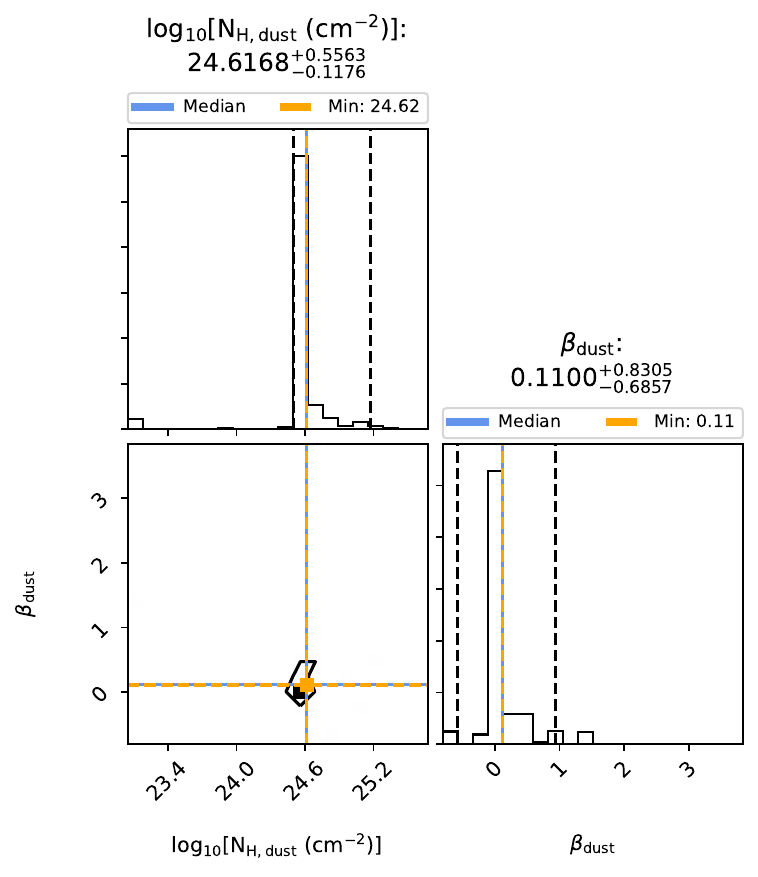}\\
   \caption{Corner plot for source A25 in Sgr~B2(M), envelope layer.}
   \label{fig:SgrB2-MErrorA25env}
\end{figure*}
\newpage
\clearpage

\begin{figure*}[!htb]
   \centering
   \includegraphics[width=0.99\textwidth]{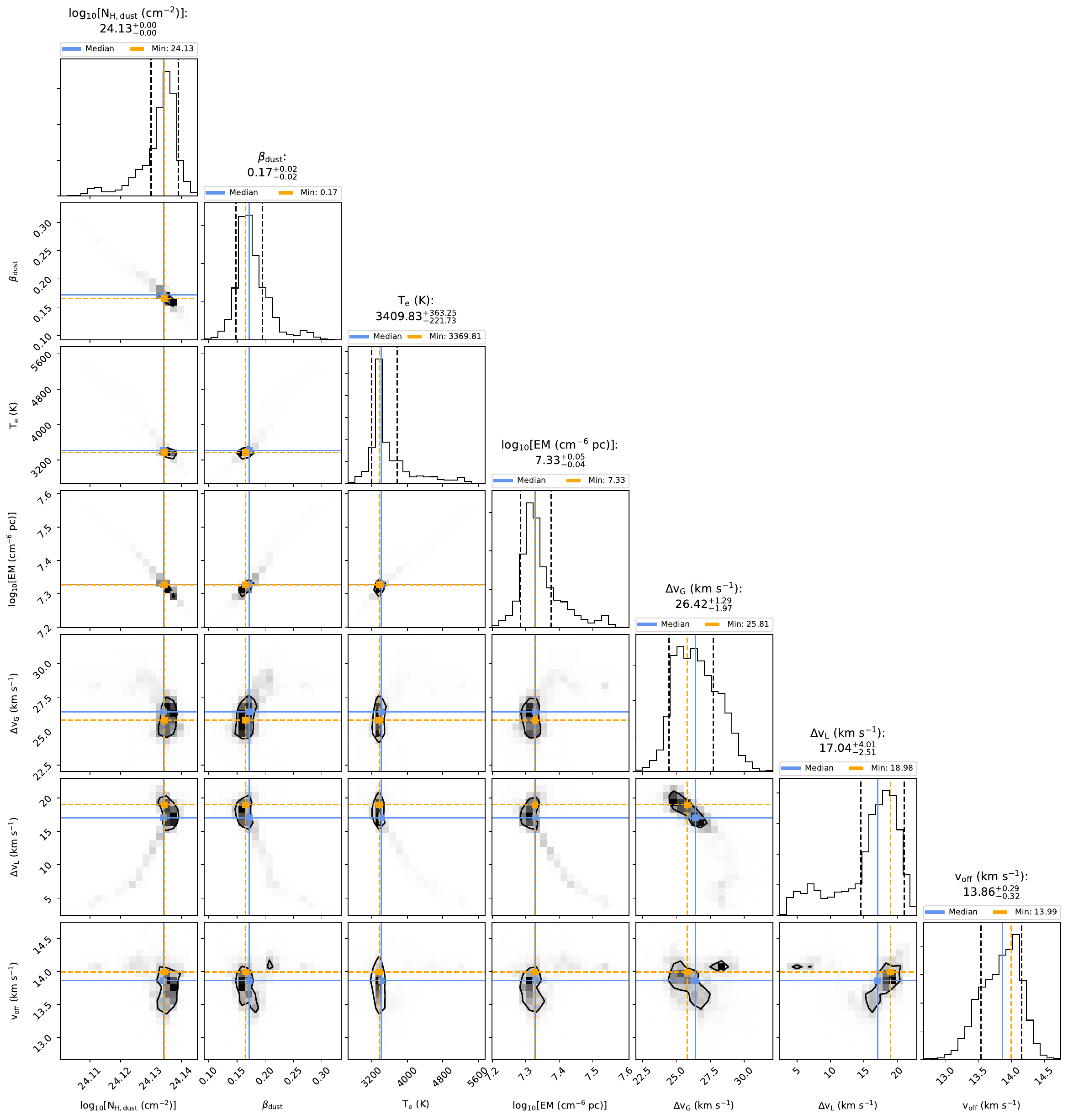}\\
   \caption{Corner plot for source A26 in Sgr~B2(M), envelope layer.}
   \label{fig:SgrB2-MErrorA26env}
\end{figure*}

\begin{figure*}[!htb]
   \centering
   \includegraphics[width=0.50\textwidth]{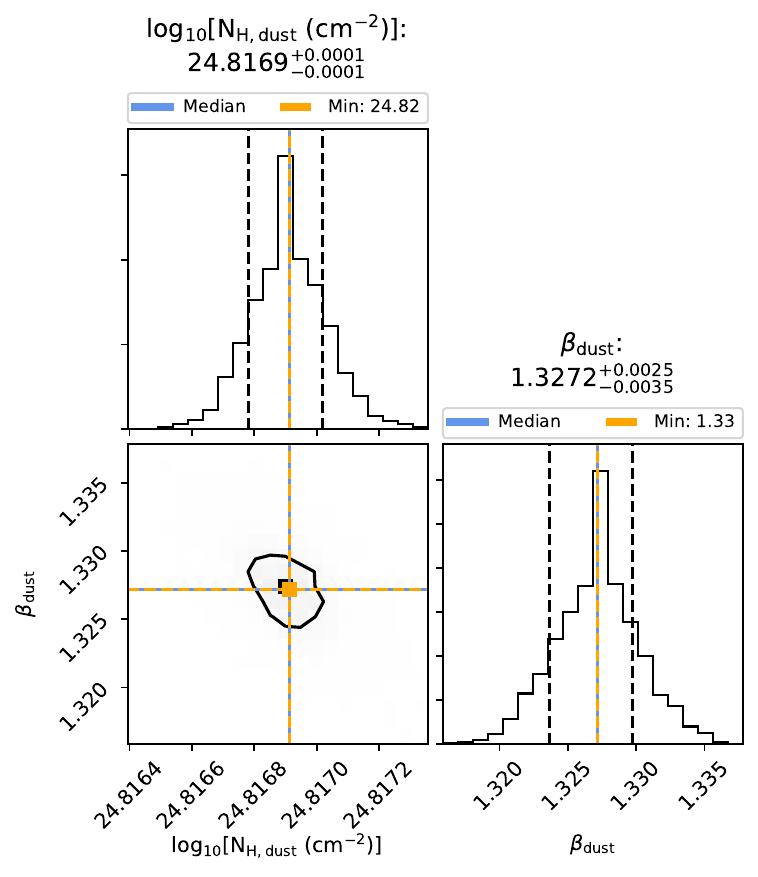}\\
   \caption{Corner plot for source A26 in Sgr~B2(M), core layer.}
   \label{fig:SgrB2-MErrorA26core}
\end{figure*}
\newpage
\clearpage

\begin{figure*}[!htb]
   \centering
   \includegraphics[width=0.50\textwidth]{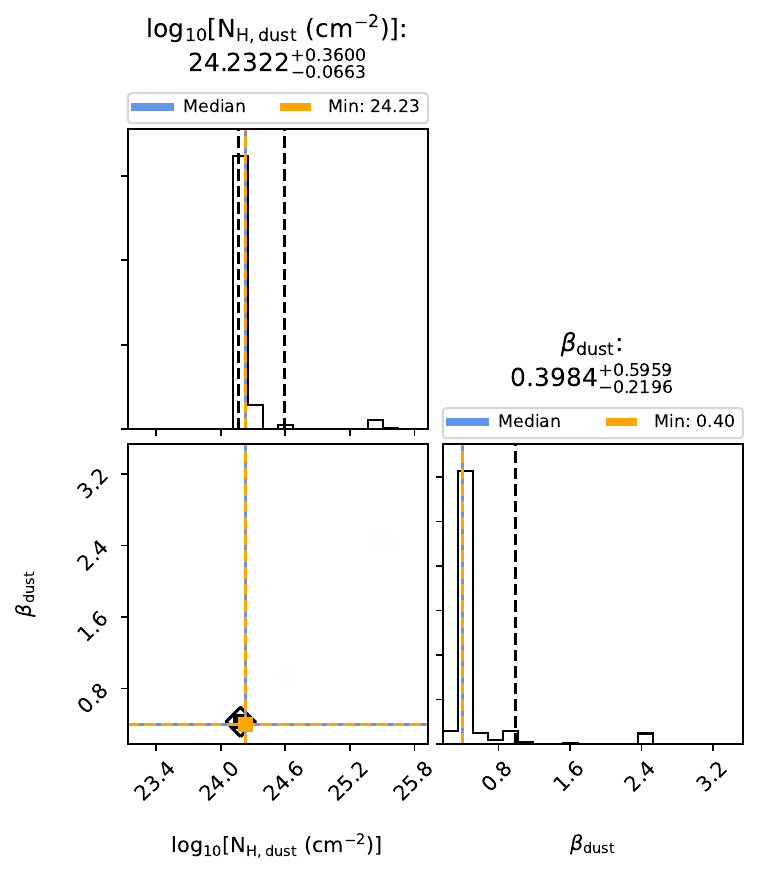}\\
   \caption{Corner plot for source A27 in Sgr~B2(M), envelope layer.}
   \label{fig:SgrB2-MErrorA27env}
\end{figure*}
\newpage
\clearpage

\subsection{Error estimation of continuum parameters for cores in Sgr~B2(N)}\label{app:ErrorSgrB2-N}

Corner plots \citepads{{corner}} showing the one and two dimensional projections of the posterior probability distributions of the continuum parameters of each core in Sgr~B2(N). On top of each column the probability distribution for each free parameter is shown together with the value of the best fit and the corresponding left and right errors. The left and right dashed lines indicate the lower and upper limits of the corresponding highest posterior density (HPD) interval, respectively. The dashed line in the middle indicates the mode of the distribution. The blue lines indicate the parameter values of the best fit. The plots in the lower left corner describe the projected 2D histograms of two parameters and the contours the HPD regions, respectively. In order to get a better estimation of the errors, we determine the error of the hydrogen column density and the emission measure on log scale and use the velocity offset (v$_{\rm off}$) related to the source velocity of v$_{\rm LSR}$ = 64~km~s$^{-1}$.

\begin{figure*}[!htb]
   \centering
   \includegraphics[width=0.50\textwidth]{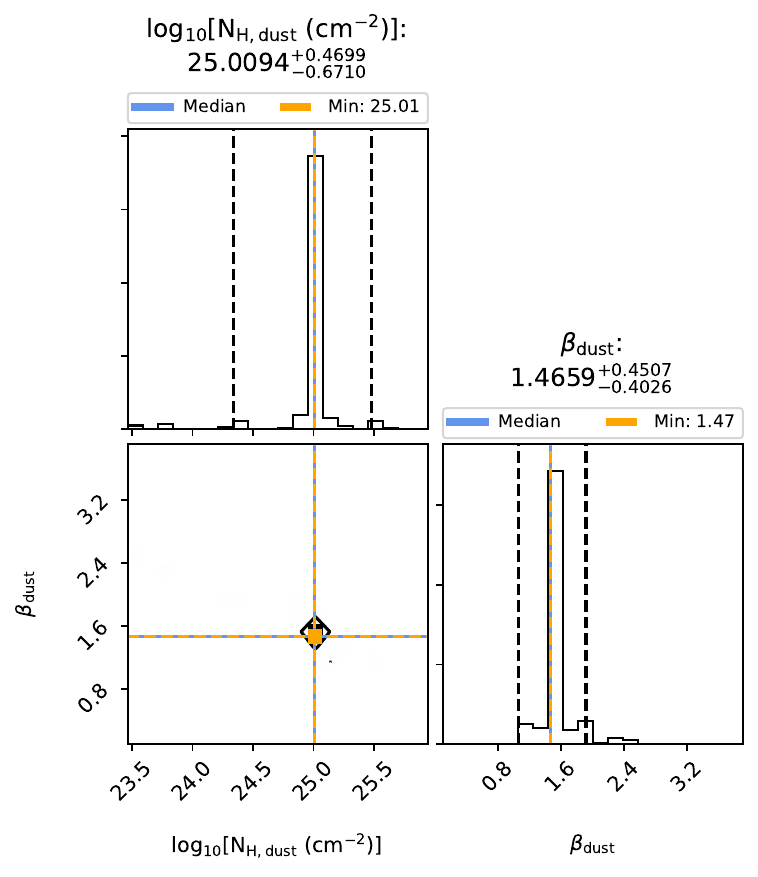}\\
   \caption{Corner plot for source A01 in Sgr~B2(N), envelope layer.}
   \label{fig:SgrB2-NErrorA01env}
\end{figure*}

\begin{figure*}[!htb]
   \centering
   \includegraphics[width=0.99\textwidth]{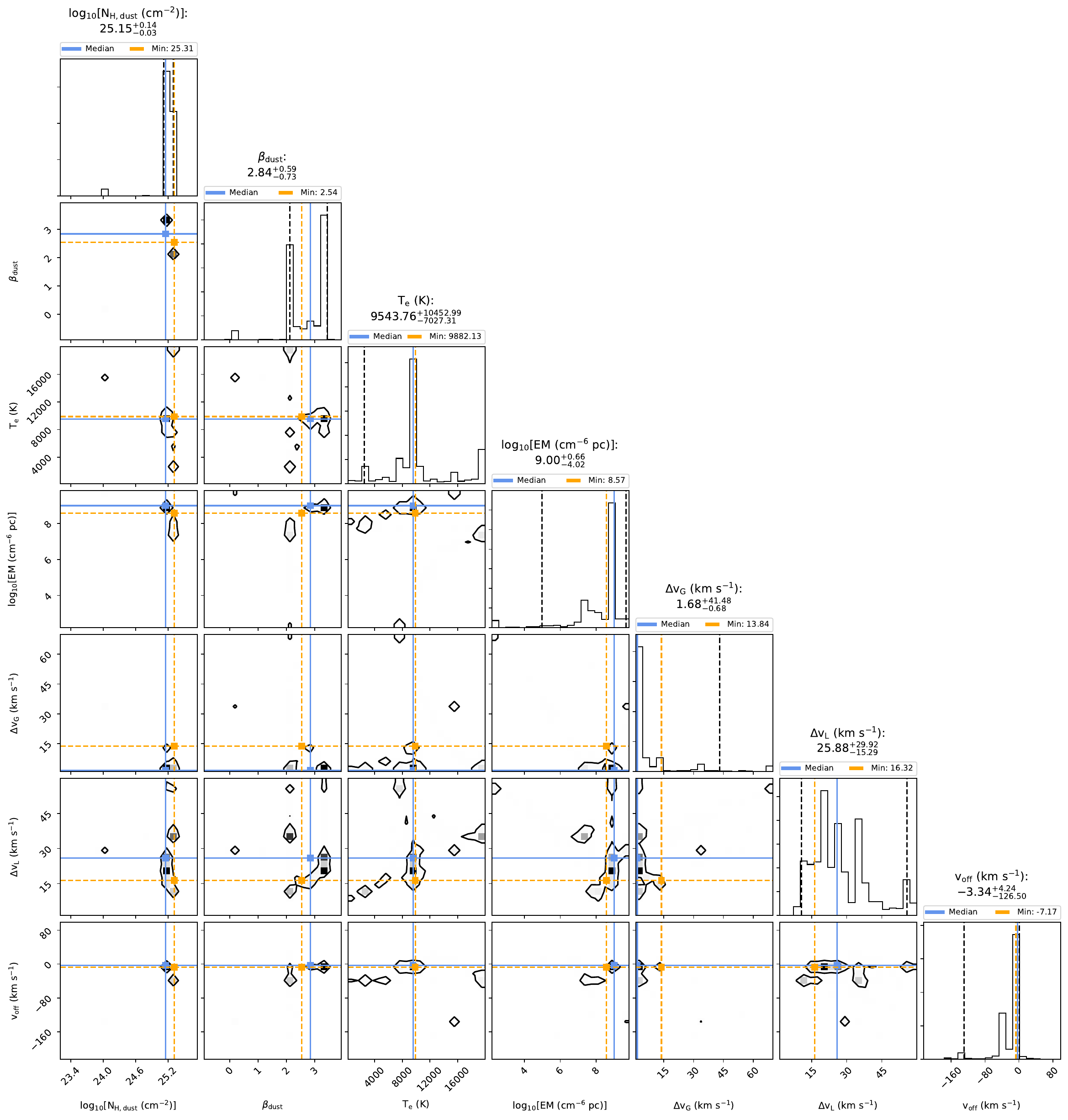}\\
   \caption{Corner plot for source A01 in Sgr~B2(N), core layer.}
   \label{fig:SgrB2-NErrorA01core}
\end{figure*}
\newpage
\clearpage

\begin{figure*}[!htb]
   \centering
   \includegraphics[width=0.50\textwidth]{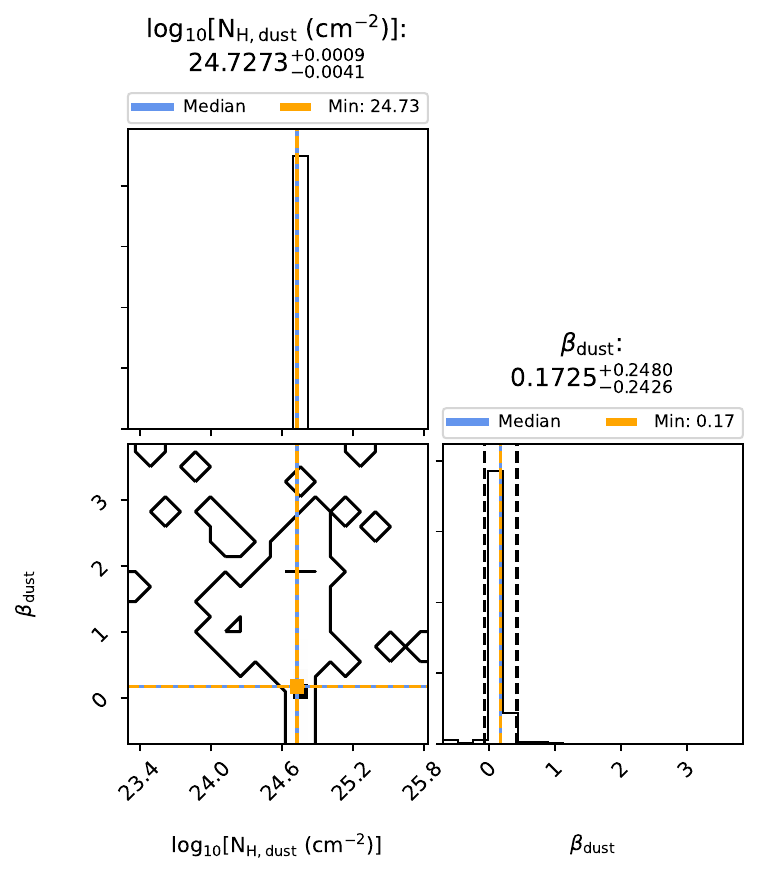}\\
   \caption{Corner plot for source A02 in Sgr~B2(N), envelope layer.}
   \label{fig:SgrB2-NErrorA02env}
\end{figure*}

\begin{figure*}[!htb]
   \centering
   \includegraphics[width=0.50\textwidth]{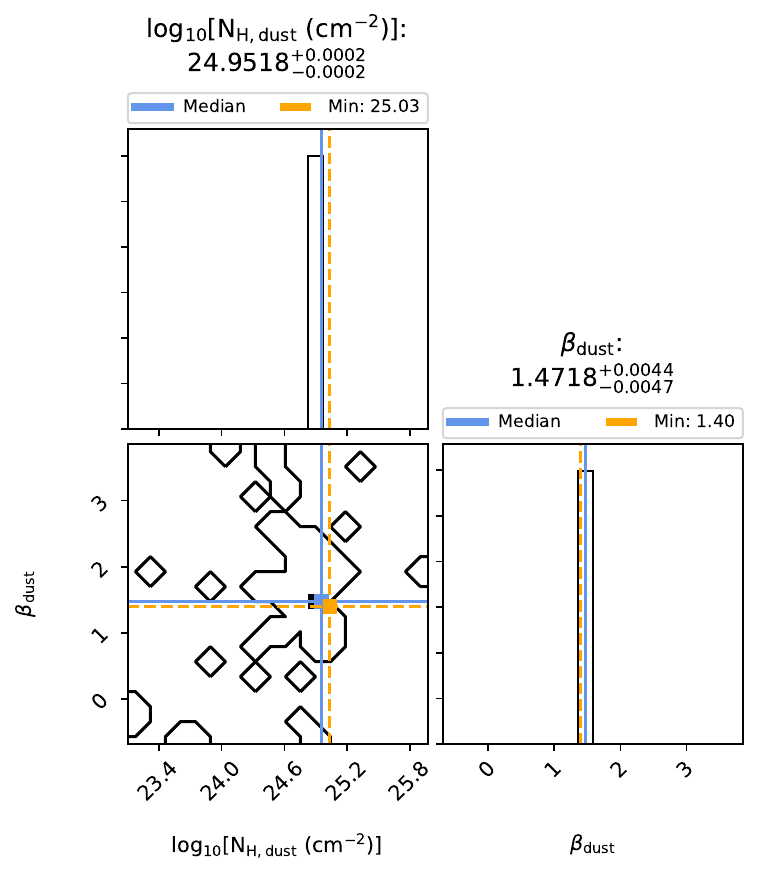}\\
   \caption{Corner plot for source A02 in Sgr~B2(N), core layer.}
   \label{fig:SgrB2-NErrorA02core}
\end{figure*}
\newpage
\clearpage

\begin{figure*}[!htb]
   \centering
   \includegraphics[width=0.50\textwidth]{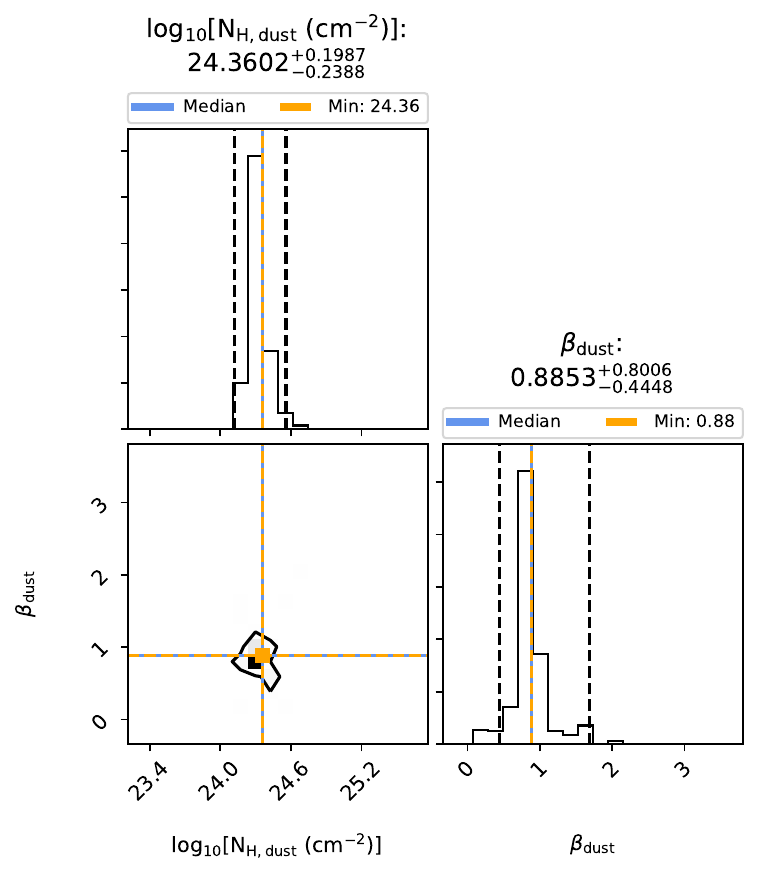}\\
   \caption{Corner plot for source A03 in Sgr~B2(N), envelope layer.}
   \label{fig:SgrB2-NErrorA03env}
\end{figure*}

\begin{figure*}[!htb]
   \centering
   \includegraphics[width=0.50\textwidth]{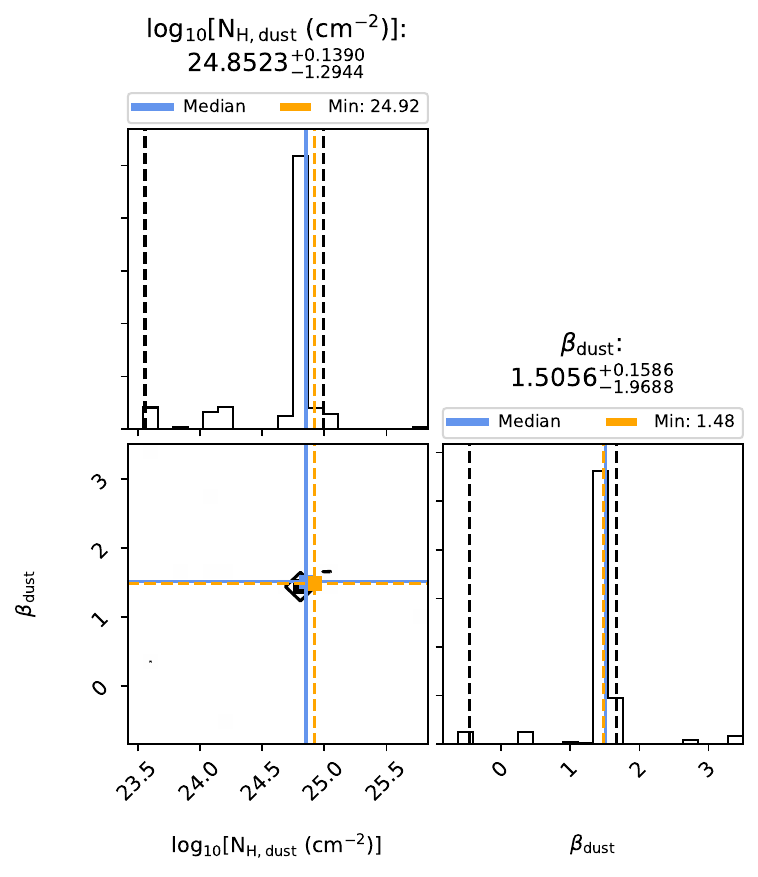}\\
   \caption{Corner plot for source A03 in Sgr~B2(N), core layer.}
   \label{fig:SgrB2-NErrorA03core}
\end{figure*}
\newpage
\clearpage

\begin{figure*}[!htb]
   \centering
   \includegraphics[width=0.50\textwidth]{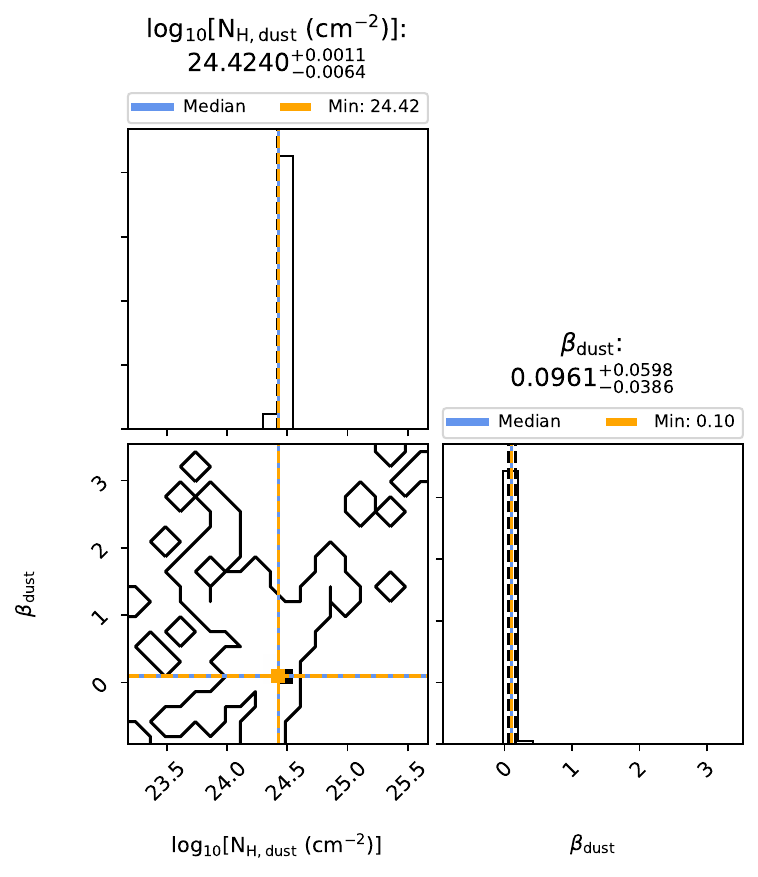}\\
   \caption{Corner plot for source A04 in Sgr~B2(N), envelope layer.}
   \label{fig:SgrB2-NErrorA04env}
\end{figure*}

\begin{figure*}[!htb]
   \centering
   \includegraphics[width=0.50\textwidth]{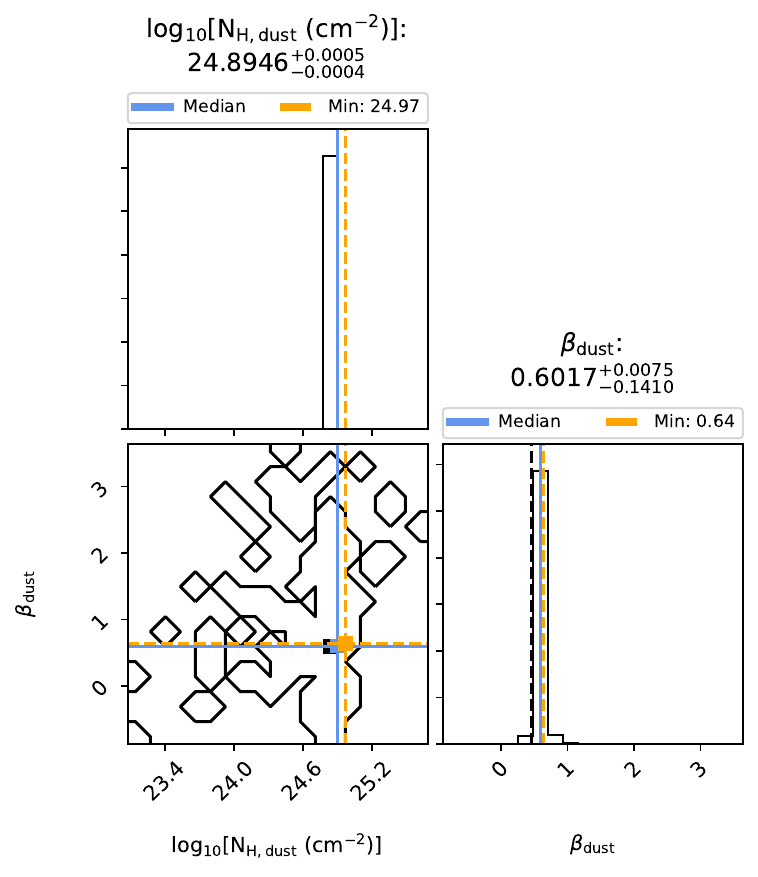}\\
   \caption{Corner plot for source A04 in Sgr~B2(N), core layer.}
   \label{fig:SgrB2-NErrorA04core}
\end{figure*}
\newpage
\clearpage

\begin{figure*}[!htb]
   \centering
   \includegraphics[width=0.50\textwidth]{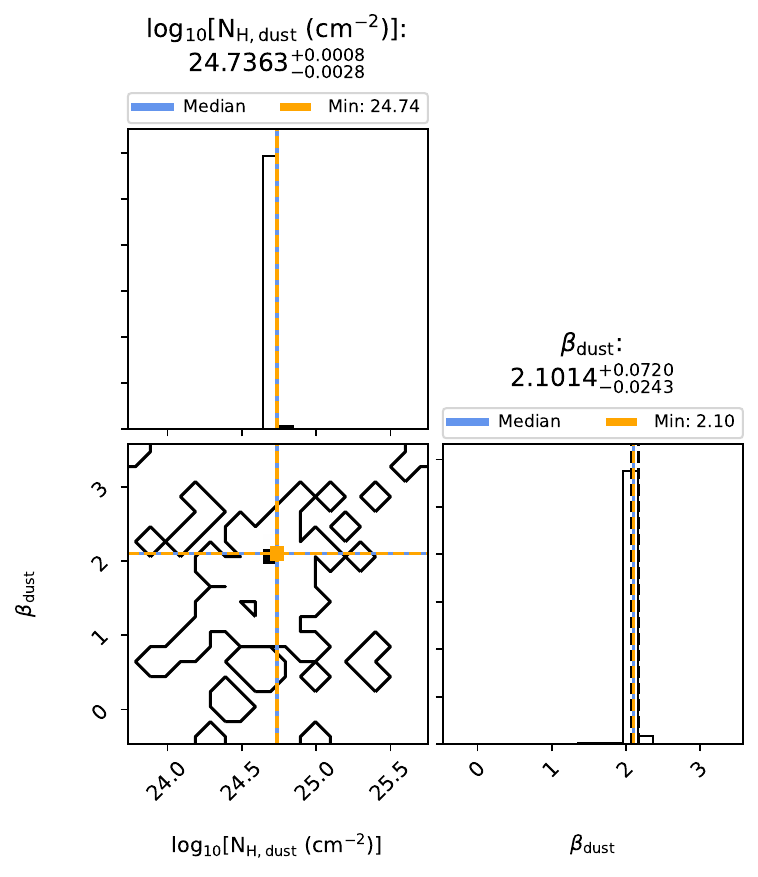}\\
   \caption{Corner plot for source A05 in Sgr~B2(N), envelope layer.}
   \label{fig:SgrB2-NErrorA05env}
\end{figure*}

\begin{figure*}[!htb]
   \centering
   \includegraphics[width=0.50\textwidth]{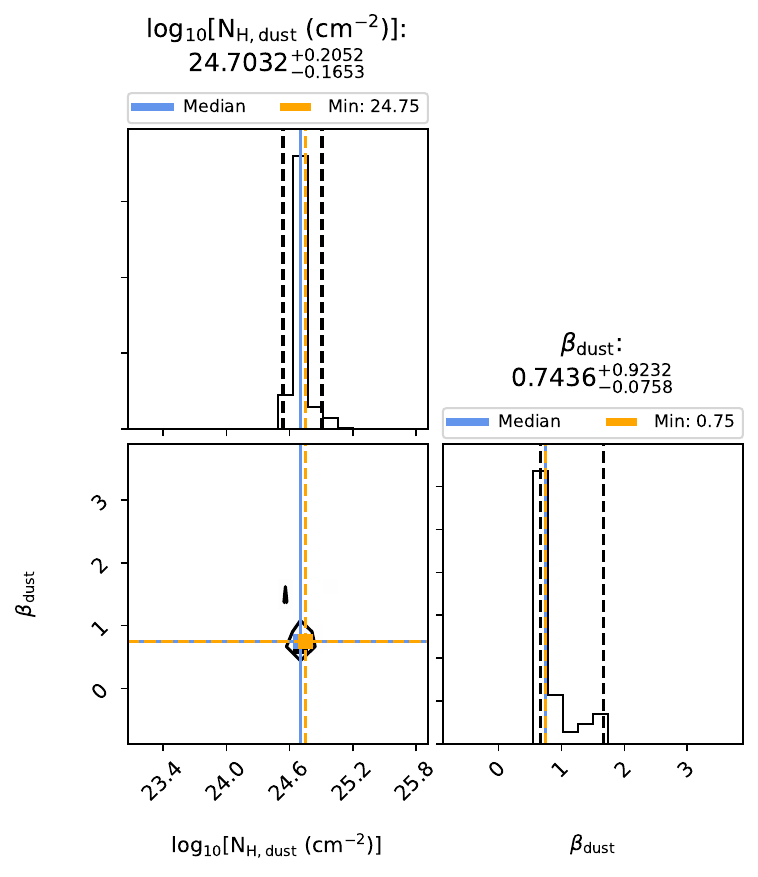}\\
   \caption{Corner plot for source A05 in Sgr~B2(N), core layer.}
   \label{fig:SgrB2-NErrorA05core}
\end{figure*}
\newpage
\clearpage

\begin{figure*}[!htb]
   \centering
   \includegraphics[width=0.50\textwidth]{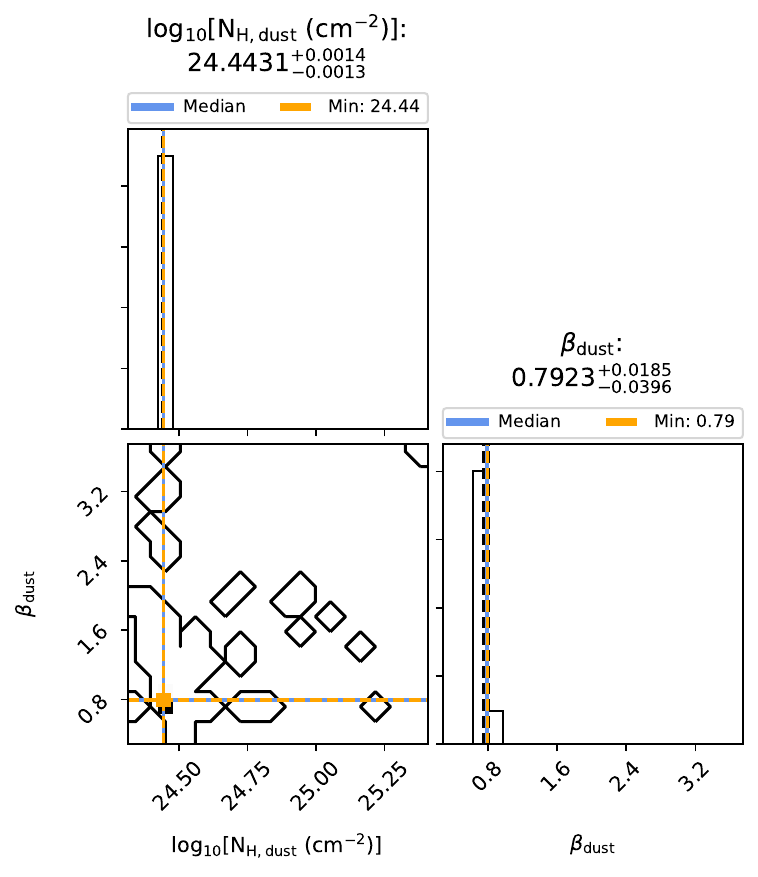}\\
   \caption{Corner plot for source A06 in Sgr~B2(N), envelope layer.}
   \label{fig:SgrB2-NErrorA06env}
\end{figure*}

\begin{figure*}[!htb]
   \centering
   \includegraphics[width=0.50\textwidth]{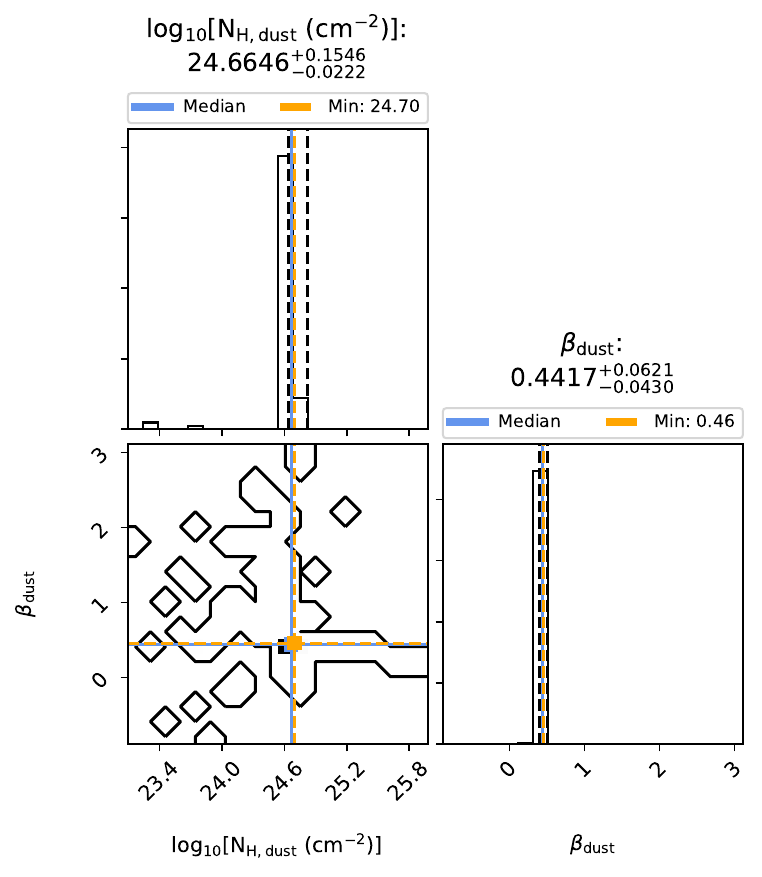}\\
   \caption{Corner plot for source A06 in Sgr~B2(N), core layer.}
   \label{fig:SgrB2-NErrorA06core}
\end{figure*}
\newpage
\clearpage

\begin{figure*}[!htb]
   \centering
   \includegraphics[width=0.50\textwidth]{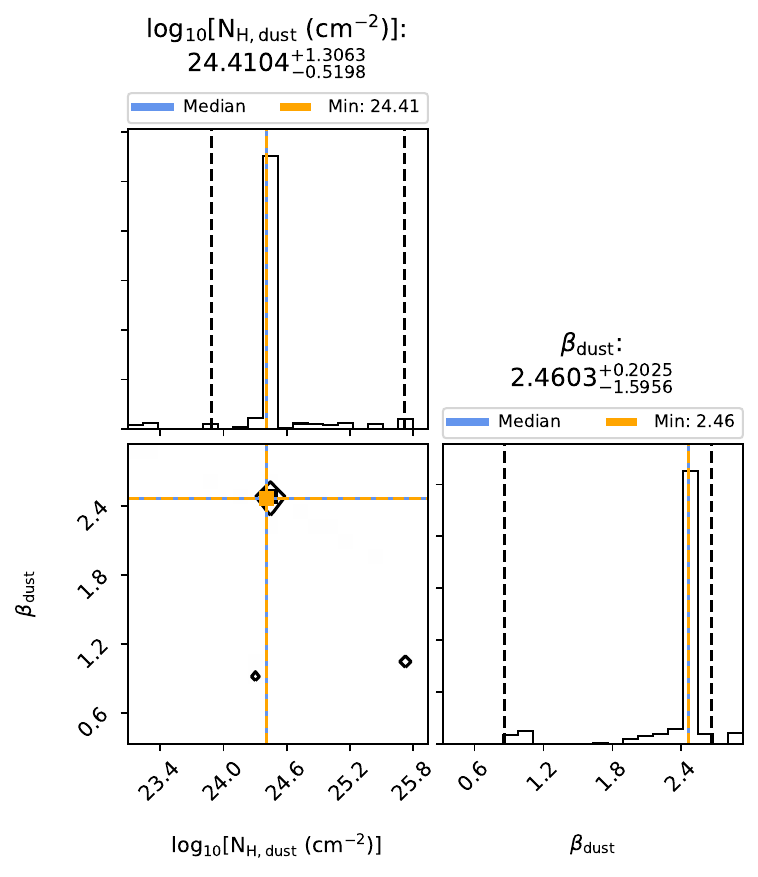}\\
   \caption{Corner plot for source A07 in Sgr~B2(N), envelope layer.}
   \label{fig:SgrB2-NErrorA07env}
\end{figure*}

\begin{figure*}[!htb]
   \centering
   \includegraphics[width=0.50\textwidth]{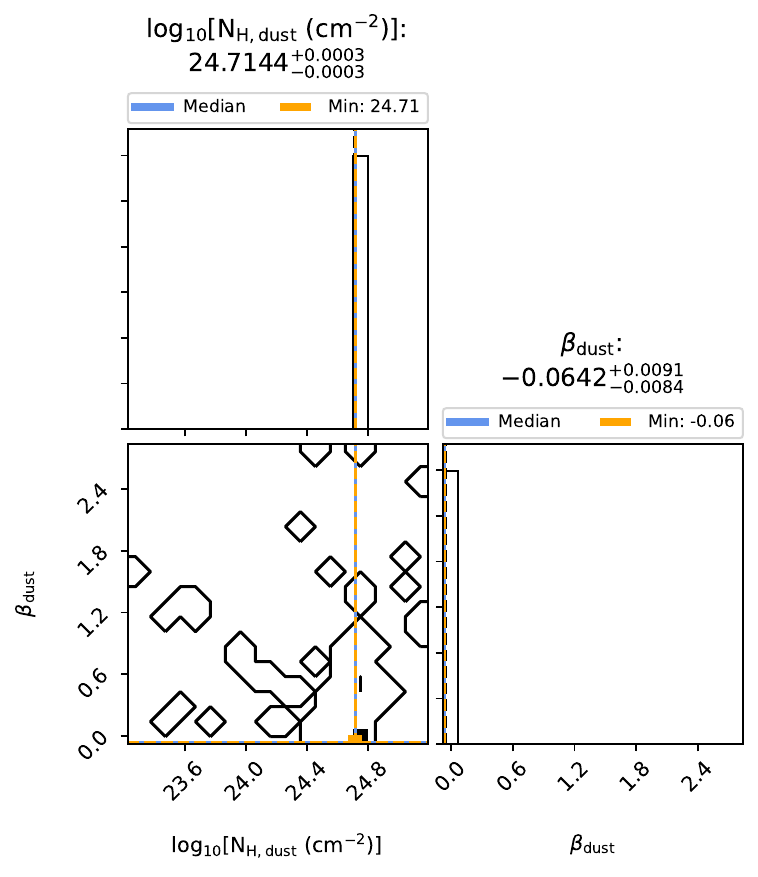}\\
   \caption{Corner plot for source A07 in Sgr~B2(N), core layer.}
   \label{fig:SgrB2-NErrorA07core}
\end{figure*}
\newpage
\clearpage

\begin{figure*}[!htb]
   \centering
   \includegraphics[width=0.50\textwidth]{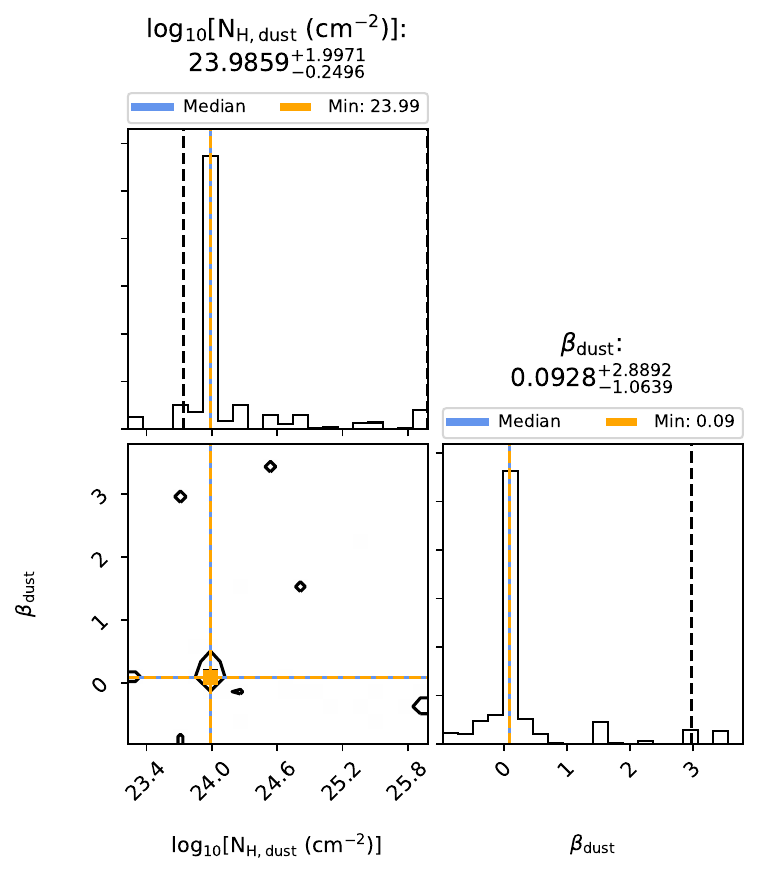}\\
   \caption{Corner plot for source A08 in Sgr~B2(N), envelope layer.}
   \label{fig:SgrB2-NErrorA08env}
\end{figure*}
\newpage
\clearpage

\begin{figure*}[!htb]
   \centering
   \includegraphics[width=0.50\textwidth]{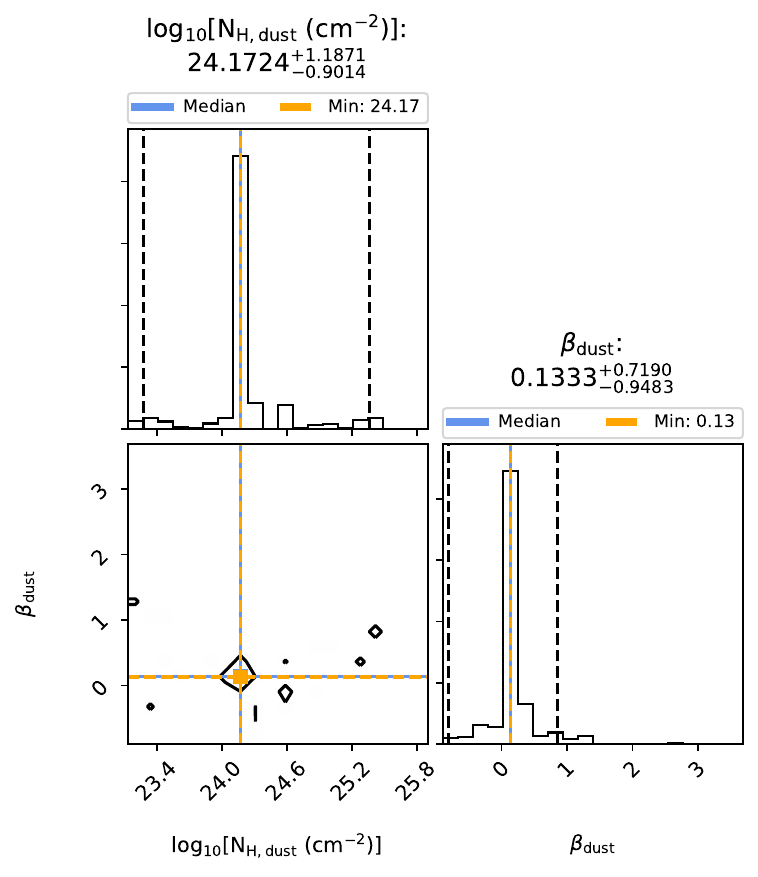}\\
   \caption{Corner plot for source A09 in Sgr~B2(N), envelope layer.}
   \label{fig:SgrB2-NErrorA09env}
\end{figure*}

\begin{figure*}[!htb]
   \centering
   \includegraphics[width=0.50\textwidth]{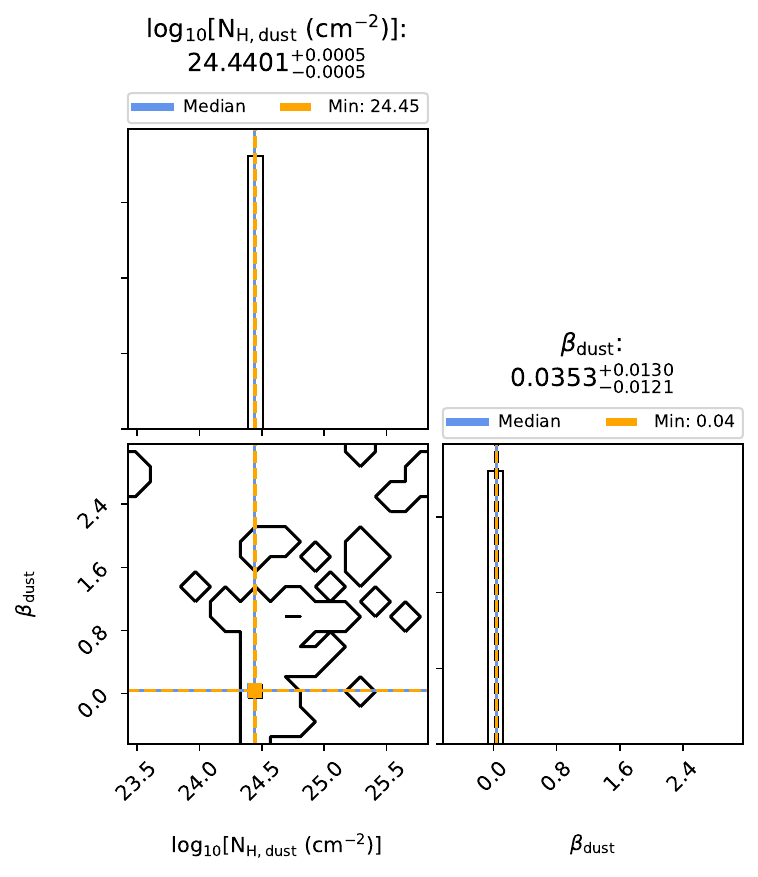}\\
   \caption{Corner plot for source A09 in Sgr~B2(N), core layer.}
   \label{fig:SgrB2-NErrorA09core}
\end{figure*}
\newpage
\clearpage

\begin{figure*}[!htb]
   \centering
   \includegraphics[width=0.50\textwidth]{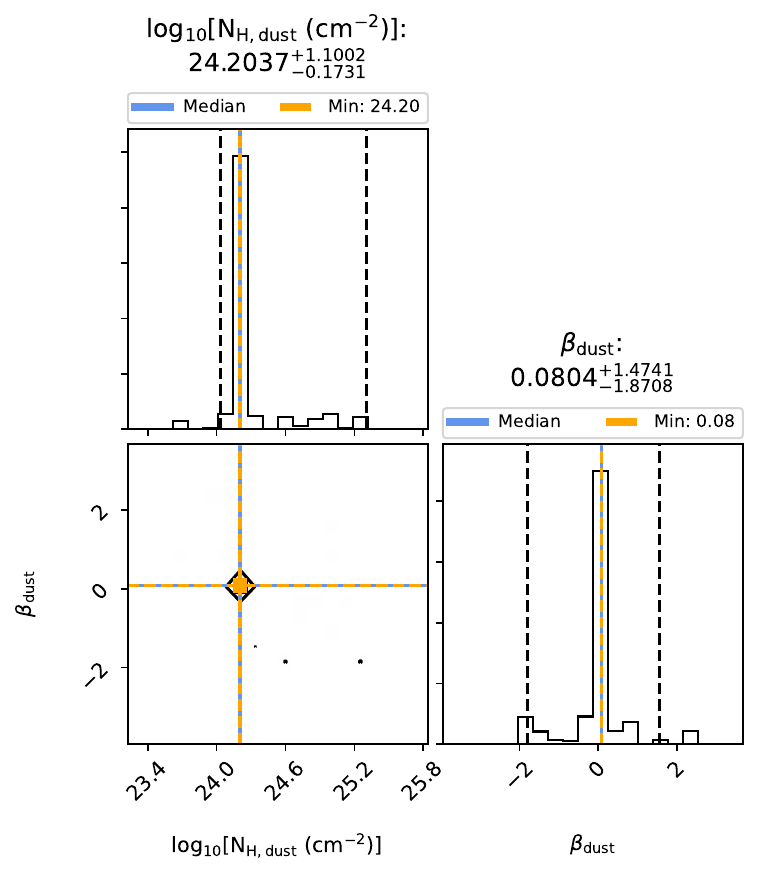}\\
   \caption{Corner plot for source A10 in Sgr~B2(N), envelope layer.}
   \label{fig:SgrB2-NErrorA10env}
\end{figure*}
\newpage
\clearpage

\begin{figure*}[!htb]
   \centering
   \includegraphics[width=0.50\textwidth]{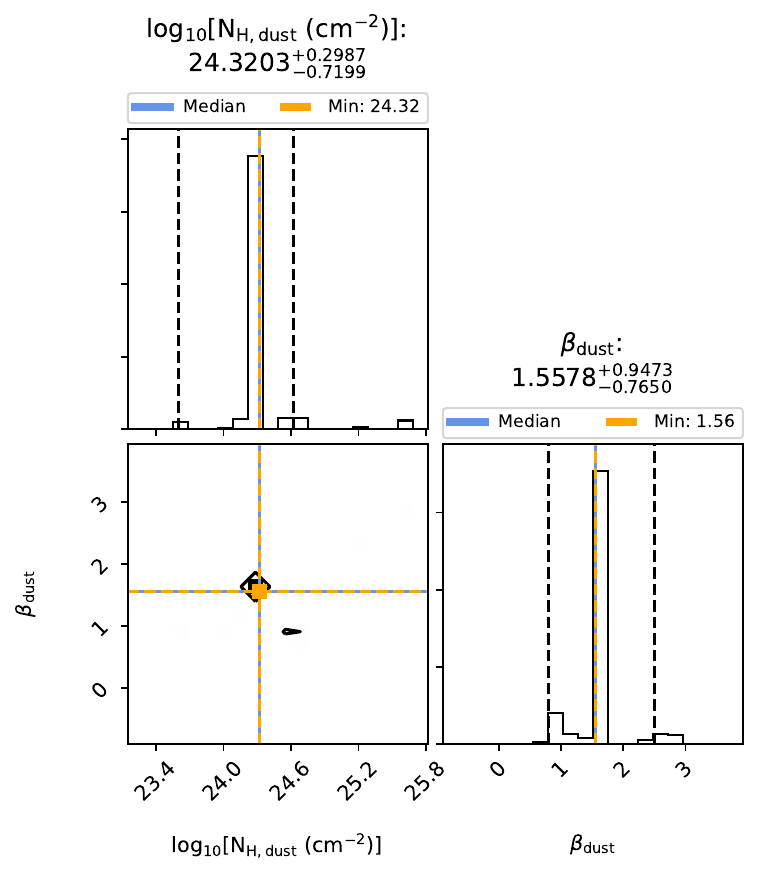}\\
   \caption{Corner plot for source A11 in Sgr~B2(N), envelope layer.}
   \label{fig:SgrB2-NErrorA11env}
\end{figure*}
\newpage
\clearpage

\begin{figure*}[!htb]
   \centering
   \includegraphics[width=0.50\textwidth]{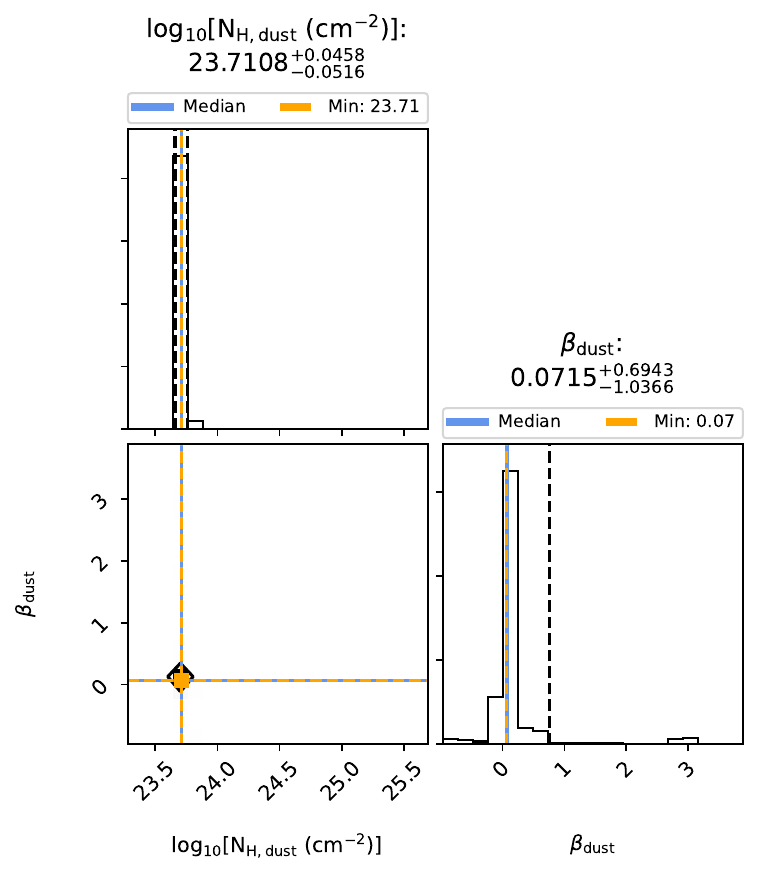}\\
   \caption{Corner plot for source A12 in Sgr~B2(N), envelope layer.}
   \label{fig:SgrB2-NErrorA12env}
\end{figure*}
\newpage
\clearpage

\begin{figure*}[!htb]
   \centering
   \includegraphics[width=0.50\textwidth]{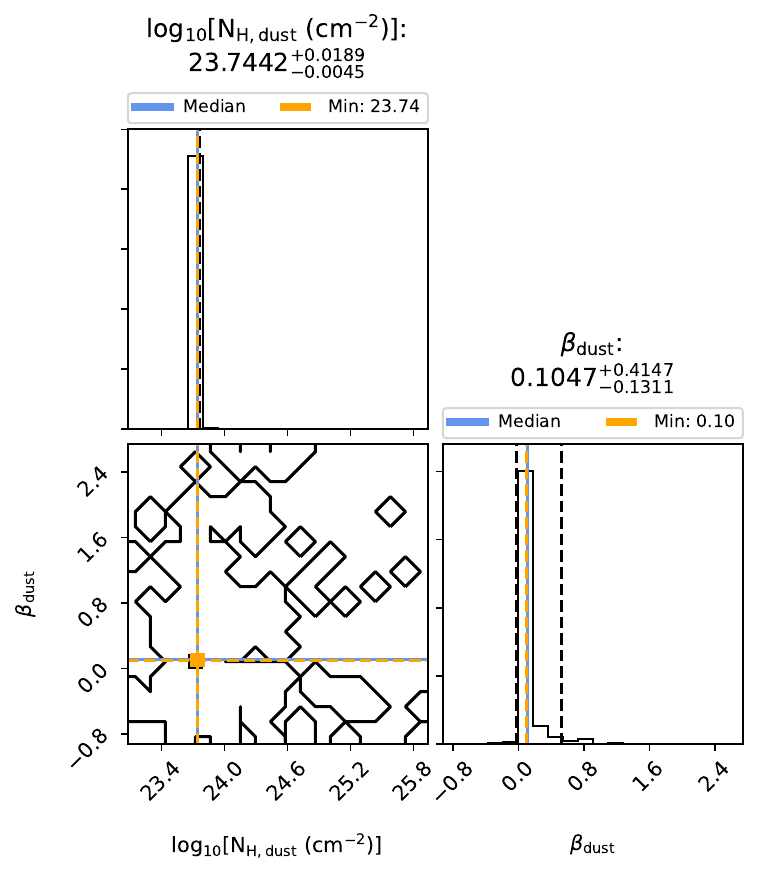}\\
   \caption{Corner plot for source A13 in Sgr~B2(N), envelope layer.}
   \label{fig:SgrB2-NErrorA13env}
\end{figure*}
\newpage
\clearpage

\begin{figure*}[!htb]
   \centering
   \includegraphics[width=0.50\textwidth]{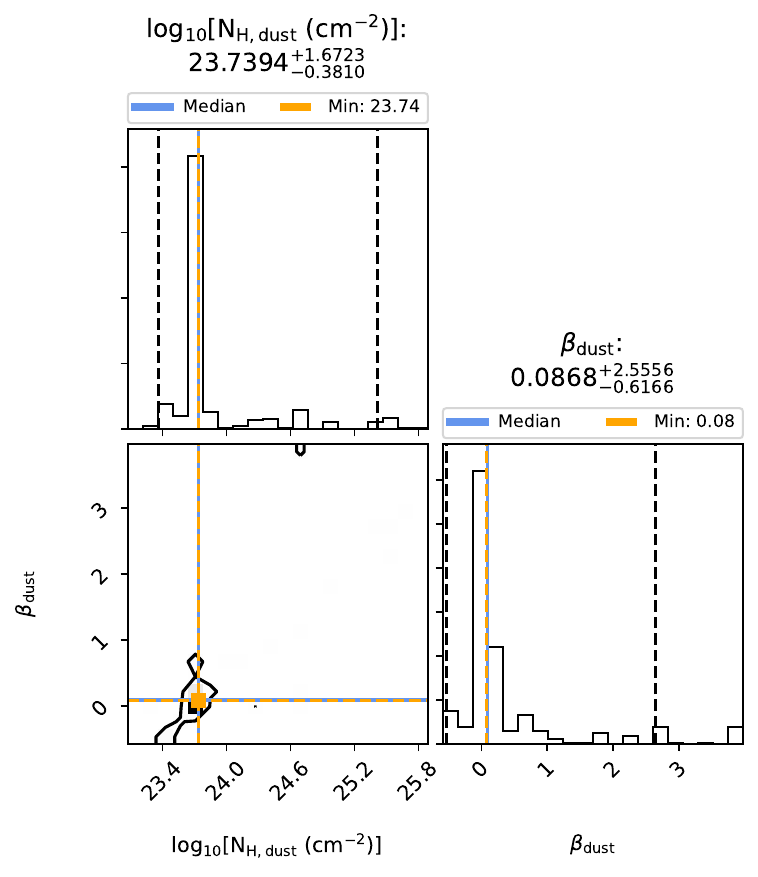}\\
   \caption{Corner plot for source A14 in Sgr~B2(N), envelope layer.}
   \label{fig:SgrB2-NErrorA14env}
\end{figure*}

\begin{figure*}[!htb]
   \centering
   \includegraphics[width=0.50\textwidth]{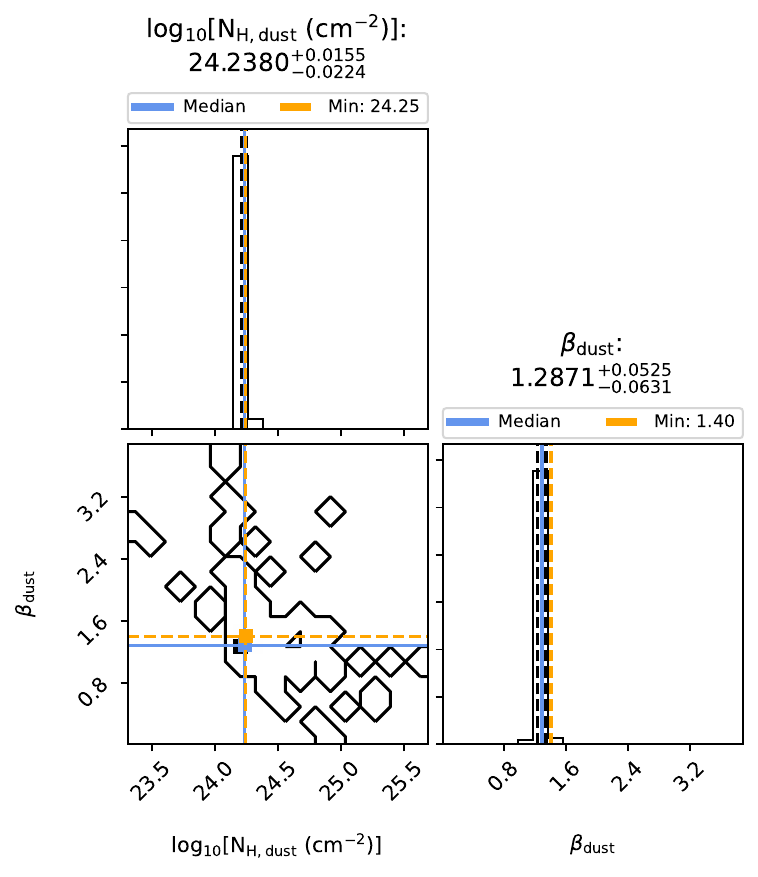}\\
   \caption{Corner plot for source A14 in Sgr~B2(N), core layer.}
   \label{fig:SgrB2-NErrorA14core}
\end{figure*}
\newpage
\clearpage

\begin{figure*}[!htb]
   \centering
   \includegraphics[width=0.50\textwidth]{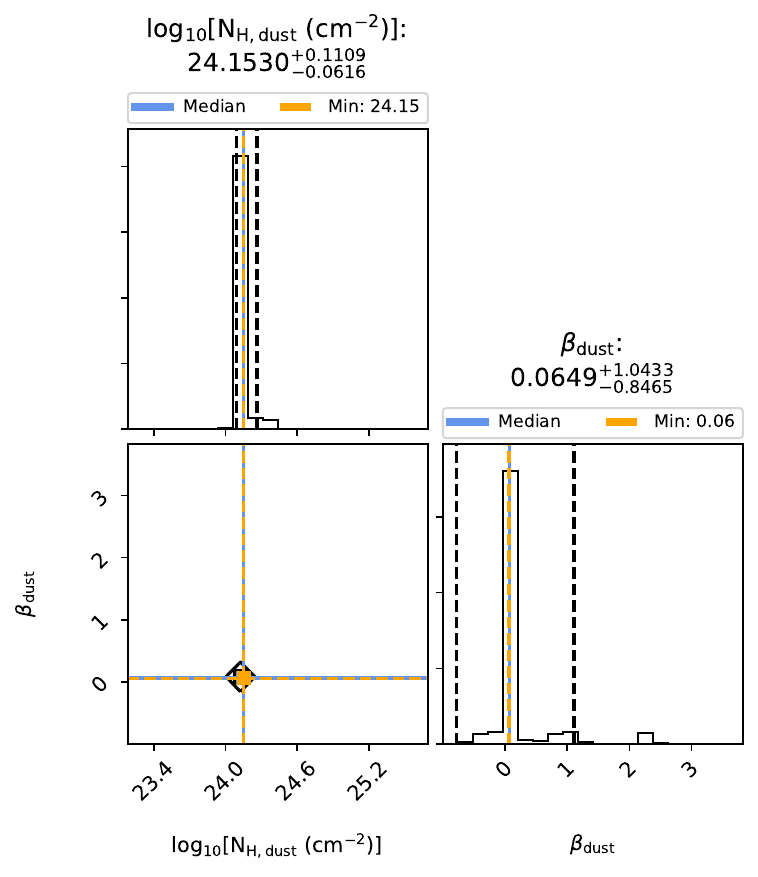}\\
   \caption{Corner plot for source A15 in Sgr~B2(N), envelope layer.}
   \label{fig:SgrB2-NErrorA15env}
\end{figure*}

\begin{figure*}[!htb]
   \centering
   \includegraphics[width=0.50\textwidth]{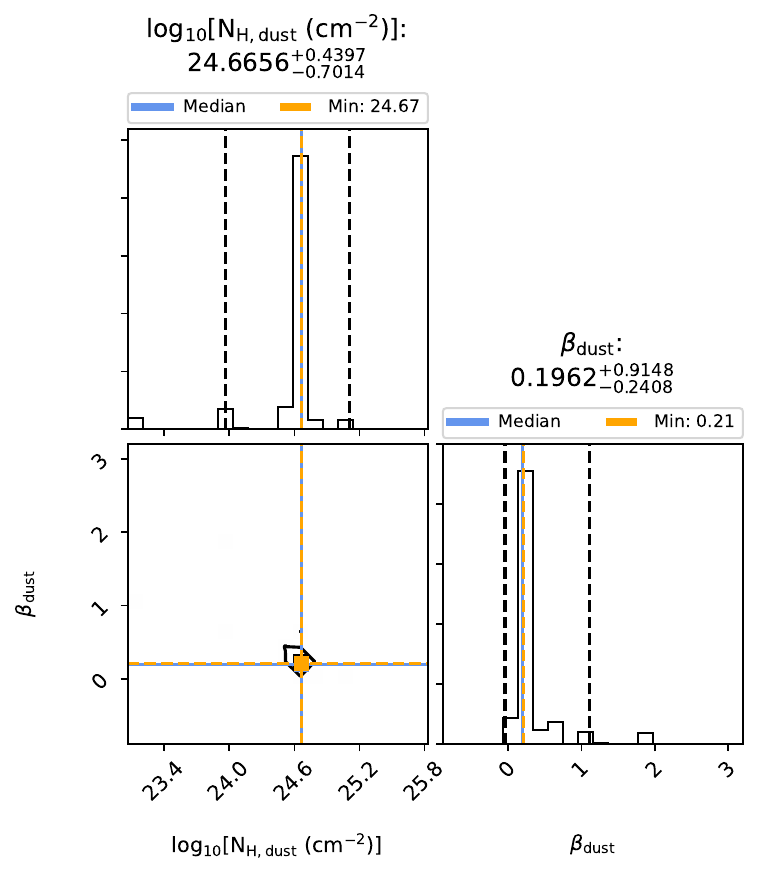}\\
   \caption{Corner plot for source A15 in Sgr~B2(N), core layer.}
   \label{fig:SgrB2-NErrorA15core}
\end{figure*}
\newpage
\clearpage

\begin{figure*}[!htb]
   \centering
   \includegraphics[width=0.50\textwidth]{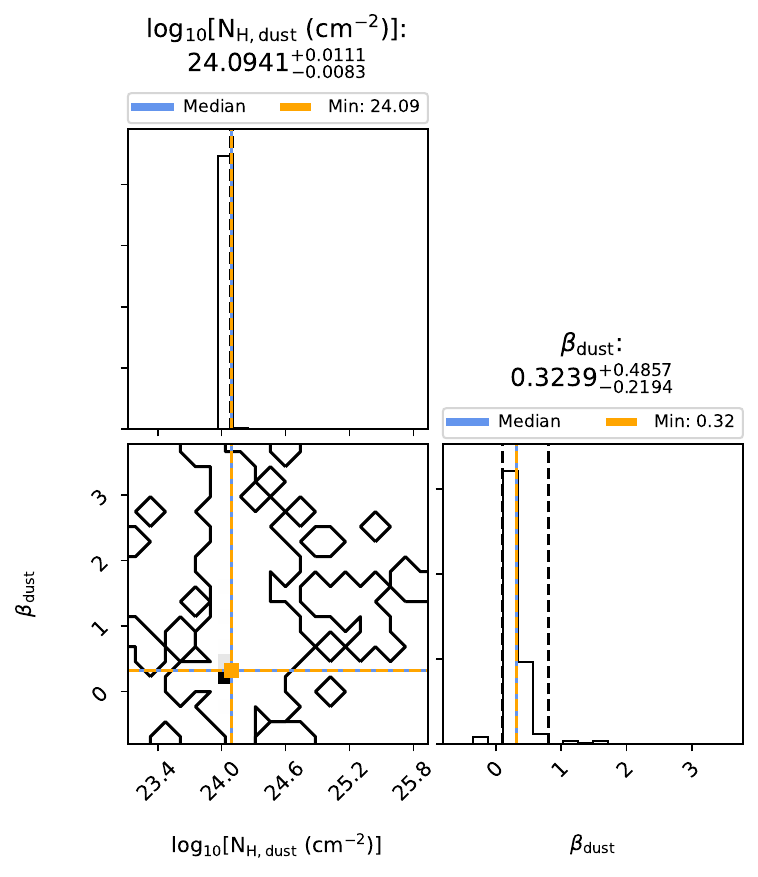}\\
   \caption{Corner plot for source A16 in Sgr~B2(N), envelope layer.}
   \label{fig:SgrB2-NErrorA16env}
\end{figure*}

\begin{figure*}[!htb]
   \centering
   \includegraphics[width=0.50\textwidth]{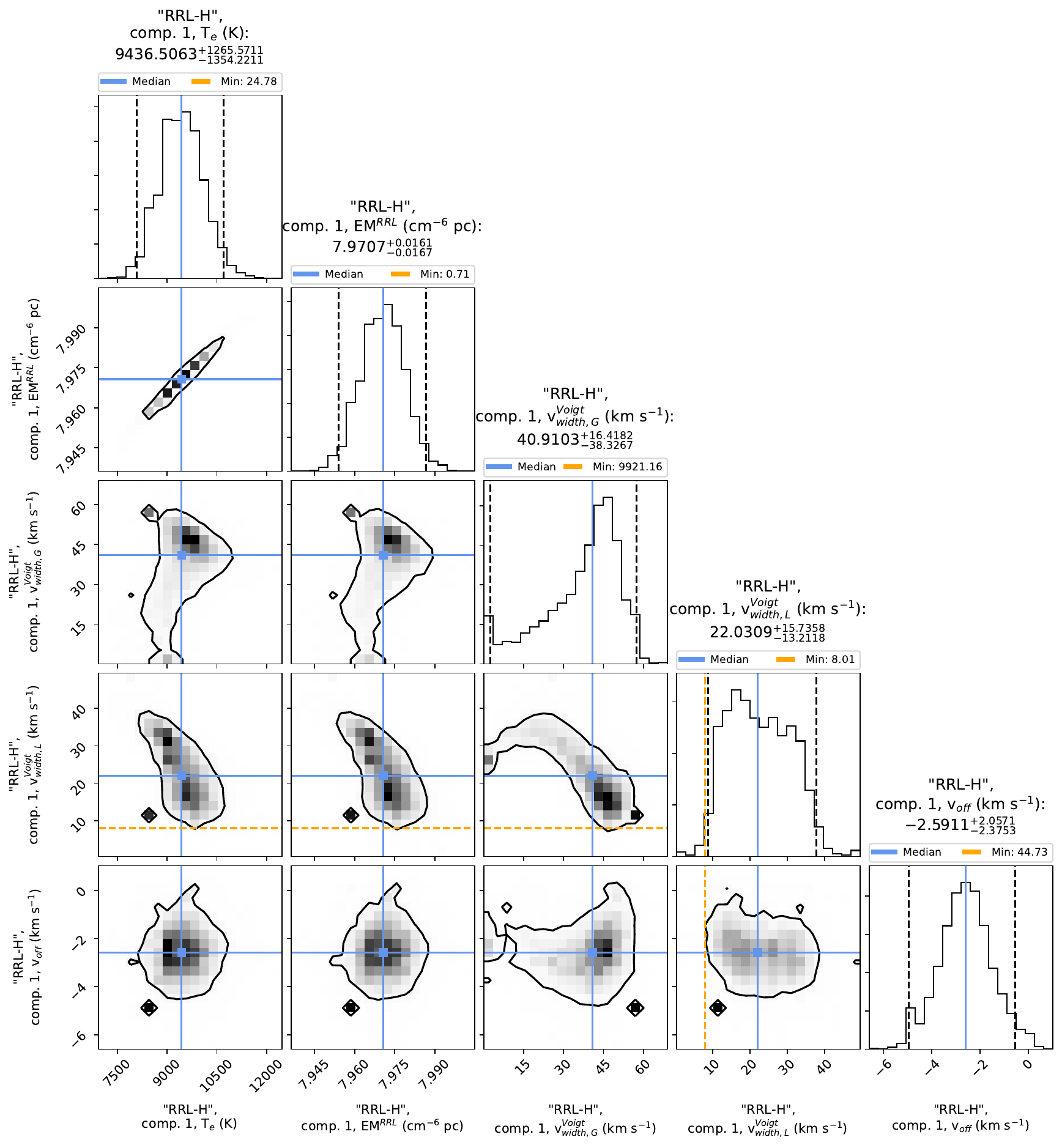}\\
   \caption{Corner plot for source A16 in Sgr~B2(N), core layer.}
   \label{fig:SgrB2-NErrorA16core}
\end{figure*}
\newpage
\clearpage

\begin{figure*}[!htb]
   \centering
   \includegraphics[width=0.50\textwidth]{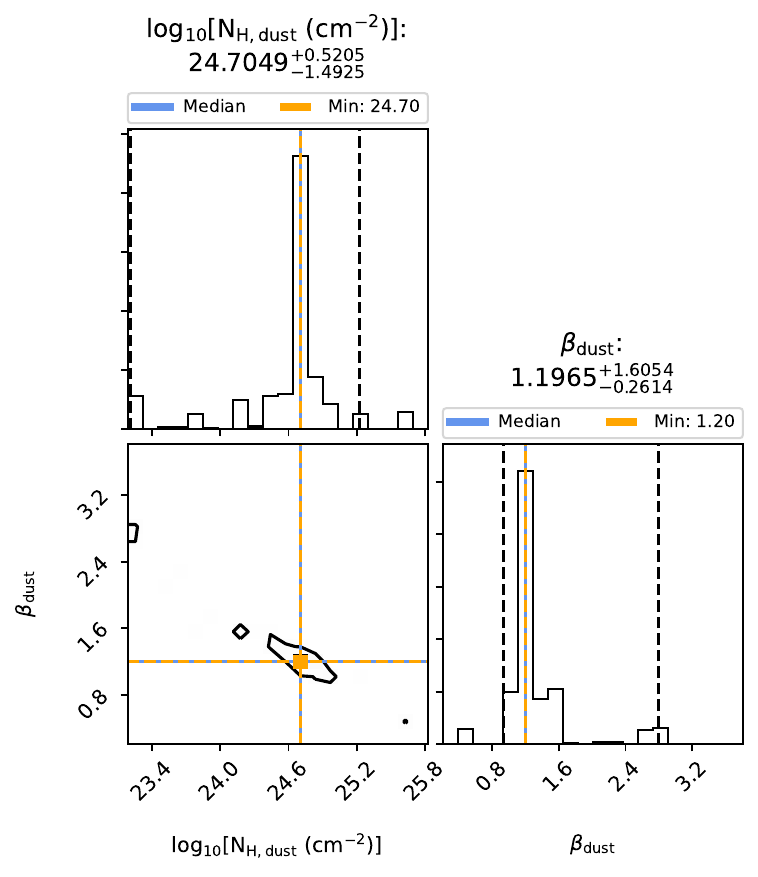}\\
   \caption{Corner plot for source A17 in Sgr~B2(N), envelope layer.}
   \label{fig:SgrB2-NErrorA17env}
\end{figure*}

\begin{figure*}[!htb]
   \centering
   \includegraphics[width=0.50\textwidth]{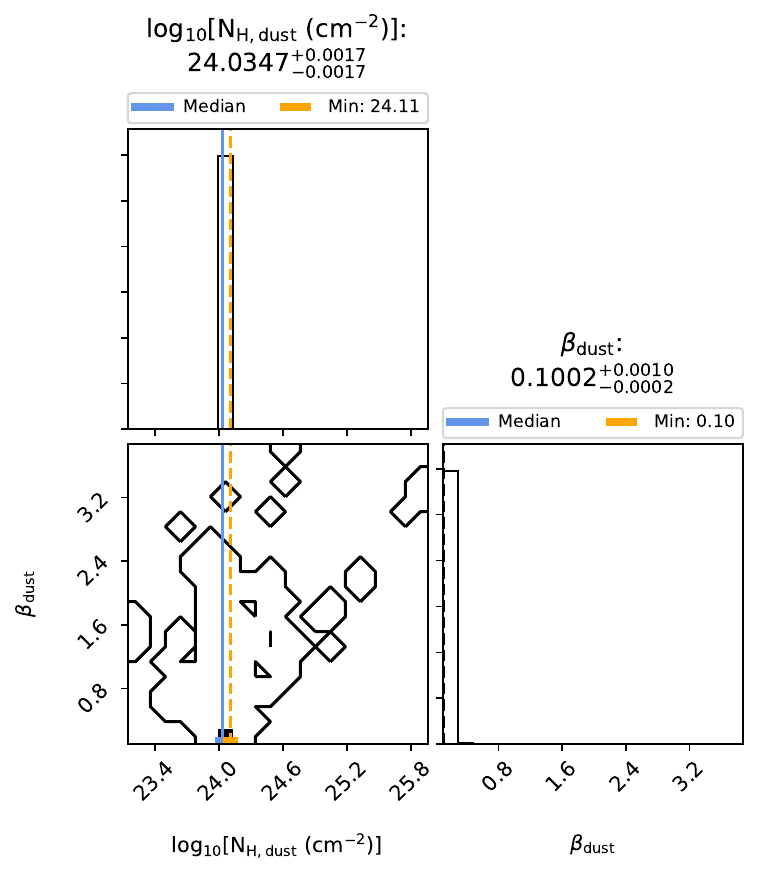}\\
   \caption{Corner plot for source A17 in Sgr~B2(N), core layer.}
   \label{fig:SgrB2-NErrorA17core}
\end{figure*}
\newpage
\clearpage

\begin{figure*}[!htb]
   \centering
   \includegraphics[width=0.50\textwidth]{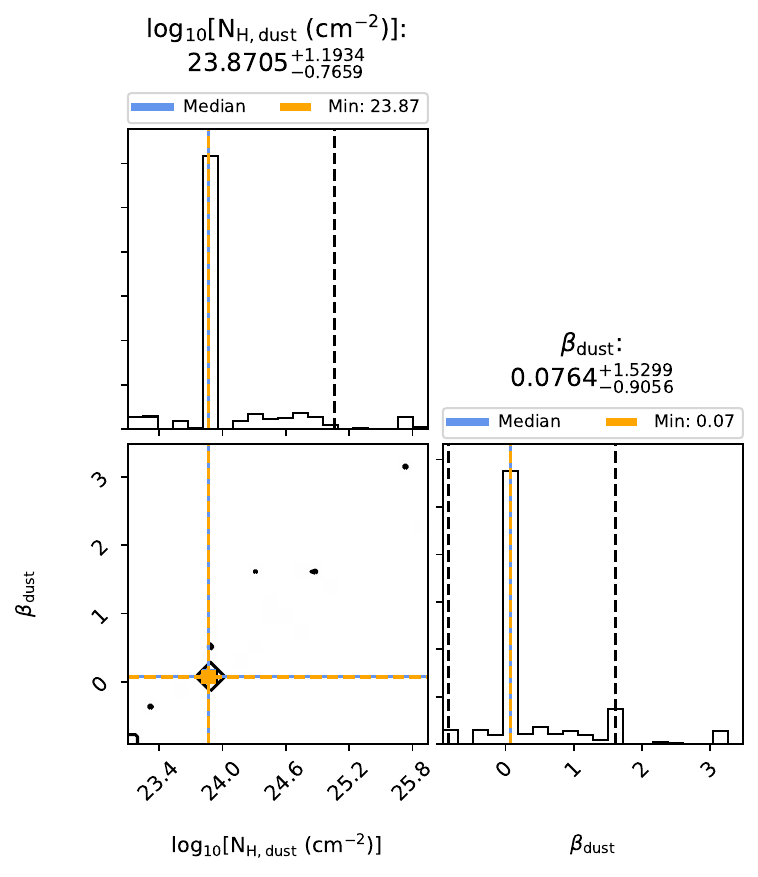}\\
   \caption{Corner plot for source A18 in Sgr~B2(N), envelope layer.}
   \label{fig:SgrB2-NErrorA18env}
\end{figure*}

\begin{figure*}[!htb]
   \centering
   \includegraphics[width=0.50\textwidth]{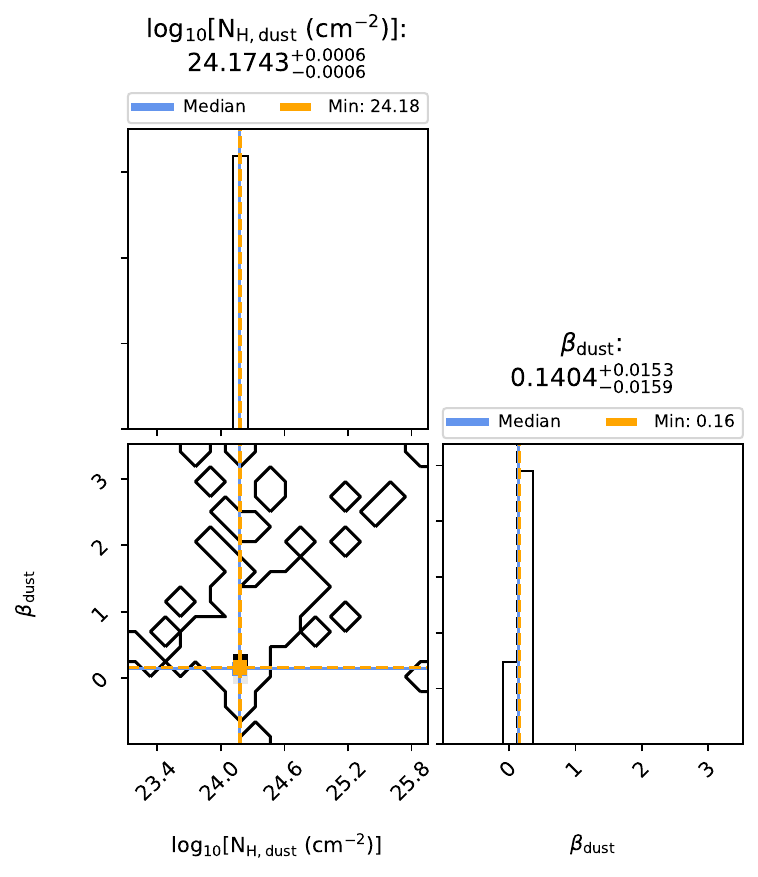}\\
   \caption{Corner plot for source A18 in Sgr~B2(N), core layer.}
   \label{fig:SgrB2-NErrorA18core}
\end{figure*}
\newpage
\clearpage

\begin{figure*}[!htb]
   \centering
   \includegraphics[width=0.50\textwidth]{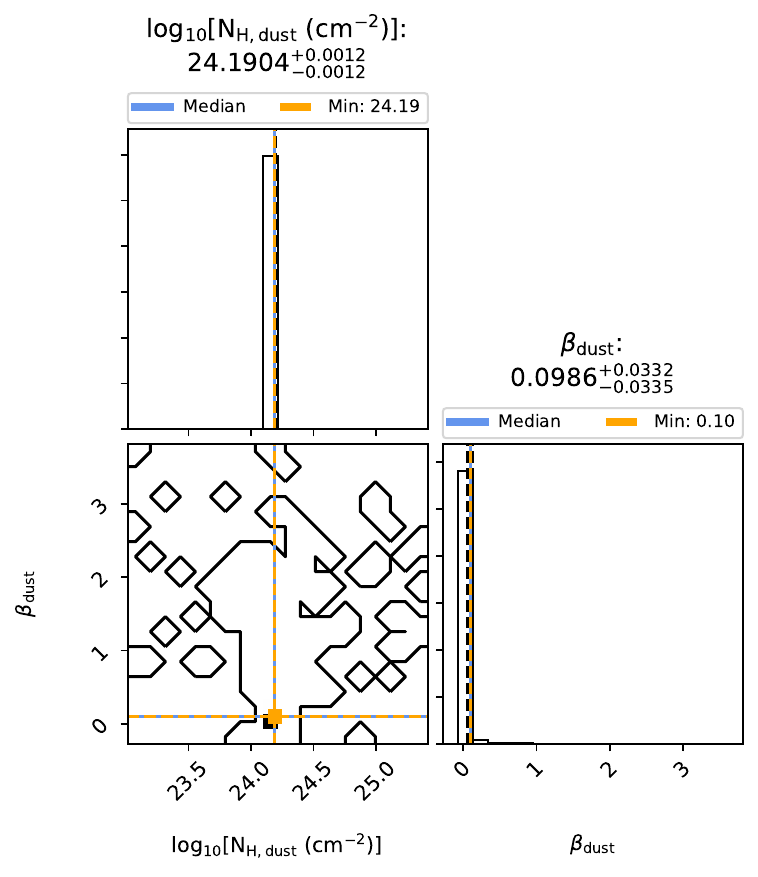}\\
   \caption{Corner plot for source A19 in Sgr~B2(N), envelope layer.}
   \label{fig:SgrB2-NErrorA19env}
\end{figure*}

\begin{figure*}[!htb]
   \centering
   \includegraphics[width=0.50\textwidth]{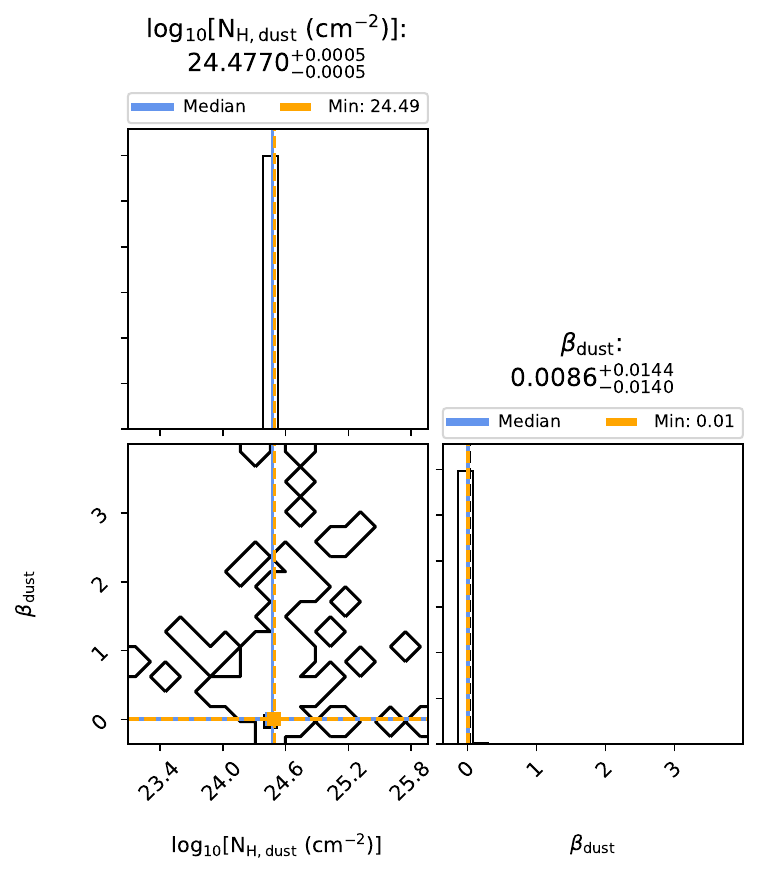}\\
   \caption{Corner plot for source A19 in Sgr~B2(N), core layer.}
   \label{fig:SgrB2-NErrorA19core}
\end{figure*}
\newpage
\clearpage

\begin{figure*}[!htb]
   \centering
   \includegraphics[width=0.50\textwidth]{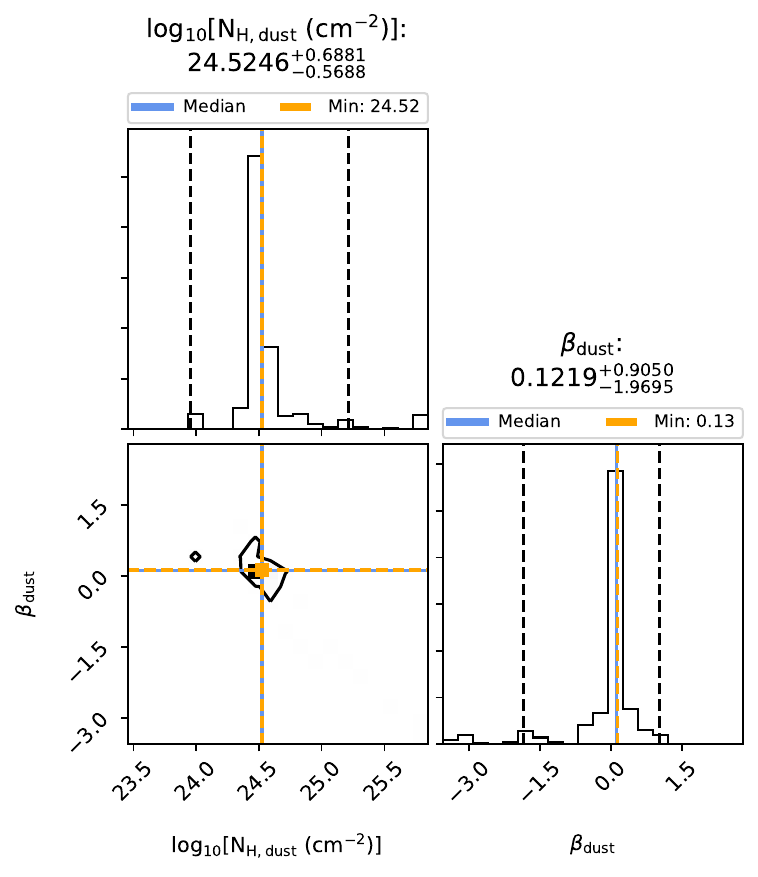}\\
   \caption{Corner plot for source A20 in Sgr~B2(N), envelope layer.}
   \label{fig:SgrB2-NErrorA20env}
\end{figure*}
\newpage
\clearpage

    \end{appendix}
\end{document}